\documentclass[sn-basic]{sn-jnl}% Basic Springer Nature Reference Style/Chemistry Reference Style

\usepackage{graphicx}%
\usepackage{multirow}%
\usepackage{amsmath,amssymb,amsfonts}%
\usepackage{amsthm}%
\usepackage{mathrsfs}%
\usepackage[title]{appendix}%
\usepackage{xcolor}%
\usepackage{textcomp}%
\usepackage{manyfoot}%
\usepackage{booktabs}%
\usepackage{algorithm}%
\usepackage{algorithmicx}%
\usepackage{algpseudocode}%
\usepackage{listings}%
\usepackage{xltabular}
\usepackage{makecell}

\newcommand*\rot{\rotatebox{90}}
\usepackage[]{mdframed}
\usepackage{lscape}
\usepackage[table]{xcolor}
\usepackage{lmodern}
\usepackage{placeins}
\usepackage{diagbox}
\usepackage{slashbox}

\mdfsetup{nobreak=true}
\newcommand{\rev}[1]{{#1}}
\newcommand{\revv}[1]{{#1}}

\newcommand{\perez}{Rodr{\'i}guez-P{\'e}rez et al.}

\theoremstyle{thmstyleone}%
\theoremstyle{thmstyletwo}%

\theoremstyle{thmstylethree}%

\begin{document}

% \title[Article Title]{An exploratory study of bug-introducing changes: what happens when bugs are introduced in open source software?}
\title[Article Title]{\rev{An Exploratory Study of Bug-Introducing Changes: Exploring Relationships in Bug-Introducing Changes Towards Causal Understanding}}

%%=============================================================%%
%% GivenName	-> \fnm{Joergen W.}
%% Particle	-> \spfx{van der} -> surname prefix
%% FamilyName	-> \sur{Ploeg}
%% Suffix	-> \sfx{IV}
%% \author*[1,2]{\fnm{Joergen W.} \spfx{van der} \sur{Ploeg} 
%%  \sfx{IV}}\email{iauthor@gmail.com}
%%=============================================================%%

\author*[1]{\fnm{Lukas} \sur{Schulte} \orcid{https://orcid.org/0000-0001-9336-2075}}\email{lukas.schulte@uni-passau.de}\equalcont{These authors contributed equally to this work.}

\author[1]{\fnm{Anamaria} \sur{Mojica-Hanke} \orcid{https://orcid.org/0000-0002-5292-2977}}\email{anamaria.mojica-hanke@uni-passau.de}\equalcont{These authors contributed equally to this work.}

\author[2]{\fnm{Mario} \sur{Linares-V\'asquez} \orcid{https://orcid.org/0000-0003-0161-2888}}\email{m.linaresv@uniandes.edu.co}

\author[1]{\fnm{Steffen} \sur{Herbold} \orcid{http://orcid.org/0000-0001-9765-2803}}\email{steffen.herbold@uni-passau.de}

\affil[1]{\orgdiv{Faculty of Computer Science and Mathematics}, \orgname{University of Passau}, \orgaddress{\street{Innstraße 41}, \city{Passau}, \postcode{94032}, \country{Germany}}}

\affil[2]{\orgdiv{Ingeniería de Sistemas}, \orgname{Universidad de Los Andes}, \orgaddress{\street{Cra 1 Nº. 18A}, \city{Bogotá}, \postcode{111711}, \country{Colombia}}}

\abstract{
    \textbf{Context:} Many studies consider the relation between individual aspects of the software engineering process and bug-introduction, e.g., software testing and code review. These studies typically only identify correlations between their set of variables without accounting for interactions with external variables, such as confounding factors.

    \textbf{Objective:} Within this study, we provide a broad empirical view on practices of software development and their relation to bug-introducing changes \rev{to enable} future work on causal relations between those aspects.

    \textbf{Method:} We consider the bugs, the type of change that introduced the bug, aspects of the build process, code review, software tests, and any other discussion related to the bug that we can identify. We use a qualitative approach that first describes variables of the development process and then groups the variables based on their relations. From these groups, we deduce how their (pairwise) interactions affect bug-introducing changes.

    \textbf{Results:} We found multiple relevant relations within the development process of bug-introducing changes. Logical groups of variables and their relations provide a framework for discovering areas of interest regarding intermediate effects in the process and confounders towards bug-introduction.

    \textbf{Conclusion:} Software engineering practices applied during the development of bug-introducing changes are interdependent. This work lays the foundation to understand \textit{why} bugs are introduced using causal modeling, discovery, and inference.
}


%%\pacs[JEL Classification]{D8, H51}

%%\pacs[MSC Classification]{35A01, 65L10, 65L12, 65L20, 65L70}

\maketitle

% #########################################################
% 
% #########################################################

\section{Introduction}

A key concern of software engineering is to ensure the quality of a software product, either by avoiding bugs during development or by identifying and fixing them afterward. Within this study, we contribute to the knowledge required for the former, i.e., to avoid bugs. Concretely, we aim to improve our understanding of bug-introducing changes, i.e., such changes to software in which a new bug is introduced into the code due to a coding mistake.\footnote{We follow terminology introduced by \citep{rodriguez-perezHowBugsAre2020a} and use the term bug-introducing change instead of the also often used term bug inducing change.} Bug-introducing changes have been intensively studied, e.g., as part of the defect prediction literature that tries to predict bug-introducing changes \citep{kameiLargescaleEmpiricalStudy2013}. Further, the literature provides a good framework that describes how bugs are introduced \citep{rodriguez-perezHowBugsAre2020a} as well as regarding developer viewpoints on avoiding bugs \citep{souzaDevelopersViewpointsAvoid2022}. There is also knowledge about the coverage of buggy code by software tests \citep{ghafariTestabilityFirst2019a}, the correlation between test-driven development, code coverage, and subsequent bug fixes~\citep{borleAnalyzingEffectsTest2018a}, the impact of code reviews \citep{kononenkoInvestigatingCodeReview2015b}, and practices used during debugging \citep{perscheidStudyingAdvancementDebugging2017, bellerDichotomyDebuggingBehavior2018}. However, this rather helps us to understand what developers tend to do in general to avoid or find bugs, not what happens when bugs are actually introduced.

Moreover, the prior work on what happens during bug-introduction is observational: since there are neither interventions nor suitable control mechanisms (e.g., considering all -- or at least many -- other quality assurance aspects that could affect the results) it is not suitable to make causal claims. For example, consider that unit tests and code review are both used together within a data set and that both correlate with fewer bugs. Without considering data about both, unit tests and code reviews, it would not be possible to make sound claims about either technique: one might be effective and causing the effect, both might be effective, each for a part of the effect, or they might only be effective if used together. Within this study, we plan to provide a first perspective on a joint consideration of many aspects of the software development process. Thereby, we can better understand correlations between techniques and, in the long run, narrow the gap between the current correlational results and causal knowledge.

Our goal is two-fold. First, we create a new kind of data set, that covers a broad range of aspects of the software development process, i.e., the involved issue tracking, code review, testing, as well as other identifiable aspects of development practices. This is in contrast to prior literature that considered specific aspects like code quality~\citep{querelWarningIntroducingCommitsVs2021a}, tests~\citep{ghafariTestabilityFirst2019a}, or code review~\citep{kononenkoInvestigatingCodeReview2015b}: the authors all used different data, which makes it challenging to understand possible interactions between the scope of their research. We use a retrospective repository mining approach - that is, we analyze the publicly available resources of the projects under study. These resources commonly include the codebase itself and the issue tracker, but also often additional tools like continuous integration system, code review tools, and mailing lists. We examine the archives of all available systems to understand the development activity that led to the bug-introducing change. Second, we analyze this data with the goal of identifying patterns - or the lack thereof - within the development practices. While the observational data we gain from this analysis itself is hardly suitable to claim causal relationships between any pattern and bugs, our data and analysis will be a valuable foundation to inductively derive evidence-based theories about the relation between development practices and bugs. This allows the confirmatory design of subsequent studies to establish causal reasons for bug-introduction. While such foundational research does not directly impact software development, it will help researchers, developers, and tool vendors to understand which development practices and processes could be targeted with the best hopes of avoiding bugs.

The contributions of our \textit{exploratory} research are the following:

\begin{itemize}
    \item A novel data set with diverse data about 71 bugs from two projects, where for each bug the development process that led to the bug-introducing commit is described.
    \item A control group of 71 commits, randomly sampled from the same projects, for which we are not aware of any bug-introduction and which can therefore serve as a baseline for comparisons between development practices.
    \item An analysis of patterns within the data set that is suitable to derive hypotheses about relationships between development practices (i.e., through logical variable groups) and bugs for theory development and confirmation.
\end{itemize}

The diversity of the data collection requires a harmonized process to ensure that all relevant information is captured, as the effort for subsequent analysis due to missed factors would be large. We therefore pre-registered the protocol of this study in advance \citep{schulteExploratoryStudyBugintroducing2023}. Pre-registering allowed us to benefit from an early peer-review process and reduce the risk of missing important aspects. Analysis approach and expectations were therefore defined beforehand and deviations were noted in Section \ref{sec:deviations}.

The remainder of this paper is structured as follows. In Section~\ref{sec:relatedWork}, we give a short overview of the related work. Next, we introduce our research question in Section~\ref{sec:research-questions}. Then, we describe our research protocol, including materials, variables, execution plan, analysis plan, and deviations from the registered report in Section~\ref{sec:research-protocol}. We then present the results of our study in Section~\ref{sec:results} as well as a discussion of said results in Section~\ref{sec:discussion}. In Section~\ref{sec:threats}, we describe the threads to validity in regard to our study, followed by the conclusion in Section~\ref{sec:conclusion}.

% #########################################################
% 
% #########################################################

\section{Related work}
\label{sec:relatedWork}

In a recent work, \cite{souzaDevelopersViewpointsAvoid2022} studied the topic of bug-introduction based on the perception of developers. In their work, they conducted a mixed-methods approach based on surveying developers' opinions about specific practices. This work is closely related to our work, in the sense that it also considers a broad perspective on bug-introducing changes, considering many aspects of software development. Overall, their study covered developers' views on the impact of documentation, the complexity of changes, the role of code review and automated testing, version control aspects like merge commits and pull requests, internal quality aspects like code smells and static analysis, the type of changes being conducted, code ownership, dependencies, programming languages, and geographic distribution of teams. Our variables are inspired by this broad perspective and explicitly cover all aspects except for the geographic distribution and type of programming language. The major difference between our work and the work by \emph{Souza et al.} is that we study a population of bugs directly, i.e., collect data about these bugs from the development history, while \emph{Souza et al.} studied this indirectly through developer opinions. These two different approaches complement each other, as we need to understand both the perception of the developers, as well as the support in the data for such perceptions.

There is also a large body of work regarding individual development practices and their relation to bug-introducing changes. For example, \cite{taoHowSoftwareEngineers2012} study the role of code understanding, \cite{ghafariTestabilityFirst2019a} and \cite{borleAnalyzingEffectsTest2018a} of test coverage, \cite{kononenkoInvestigatingCodeReview2015b} of code review, and \cite{querelWarningIntroducingCommitsVs2021a} of static analysis tools. Such studies usually find that practices are, to some degree, negatively correlated with bug-introducing changes, i.e., that there are fewer bugs when followed.\footnote{\cite{borleAnalyzingEffectsTest2018a} is the notable exception in this list, who found no relationship between test coverage and subsequent bug fixes for most settings.} Other work studies aspects of the code base, like the values of code metrics and internal code quality~\citep{trautschStaticSourceCode2020a}. While it is certainly not unreasonable to assume that each of these practices has a positive impact on software development and bug-introducing changes, we cannot understand how the practices interact with each other by studying them individually. This is where our study deviates from the above prior work: we provide a broader view with the goal of being able to observe interactions between the different practices (i.e., through logical variable
groups) to also enable an understanding regarding how they are working together.

The number of studies that employ causal modeling and inference techniques in the field of software engineering, and therefore can make sound causal claims, is small. \cite{furiaCausalAnalysisEmpirical2024a} conducted a comprehensive study into causal analysis of empirical software engineering data from a Bayesian perspective, in which they create a causal model of programmer performance in a coding contest. Their study demonstrates the effectiveness of causal analysis for identifying confounders and provides a practical framework for other studies to follow. Most recently, \cite{furiaMitigatingOmittedVariable2025a} introduced guidelines for software engineering researchers to mitigate omitted variable bias through causal analysis. Further, \cite{frattiniApplyingBayesianData2025} applied causal modeling and causal inference through Bayesian data analysis to conceptually replicate a software engineering study by \cite{femmerImpactPassiveVoice2014} with refined causal assumptions. Outside explicitly Bayesian data analysis, \cite{kazmanCausalModelingDiscovery2017a} employ the PC algorithm \citep{spirtesCausationPredictionSearch2000a} for causal discovery of a model that encodes the effect of architectural design flaws on bugs. Also, \cite{huPracticalApproachExplaining2023a} leverage the Greedy Fast Causal Inference algorithm \citep{ogarrioHybridCausalSearch2016a} for causal discovery of the cause of defect proneness of commits based on issue and commit information. While our study applies neither causal modeling nor inference, it provides us with a basis to do so in future work and encourages considerations about potential causal effects in the software engineering process from outside the scope of related studies.

% #########################################################
% 
% #########################################################

\section{Research question}
\label{sec:research-questions}

The goal of our study is to contribute to finding an answer to the following research question.

\begin{itemize}
    \item[] \textbf{RQ:} Which software engineering practices are applied during the development process of bug-introducing code of open-source projects, and how does this relate to non-bug-introducing changes?
\end{itemize}

For our research question, we define software engineering practices as consisting of three components: \textit{Input}, \textit{Technique}, and \textit{Purpose} \citep{mojica-hankePerspectiveSoftwareEngineering2024a}. Specifically, we focus on the \textit{Input}, which refers to the input required to execute the practice, and the \textit{Technique}, which refers to the way possible actions are executed in the software engineering process \citep{mojica-hankePerspectiveSoftwareEngineering2024a}. This selection aligns well with the way related work approaches software engineering practices, such as the presence of warnings in version control systems \citep{querelWarningIntroducingCommitsVs2021a} as an \textit{Input}, and implicit \textit{Techniques} like ``Testability first!" \citep{ghafariTestabilityFirst2019a} or the impact of the number of participants in code reviews on bug-introduction \citep{kononenkoInvestigatingCodeReview2015b}. Ultimately, we also measure \textit{Purpose}, i.e., the reason why a certain practice is applied \citep{mojica-hankePerspectiveSoftwareEngineering2024a}. In the context of our paper, this \textit{Purpose} is the prevention of bug-introduction and is encoded in the difference between variable pairs from the bug and control group.

In comparison to prior work that either focused on the effectiveness of specific practices (e.g., \cite{ghafariTestabilityFirst2019a, kononenkoInvestigatingCodeReview2015b}) or on information from a specific source (e.g., \cite{querelWarningIntroducingCommitsVs2021a}), our approach does not restrict the practices we consider or the data sources from which we collect information. We try to collect as much information as possible about bug-introducing changes, and additionally create a control group of non-bug-introducing changes to compare against.
This approach provides a new perspective on bug-introducing changes by examining a wide set of granular variables and their pairwise relationships to bug introduction, rather than focusing on specific aspects in isolation.
These variables represent \textit{Inputs} that describe the extent to which \textit{Techniques} are applied. By logically associating these variables, we create an abstraction layer that describes groups of practices. We refer to these logical variable groups as \textit{logical groups} and use them to address the research question in Section~\ref{sub-sec:discussion-research-question}.

% #########################################################
% 
% #########################################################

\section{Research protocol}
\label{sec:research-protocol}

This section discusses the research protocol, which was pre-registered \citep{schulteExploratoryStudyBugintroducing2023} for this study. The structure and contents are closely aligned with the pre-registration. All deviations are summarized in Section \ref{sec:deviations}.

% We now define the subjects, variables, execution plan, and analysis plan of our research protocol. 

% #########################################################
% 
% #########################################################

\subsection{Subjects}

The subjects of our investigation are bug-introducing changes. Since noise within automatically collected defect data is a well-known problem \citep[see, e.g.,][]{herboldProblemsSZZFeatures2022b, rodriguez-perezHowBugsAre2020a}, we use a manually validated data set. While there are several manually validated data sets for bugs \citep[e.g., Defects4J and LLTC4J,][]{justDefects4JDatabaseExisting2014, herboldFinegrainedDataSet2022a}, we are aware of only one data set for bug-introducing changes: the data collected by \cite{rodriguez-perezHowBugsAre2020a} about the Nova module of OpenStack and Elasticsearch (see Table~\ref{tbl:subjects}). For both projects, \perez{} collected detailed data about the introducing changes for 60 bugs, i.e., 120 bugs in total. These bugs are further divided into intrinsic bugs (i.e., introduced by changes to the source code) and extrinsic bugs (i.e., introduced by external factors, such as bugs in dependencies or requirements changes). Within our work, we focus only on intrinsic bugs, as extrinsic bugs are not the direct responsibility of the developers of a project and there is no bug-introducing change~\citep{herboldProblemsSZZFeatures2022b}. Furthermore, \perez{} could not determine the bug-introducing changes for all intrinsic bugs. As a result, only a subset of 34 bugs for Nova and 38 bugs for Elasticsearch is suitable for our study, i.e., the intrinsic bugs for which the bug-introducing changes were identified. During the initial review of the data, we had to exclude one more intrinsic bug from the Nova project, due to the revision hash of the bug-introducing commit being equal to the bug-fixing commit (NOVA 1481164).

For the control group (CG), we sample commits that are not bug-introducing. First, we identify bug-introducing commits using a standard SZZ algorithm~\citep{sliwerskiWhenChangesInduce2005a}, except that we ignore whitespace changes. We identify bug fixes through both keywords and links to bug issues~\citep{trautschStaticSourceCode2020a}. This likely avoids false negatives at the cost of false positives~\citep{herboldProblemsSZZFeatures2022b}, and allows us to restrict the sample to commits for which no (known and fixed) bugs were introduced without manual validation. For each bug in our data, we randomly select one non-bug-introducing commit from within 30 days (i.e., one month) before or after the bug-introducing commit. This approach allows us to pair each bug-introducing commit with a non-bug-introducing commit from approximately the same time period. In cases where multiple bug-introducing commits have overlapping time windows, we ensure that no duplicates are sampled. We further ensure that all sampled commits were later merged into the main branch and exclude merge commits from the ones sampled.

As \cite{rodriguez-perezHowBugsAre2020a} explain, Nova and Elasticsearch are good and interesting subjects for a study, due to the diversity of the people involved, the number of developers and companies contributing to the projects, the long development history, and the rich information available about the history of these projects beyond the source code.

\begin{table*}[]
    \centering
    \begin{tabular}{@{}l|lll|l@{}}
        \toprule
        Project       & \#Intrinsic Bugs & \#Extrinsic Bugs & \#Total & \#Bugs included \\ \midrule
        Nova          & 34               & 12               & 57      & 33              \\
        Elasticsearch & 38               & 5                & 59      & 38              \\ \midrule
        Total         & 72               & 17               & 116     & 71              \\ \bottomrule
    \end{tabular}
    \caption{Subjects of our study. \textit{\#Bugs included} shows the number of intrinsic bugs for which sufficient data about their introduction is available from the prior work by \cite{rodriguez-perezHowBugsAre2020a}.}
    \label{tbl:subjects}
\end{table*}

% #########################################################
% 
% #########################################################

\subsection{Variables}
\label{sec:participantstags}

At the core of our study is our intent to capture extensive information on both the development process-related information about the bug-introducing change and general information about the bug itself. We achieve this by using a large set of variables whose values are derived from different information sources. Regardless of the information source, we have three kinds of variables:

\begin{enumerate}
    \item Identifiers - such as issues and commits - which serve as important metadata to establish trace links to our data sources and provide support for gathering additional information for future studies.
    \item Variables collected by automation that either capture aspects of the development process like the number of developers involved or serve as a proxy for such aspects, e.g., the percentage of lines responsible for the bug-introduction as proxy for estimating if the bug-introduction is a small aspect of the overall change or a major part of the overall development activity.
    \item Manually coded variables, which involve researchers analyzing data to describe specific aspects of the development activity. The reasons for manual intervention are due to the task being too complex to automate (e.g., to codify the type of bug) or because manual inference is more efficient (e.g., to determine if there were test changes). Where viable, we support the manual coding with automatically collected data, e.g., labels from an issue tracker were collected automatically and then manually grouped by a researcher.
\end{enumerate}

Our dependent variable describes whether a commit was bug-introducing or not. Thus, its value determines if an introducing commit is part of the 71 bugs from \cite{rodriguez-perezHowBugsAre2020a} or if it was sampled for the control group.

Table~\ref{tbl:all-variables} lists the independent variables of our study. For each variable, we give the ID, the name, the scale, and the description. The table is divided into compartments, such that each compartment contains related variables, i.e., about the fixed bug (variables \texttt{F1-F9}), the introducing issues (\texttt{I1-I10}),\footnote{Issues that were worked on when the bug was introduced.} the mailing list (\texttt{ML1-ML6}), potential information from other media (\texttt{O1-O4}), the use of a continuous integration system (\texttt{CI1-CI3}), the build process (\texttt{B1-B3}), the commit that introduced the change (\texttt{C1-C22}), and the code review (\texttt{R1-R10}). For the control group, we cannot measure the variables \texttt{F1-F8}, \texttt{C3-C6}, \texttt{C11}, \texttt{C12}, \texttt{C16}, \texttt{C18}, \texttt{C20}, \texttt{C22}, \texttt{M5}, and \texttt{M6}, as they are directly related to the bugs, which do not exist in our control group by definition.

In addition to the variables, we collect the raw data directly from the repositories that we use to mine the information. This ensures that our manual coding can be reproduced and facilitates that others may gather additional information for subsequent studies. In line with our above variables, we collect the following raw data:

\begin{itemize}
    \item bug issue and discussion
    \item source code of the introducing commit and fixing commit
    \item introducing issues and discussions
    \item code reviews and pull requests
    \item build logs from the continuous integration system
    \item reports regarding test results and test coverage
    \item related mailing list posts
    \item related data from other media (documentation and release information in wikis, specifications, and related bugs that are referenced by the initial bug)
    \item developer / discussant types following \cite{joblinClassifyingDevelopersCore2017a, bockAutomaticCoreDeveloperIdentification2023a}
\end{itemize}

\renewcommand{\arraystretch}{1.1}
\begin{longtable}{>{\raggedright\arraybackslash}p{2.8cm}|>{\raggedright\arraybackslash}p{1.25cm}|p{7.6cm}}
    \toprule
    \textbf{ID - Name}                                                                     & \textbf{Scale}  & \textbf{Description}                                                                                                                                                                                                                                                                                                                                                                                                                                                                                                                                                                                                                                                         \\
    \midrule
    \endfirsthead
    \toprule
    \textbf{ID - Name}                                                                     & \textbf{Scale}  & \textbf{Description}                                                                                                                                                                                                                                                                                                                                                                                                                                                                                                                                                                                                                                                         \\
    \midrule
    \endhead
    \midrule
    \multicolumn{3}{r}{Continued on next page}                                                                                                                                                                                                                                                                                                                                                                                                                                                                                                                                                                                                                                                                                                                                              \\
    \midrule
    \endfoot
    \endlastfoot
    \toprule
    \textbf{F1} \newline \texttt{Bug ID}                                                   & nominal         & Identifier of the bug within the issue tracking system.                                                                                                                                                                                                                                                                                                                                                                                                                                                                                                                                                                                                                      \\
    \textbf{F2} \newline \texttt{Bug severity}                                             & nominal         & Maximum severity assigned to the bug, e.g., high, medium, low.                                                                                                                                                                                                                                                                                                                                                                                                                                                                                                                                                                                                               \\
    \textbf{F3} \newline \texttt{Bug labels}                                               & list of nominal & Additional labels of the bug, e.g., components or similar.                                                                                                                                                                                                                                                                                                                                                                                                                                                                                                                                                                                                                   \\
    \textbf{F4} \newline \texttt{\#Bug discussants}                                        & integer         & Number of people involved in the reporting and discussion of the bug.                                                                                                                                                                                                                                                                                                                                                                                                                                                                                                                                                                                                        \\
    \textbf{F4.1} \newline \texttt{Bug discussant roles}                                   & list of nominal & Whether the bug discussants were core developers, peripheral developers or not involved in the project before.                                                                                                                                                                                                                                                                                                                                                                                                                                                                                                                                                               \\
    \textbf{F4.2} \newline \texttt{Bug reporter role}                                      & nominal         & Whether the highest-ranking bug reporter was of rank core developer, peripheral developer or not involved in the project before.                                                                                                                                                                                                                                                                                                                                                                                                                                                                                                                                             \\
    \textbf{F4.3} \newline \texttt{Bug assignee role}                                      & nominal         & Whether the highest-ranking bug assignee was of rank core developer, peripheral developer or not involved in the project before                                                                                                                                                                                                                                                                                                                                                                                                                                                                                                                                              \\
    \textbf{F4.4} \newline \texttt{Bug creator role}                                       & nominal         & Whether the highest-ranking bug creator was of rank core developer, peripheral developer or not involved in the project before                                                                                                                                                                                                                                                                                                                                                                                                                                                                                                                                               \\
    \textbf{F5}\text{*} \newline \texttt{Aspects of the bug discussion}                    & list of nominal & Aspects that were discussed in the issue comments of the bug, e.g., how to fix the bug or related features.  Stored as a list, as discussion may have multiple aspects.                                                                                                                                                                                                                                                                                                                                                                                                                                                                                                      \\
    \textbf{F6}\text{*} \newline \texttt{Type of bug}                                      & list of nominal & Type of main bug that was introduced following the taxonomy proposed by \cite{catolinoNotAllBugs2019} as one of the following: configuration issue, network issue, database-related issue, GUI-related issue, performance issue, permission/deprecation issue, security issue, program anomaly issue, and test code-related issue.                                                                                                                                                                                                                                                                                                                                           \\
    \textbf{F7}\text{*} \newline \texttt{Bug has duplicate}                                & boolean         & Whether the bug has any duplicate within the project.                                                                                                                                                                                                                                                                                                                                                                                                                                                                                                                                                                                                                        \\
    \textbf{F8} \newline \texttt{Bug reopened}                                             & boolean         & Whether the bug was reopened at any point.                                                                                                                                                                                                                                                                                                                                                                                                                                                                                                                                                                                                                                   \\
    \textbf{F9} \newline \texttt{\#Bug comments}                                           & integer         & Number of comments in the discussion section of the bug.                                                                                                                                                                                                                                                                                                                                                                                                                                                                                                                                                                                                                     \\
    \midrule
    \textbf{I1} \newline \texttt{Introducing issue IDs}                                    & list of nominal & Identifier of the issues as part of which the development happened, that introduced the bug.                                                                                                                                                                                                                                                                                                                                                                                                                                                                                                                                                                                 \\
    \textbf{I2}\text{*} \newline \texttt{Introducing issue types}                          & list of nominal & Types of the introducing issues, e.g., bug or improvement.                                                                                                                                                                                                                                                                                                                                                                                                                                                                                                                                                                                                                   \\
    \textbf{I3} \newline \texttt{Introducing issue severities}                             & list of nominal & Severities of the introducing issues.                                                                                                                                                                                                                                                                                                                                                                                                                                                                                                                                                                                                                                        \\
    \textbf{I4} \newline \texttt{Introducing issue labels}                                 & list of nominal & Additional labels of the introducing issues.                                                                                                                                                                                                                                                                                                                                                                                                                                                                                                                                                                                                                                 \\
    \textbf{I5} \newline \texttt{\#Introducing issue discussants}                          & integer         & Number of people involved in the reporting and discussion of the issues.                                                                                                                                                                                                                                                                                                                                                                                                                                                                                                                                                                                                     \\
    \textbf{I5.1} \newline \texttt{Issue discussant roles}                                 & list of nominal & Whether the issue discussants were core developers, peripheral developers or not involved in the project before.                                                                                                                                                                                                                                                                                                                                                                                                                                                                                                                                                             \\
    \textbf{I5.2} \newline \texttt{Issue reporter role}                                    & nominal         & Whether the highest-ranking issue reporter was a core developer, peripheral developer or not involved in the project before.                                                                                                                                                                                                                                                                                                                                                                                                                                                                                                                                                 \\
    \textbf{I5.3} \newline \texttt{Issue assignee role}                                    & nominal         & Whether the highest-ranking issue assignee was a core developer, peripheral developer or not involved in the project before.                                                                                                                                                                                                                                                                                                                                                                                                                                                                                                                                                 \\
    \textbf{I5.4} \newline \texttt{Issue creator role}                                     & nominal         & Whether the highest-ranking issue creator was a core developer, peripheral developer or not involved in the project before.                                                                                                                                                                                                                                                                                                                                                                                                                                                                                                                                                  \\
    \textbf{I6}\text{*} \newline \texttt{Aspects of introducing issue discussion}          & list of nominal & Aspects that were discussed in the comments of the introducing issue, e.g., discussion of solution variants. Stored as a list, as a discussion may have multiple aspects.                                                                                                                                                                                                                                                                                                                                                                                                                                                                                                    \\
    \textbf{I7} \newline \texttt{\#Introducing issue commits}                              & integer         & Number of commits in which the introducing issues were addressed.                                                                                                                                                                                                                                                                                                                                                                                                                                                                                                                                                                                                            \\
    \textbf{I8}\text{*} \newline \texttt{Introducing issue has wiki/specification}         & boolean         & Whether the issues directly relates to any wiki or specification.                                                                                                                                                                                                                                                                                                                                                                                                                                                                                                                                                                                                            \\
    \textbf{I9} \newline \texttt{Introducing issue reopened}                               & boolean         & Whether the issues were reopened at any time.                                                                                                                                                                                                                                                                                                                                                                                                                                                                                                                                                                                                                                \\
    \textbf{I10} \newline \texttt{\#Introducing issue comments}                            & integer         & Number of comments in the discussion section of the introducing issues.                                                                                                                                                                                                                                                                                                                                                                                                                                                                                                                                                                                                      \\
    \midrule
    \textbf{ML1}\text{*} \newline \texttt{ML bug discussion IDs}                           & nominal         & Identifier of emails related to the discussion of the bug (i.e. the bug fixing commit and the bug issue) on the Mailing List (ML).                                                                                                                                                                                                                                                                                                                                                                                                                                                                                                                                           \\
    \textbf{ML2} \newline \texttt{\#ML bug discussants}                                    & integer         & Number of people involved in the discussion of the bug on the ML.                                                                                                                                                                                                                                                                                                                                                                                                                                                                                                                                                                                                            \\
    \textbf{ML2.1} \newline \texttt{ML bug discussant roles}                               & list of nominal & Whether the discussants in the ML of the bug were core developers, peripheral developers or not involved in the project before.                                                                                                                                                                                                                                                                                                                                                                                                                                                                                                                                              \\
    \textbf{ML3}\text{*} \newline \texttt{Aspects of ML bug discussion}                    & list of nominal & Aspects that were discussed in the ML of the bug.                                                                                                                                                                                                                                                                                                                                                                                                                                                                                                                                                                                                                            \\
    \textbf{ML4}\text{*} \newline \texttt{ML introducing change discussion IDs}            & nominal         & Identifier of emails related to the discussion of the introducing change (i.e. the introducing commit or the introducing issue).                                                                                                                                                                                                                                                                                                                                                                                                                                                                                                                                             \\
    \textbf{ML5} \newline \texttt{\#ML introducing change discussants}                     & integer         & Number of people involved in the discussion of the introducing change on the ML.                                                                                                                                                                                                                                                                                                                                                                                                                                                                                                                                                                                             \\
    \textbf{ML5.1} \newline \texttt{ML introducing change discussant roles}                & list of nominal & Whether the discussants in the ML of the introducing change were core developers, peripheral developers or not involved in the project before.                                                                                                                                                                                                                                                                                                                                                                                                                                                                                                                               \\
    \textbf{ML6}\text{*} \newline \texttt{Aspects of ML introducing change discussion}     & nominal         & Aspects that were discussed in the ML of the introducing change.                                                                                                                                                                                                                                                                                                                                                                                                                                                                                                                                                                                                             \\
    \midrule
    \textbf{O1}\text{*} \newline \texttt{Identifier of related information in other media} & list of nominal & Identifier of all discussions about the bug that we could locate in other media of the project.                                                                                                                                                                                                                                                                                                                                                                                                                                                                                                                                                                              \\
    \textbf{O2}\text{*} \newline \texttt{Type of other media in which bug was discussed}   & list of nominal & Types of other media that were used, e.g., developer blogs, forums, and chats.                                                                                                                                                                                                                                                                                                                                                                                                                                                                                                                                                                                               \\
    \textbf{O3} \newline \texttt{\#Discussants in other media}                             & integer         & Number of discussants involved in other media.                                                                                                                                                                                                                                                                                                                                                                                                                                                                                                                                                                                                                               \\
    \textbf{O3.1} \newline \texttt{Discussant roles in other media}                        & list of nominal & Whether the discussants involved in other media were core developers, peripheral developers or not involved in the project before.                                                                                                                                                                                                                                                                                                                                                                                                                                                                                                                                           \\
    \textbf{O4}\text{*} \newline \texttt{Aspects of discussions in other media}            & list of nominal & Aspects that were discussed in other media.                                                                                                                                                                                                                                                                                                                                                                                                                                                                                                                                                                                                                                  \\
    \textbf{O5}\text{*} \newline \texttt{Buggy files present in the BIC}                   & nominal         & Whether the buggy files identified by \perez~ are present in the bug-introducing commit.                                                                                                                                                                                                                                                                                                                                                                                                                                                                                                                                                                                     \\
    \midrule
    \textbf{CI1} \newline \texttt{CI build ID}                                             & nominal         & Identifier of a build in the continuous integration (CI) system.                                                                                                                                                                                                                                                                                                                                                                                                                                                                                                                                                                                                             \\
    \textbf{CI2} \newline \texttt{CI build status}                                         & nominal         & Status of the CI build.                                                                                                                                                                                                                                                                                                                                                                                                                                                                                                                                                                                                                                                      \\
    \textbf{CI3} \newline \texttt{CI build status change}                                  & boolean         & Whether the status of the CI build is the same as for the previous commit.                                                                                                                                                                                                                                                                                                                                                                                                                                                                                                                                                                                                   \\
    \midrule
    \textbf{B1}\text{*} \newline \texttt{Build tools}                                      & list of nominal & Tooling used for building the software, e.g., Gradle or Maven. List in case multiple tools are used as part of a tool chain.                                                                                                                                                                                                                                                                                                                                                                                                                                                                                                                                                 \\
    \textbf{B2}\text{*} \newline \texttt{Dependency resolution}                            & nominal         & How external dependencies are resolved, e.g., to imported libraries. Possible values are remote repositories (e.g., maven central), project cache (i.e., a cache of dependencies that is maintained by the project), and mixed.                                                                                                                                                                                                                                                                                                                                                                                                                                              \\
    \textbf{B3}\text{*} \newline \texttt{Build practices}                                  & list of nominal & The practices that are used as part of the build process as a list of nominal values. We pre-define a set of categories as part of this registration. This set is inductively extended as part of the data collection, if necessary. The pre-defined categories are: usage of bug finding static analysis tool (e.g., SpotBugs), usage of style checking static analysis tools (e.g., Checkstyle), execution of automated unit and integration tests, execution and reporting of automated system tests, execution and reporting of other automated tests (e.g., performance tests), measurement and reporting of code coverage, and generation of source code documentation \\
    \midrule
    % variables about the code
    \textbf{C1} \newline \texttt{Introducing commit}                                       & nominal         & Revision hash of the bug-introducing change.                                                                                                                                                                                                                                                                                                                                                                                                                                                                                                                                                                                                                                 \\
    \textbf{C2.1}\text{*} \newline \texttt{Commit type (current)}                          & nominal         & How the commit was integrated into the main branch as one of the following: direct commit, merge of pull request, merge of textual diff provided, and other methods.                                                                                                                                                                                                                                                                                                                                                                                                                                                                                                         \\
    \textbf{C2.2}\text{*} \newline \texttt{Commit types (children)}                        & list of nominal & How the child commits were integrated into the main branch between the following: direct commit, merge of pull request, merge of textual diff provided, and other methods.                                                                                                                                                                                                                                                                                                                                                                                                                                                                                                   \\
    \textbf{C2.3}\text{*} \newline \texttt{Commit types (parents)}                         & list of nominal & How the parent commits were integrated into the main branch between the following: direct commit, merge of pull request, merge of textual diff provided, and other methods.                                                                                                                                                                                                                                                                                                                                                                                                                                                                                                  \\
    \textbf{C3} \newline \texttt{Number of files changes}                                  & integer         & Number of files that were changed.                                                                                                                                                                                                                                                                                                                                                                                                                                                                                                                                                                                                                                           \\
    \textbf{C4} \newline \texttt{Commit file bugginess}                                    & float           & Percentage of changed files that were responsible for the bug introduction.                                                                                                                                                                                                                                                                                                                                                                                                                                                                                                                                                                                                  \\
    \textbf{C5} \newline \texttt{Commit total line bugginess}                              & float           & Percentage of code lines within the commit that are bug-introducing.                                                                                                                                                                                                                                                                                                                                                                                                                                                                                                                                                                                                         \\
    \textbf{C6} \newline \texttt{Commit file line bugginess}                               & float           & Percentage of lines within the files that have bug-introducing changes.                                                                                                                                                                                                                                                                                                                                                                                                                                                                                                                                                                                                      \\
    \textbf{C7} \newline \texttt{Committer type}                                           & nominal         & Whether the developer was a core developer, peripheral developer, or not involved in the project before (i.e., first-time committer).                                                                                                                                                                                                                                                                                                                                                                                                                                                                                                                                        \\
    \textbf{C8}\text{*} \newline \texttt{Multiple concerns}                                & boolean         & Whether the commit addressed multiple concerns, e.g., because multiple intends are stated in the commit messages or there are changes beyond the indent based on the commit message.                                                                                                                                                                                                                                                                                                                                                                                                                                                                                         \\
    \textbf{C9}\text{*} \newline \texttt{Commit types}                                     & list of nominal & Commit type according to Swanson~\citep{swansonDimensionsMaintenance1976a} as perfective, corrective, adaptive. Encoded as a list, in case a commit tangles multiple concerns (\revv{e.g.,} a bug fix and a feature addition at the same time).                                                                                                                                                                                                                                                                                                                                                                                                                              \\
    \textbf{C10}\text{*} \newline \texttt{Test changes}                                    & boolean         & Whether any automated tests were added or changed.                                                                                                                                                                                                                                                                                                                                                                                                                                                                                                                                                                                                                           \\
    \textbf{C11}\text{*} \newline \texttt{Bug-covering test changes}                       & nominal         & Whether any tests covering the bug were added or changed. None, if there are no such changes; partial, if there are changes but not all introducing lines are covered; and covered, if all introducing lines are covered.                                                                                                                                                                                                                                                                                                                                                                                                                                                    \\
    \textbf{C12}\text{*} \newline \texttt{Bug-covering changes prior or subsequent commit} & list of nominal & Same as bug-covering test changes, except that neighboring commits (parents and children) and other commits for introducing issues are considered as well.                                                                                                                                                                                                                                                                                                                                                                                                                                                                                                                   \\
    \textbf{C13}\text{*} \newline \texttt{Test failures}                                   & boolean         & Whether there are any test failures reported.                                                                                                                                                                                                                                                                                                                                                                                                                                                                                                                                                                                                                                \\
    % C14 \newline \texttt{Bug-covering test failures} & boolean & Whether there are any test failures reported that cover the bug-introducing change. \\
    \textbf{C15}\text{*} \newline \texttt{Documentation changes}                           & boolean         & Whether any documentation was changed.                                                                                                                                                                                                                                                                                                                                                                                                                                                                                                                                                                                                                                       \\
    \textbf{C16}\text{*} \newline \texttt{Bug affecting documentation}                     & boolean         & Whether any documentation directly related to the lines in which the bug was introduced changed, e.g., the comment of a method with bug-introducing code or in-line comments between bug-introducing code lines.                                                                                                                                                                                                                                                                                                                                                                                                                                                             \\
    \textbf{C17}\text{*} \newline \texttt{Refactorings}                                    & boolean         & Whether any refactorings were conducted as part of the bug-introducing change.                                                                                                                                                                                                                                                                                                                                                                                                                                                                                                                                                                                               \\
    \textbf{C18}\text{*} \newline \texttt{Bug-introducing refactorings}                    & boolean         & Whether there are any refactorings within the bug-introducing code.                                                                                                                                                                                                                                                                                                                                                                                                                                                                                                                                                                                                          \\
    \textbf{C19}\text{*} \newline \texttt{Design changes}                                  & boolean         & Whether the change includes changes to the design (or architecture) of the software, e.g., by adding new classes or packages.                                                                                                                                                                                                                                                                                                                                                                                                                                                                                                                                                \\
    \textbf{C20}\text{*} \newline \texttt{Bug-introducing design changes}                  & boolean         & Whether there are any design changes as part of the bug-introducing code.                                                                                                                                                                                                                                                                                                                                                                                                                                                                                                                                                                                                    \\
    \textbf{C21}\text{*} \newline \texttt{Change of external dependencies }                & boolean         & Whether there are any changes to the external dependencies, e.g., updates, additions, or removal of external libraries.                                                                                                                                                                                                                                                                                                                                                                                                                                                                                                                                                      \\
    \textbf{C22}\text{*} \newline \texttt{Bug-introducing dependency changes }             & boolean         & Whether there are any changes to the external dependencies that affect the bug-introducing code.                                                                                                                                                                                                                                                                                                                                                                                                                                                                                                                                                                             \\
    \midrule
    % Variables related to code review
    \textbf{R1} \newline \texttt{Code review ID}                                           & list of nominal & Identifier of the review within the code review tool.                                                                                                                                                                                                                                                                                                                                                                                                                                                                                                                                                                                                                        \\
    \textbf{R2}\text{*} \newline \texttt{Review tool}                                      & nominal         & Tooling that was used to conduct the main code review, e.g., Gerrit, GitHub pull request, or textual comments in the issue based on code provided within a branch or textual diff. One list of nominals for each system, as we may have multiple comments in multiple locations.                                                                                                                                                                                                                                                                                                                                                                                             \\
    \textbf{R3} \newline \texttt{\#Reviewers}                                              & integer         & Number of developers that reviewed code.                                                                                                                                                                                                                                                                                                                                                                                                                                                                                                                                                                                                                                     \\
    \textbf{R4} \newline \texttt{Reviewer type}                                            & list of nominal & Whether the reviewers were core developers, peripheral developers, or not involved in the project before.                                                                                                                                                                                                                                                                                                                                                                                                                                                                                                                                                                    \\
    \textbf{R5} \newline \texttt{\#Review comments}                                        & integer         & Number of comments left by reviewers.                                                                                                                                                                                                                                                                                                                                                                                                                                                                                                                                                                                                                                        \\
    \textbf{R6} \newline \texttt{Rounds of code review}                                    & integer         & Rounds of subsequent code review. A new round starts when the author provides and update of the code and requests reviewers to reassess the code.                                                                                                                                                                                                                                                                                                                                                                                                                                                                                                                            \\
    \textbf{R7}\text{*} \newline \texttt{Caused changes}                                   & boolean         & Whether the code review led to any changes in the code.                                                                                                                                                                                                                                                                                                                                                                                                                                                                                                                                                                                                                      \\
    \textbf{R8}\text{*} \newline \texttt{Aspects of review comments}                       & list of nominal & Aspects discussed as part of the code review, e.g., current bugs, style changes, efficiency, or similar.                                                                                                                                                                                                                                                                                                                                                                                                                                                                                                                                                                     \\
    \textbf{R9}\text{*} \newline \texttt{Bug covering review comments}                     & boolean         & Whether the review contains any comments that cover the bug introducing files.                                                                                                                                                                                                                                                                                                                                                                                                                                                                                                                                                                                               \\
    \textbf{R10}\text{*} \newline \texttt{Bug covering review changes}                     & boolean         & Whether the review caused any bug introducing files to change.                                                                                                                                                                                                                                                                                                                                                                                                                                                                                                                                                                                                               \\
    \textbf{R11}\text{*} \newline \texttt{Review branch}                                   & list of nominal & Type of the branch (main or secondary) on which the reviews were done.                                                                                                                                                                                                                                                                                                                                                                                                                                                                                                                                                                                                       \\
    \bottomrule
    \caption{The 81 variables used in our study. The \text{*} marks manually coded variables.}
    \label{tbl:all-variables}
\end{longtable}

\renewcommand{\arraystretch}{1}

% #########################################################
% 
% #########################################################

\subsection{Execution Plan}
\label{sec:execution-plan}

Where possible, collection of the identifiers and automatically collected variables is supported by the SmartSHARK platform~\citep{trautschAddressingProblemsReplicability2018, trautschSmartSHARKEcosystemSoftware2020}. For most of the required automatically collected data, the tooling was already available. For the remaining, we extend the available tooling appropriately, \revv{e.g.,} to collect data from the \textit{Launchpad} issue tracker and the \textit{Gerrit} review system used by Nova, and invested manual effort into tracking down required identifiers, \revv{e.g.,} identifying relevant mailing list message IDs. Manual collection of identifiers and extension of tooling are conducted in parallel by two researchers. In terms of effort, we estimate 3 weeks (assuming \rev{40h per week), i.e., about} 50 minutes of identifier collection per introducing commit (bug and control group).

The coding of the 40 manual variables is conducted via prepared spreadsheets, such that we have two columns per variable: one for the determined value of the variable (e.g., whether bug-covering tests were added) and, where possible, one for the source of this information (e.g., the concrete test cases). While we have a fixed set of possible values for some variables (e.g., none, partial, and all for bug-covering test changes), we use inductive coding~\citep{thomasGeneralInductiveApproach2006} for the variables where we believe that an a-priori definition of all possible nominal values is not feasible or does not fit our purpose (e.g., when encoding discussion topics). The codes for the nominal variables and their descriptions are maintained in a separate spreadsheet that serves as a common codebook. Whenever data for a variable is not available (e.g., because there is no introducing issue or there was no code review) the values of all related variables are set to ``Not Available (NA)'' or left empty.

The manual data collection is conducted by two researchers and is done in three phases: \textit{1)} initial labeling; \textit{2)} unification of the codebook; and \textit{3)} resolution of disagreements. In the first phase, the two researchers collect the data for each bug- and non-bug-introducing commit by filling out the spreadsheet. Whenever they add a new code as a value for one of the variables, they add this to the common codebook. The encoding for a single bug requires significant time, as different information sources (e.g., the fixing issue and code review) need to be considered and the values for 62 variables need to be determined, most of them manually. In terms of effort, we estimate an average of 5 to 7 hours per introducing commit, meaning that each of the two researchers spends between 710 and 994 hours collecting data, i.e., about 18–25 weeks of work.

In the second phase, duplicate and closely related codes were unified within the common codebook. This encompasses finding common names for the same concept (e.g., system testing and system test) as well as finding an agreed upon and consistent level of abstraction (e.g., continuous integration instead of GitHub Action and Travis CI). All changes to the codebook are recorded and implemented through a mapping of the original codes to the new codes. Where possible, the mapping will be automatically applied before the discussion of disagreements in step \textit{3)}. If this is not possible (e.g., because one researcher used the category test, but it was decided to use more fine-grained categories like integration test and system test instead), the entry will be marked, and the mapping will be resolved during the final resolution of disagreements.

In the third phase, disagreements between labels are resolved. The two researchers consider the individual labels for each bug. In the case where the labels agree, it is used as the label in the final data set. In case of disagreements, the researchers analyze the cause for this (e.g., different interpretation of the context of discussions) and come to a common resolution. If no solution can be achieved by the two researchers, a third researcher is being involved. The outcome for a resolution is recorded. The trace links to the sources of the labels aid the researchers with the resolution of disagreements. We determine the inter-rater agreement using Krippendorff's $\alpha$ \citep{krippendorffContentAnalysisIntroduction2019}, as an indicator for the reliability of the data. The computation of Krippendorff's $\alpha$ is based on the harmonized codes to avoid an overestimation of the disagreement because the same concepts were found and just recorded using diverging codes.

The estimate of the time effort for the second and third step lies at around 14 weeks. The total estimate of the time effort lies between 35 and 40 weeks of work.
% 2nd and 3rd phase: 3-4 months, so like 14 weeks

% #########################################################
% 
% #########################################################

\subsection{Study Design}
\label{sec:study-design}

The goal of our analysis is to find patterns within the variables and, at the same time, to understand whether the conclusions of prior work may have been adversely affected by correlations between different aspects of a development process. This analysis will also be the building block for a larger theory of the impact of development practices when used together. To achieve this, we focus on relationships between the variables to understand whether the use of practices correlates with each other. Building upon this, we further examine if there are observable differences between the bug-introducing changes and the control group. Unfortunately, this is non-trivial because we have a large number of nominal variables. This precludes the use of, e.g., correlation heatmaps to get a simple visual overview of the relationships that could further be formalized by considering the strengths of the correlations. In addition, given that we have some variables that are actually lists of nominals, this gets even more difficult. Further, some variables only exist for bugs, but not in the control group. To work around these challenges, we conduct a qualitative analysis of the data based on the following strategy.

We start by analyzing all pairs of variables, except for the identifier variables. For each of the pairs, we provide a short written description (e.g., no relationship, often occurs together). We consider their relationships within the set of bugs, the control group, and between both groups.

For pairs of numeric variables, we use Spearman's $\rho$ to measure the correlation and test for the significance of the correlation, with a significance level of $\alpha=0.05$.

For pairs of nominal variables (incl. booleans), we use cross-tabulation to analyze the relationships. We augment this with Fisher's exact tests to assess significance and report results as significant if they meet a threshold of $\alpha=0.05$ without correction for repeated tests.

For pairs of nominal and numeric variables, we examine the distributions of the numeric variables per nominal category. Since we have at least one category with a sample size of 35 or less\footnote{We have 71 bugs and at least two categories, thus, one of them must have a count of at most 71/2.}, we do not conduct statistical tests for differences between distributions here at all. Instead, we conduct a visual comparison that accounts for the number of observations per category. We further aid the visual comparison through additional box plots visualizing the absence of categories to contrast one variable versus the rest. In the plots, those are identifiable by the inverse labels with the suffix \textit{[not]}.

Lists of nominals are being expanded, and each entry will be treated as an individual boolean one-hot encoded variable for this analysis. This gives us $\frac{\#Variables\cdot(\#Variables-1)}{2}$ pair-wise comparisons (the upper triangle without a diagonal of a square matrix). We further made accommodations for the axial codes of the discussion aspect variables by considering each of the three axes individually against all other variables. In total, there are $3552$ possible comparisons. Since the visual comparison can be affected by data sparsity, we apply a frequency threshold for the number of observations per category, set at 8 observations (i.e., at least 10\% of 71 possible observations). \rev{We provide a sensitivity analysis regarding the effect of different frequency thresholds in section \ref{subsub-sec:frequency-sensitivity}.}

As next step, we will sort the results of the pairs into six groups:
\begin{itemize}
    \item \textbf{G1:} Variables that are likely unrelated in both the bug and the control group. When studying the relation between variables and bug-introduction, it is unproblematic whether these pairs are considered together. While individually these variables might correlate with bug-introduction, the absence of a within-sample group relation of pairs provides no basis to suspect an effect between the pair's variables that confounds on bug-introduction.
    \item \textbf{G2:} Variables that are likely related, and the relation is the same in the bugs and in the control group. Whether they are considered together is unproblematic when only the outcome, bug-introduction, is studied. However, they transmit effects within the software engineering process and can be mediators, forges, and confounders for intermediate variables.
          % For such pairs, it should be unproblematic whether they are considered together when we study the relationship to bug-introduction. This is because their relation should not have an effect on the outcome, since such an effect on bug-introduction should materialize as a difference between the groups. 
    \item \textbf{G3:} The relationship between the variables is different for the bugs and the control group. For such pairs, there is a large risk of problems, when they are not considered together when analyzing their relationship to bug-introduction, as the pair of variables has a signal related to bug introduction, i.e., the difference in their relationship. The effect of this signal can only be understood if these variables are considered together in future studies.
    \item \textbf{G31:} Variables in one of the bug or control group are likely unrelated, with the other containing only infrequent observations or no observations at all. Given more data, we would be able to assign these variables to either group G1 or G3, depending on the then fully observable relation.
    \item \textbf{G32:} Variables in one of the bug or control group are likely related, with the other group containing only infrequent observations or no observation at all. Given more data, we would be able to assign these variables to either group G2 or G3, depending on the then fully observable relation.
    \item \textbf{Error:} Both the bug and control group have only infrequent observations or no observations at all. We cannot make any claim regarding their potential relations.
\end{itemize}

\begin{figure}
    \centering
    \includegraphics[width=\textwidth]{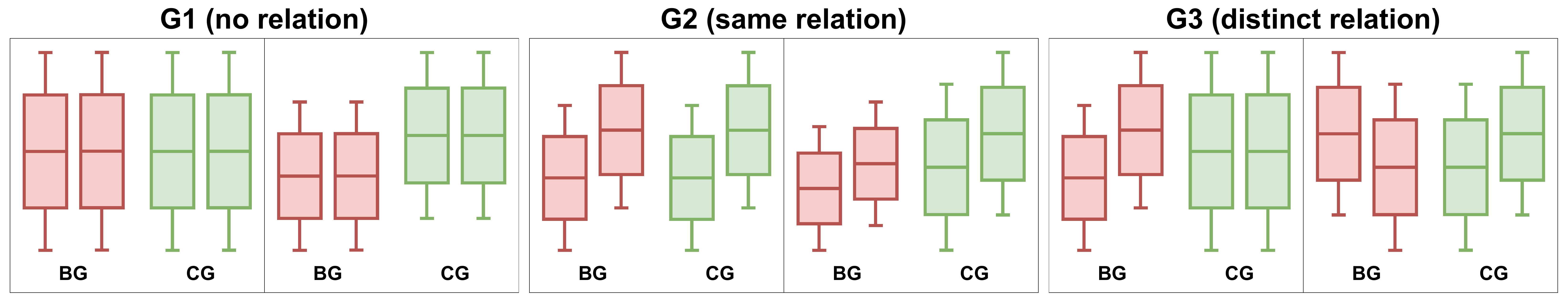}
    \caption{Abstract overview of the visual groups. Bug group (BG) in red, control group (CG) in green.}
    \label{fig:visual-groups-abstract}
\end{figure}

For the visual comparison, we referred to the abstract representation of groups in Fig. \ref{fig:visual-groups-abstract} for guidance. Section \ref{sub-sub-sec:nom-num} operationalizes the groupings with examples for all valid groups and visually explains how they compare.

Finally, we analyze the groups of relations to derive insights into our research question, i.e., which practices are used and how they relate to each other and to bug introduction. As part of this analysis, we also consider the amount of significant results of the Fisher's exact and the Spearman's $\rho$ test. If we observe that we have about 5\% (or less) significant results, this would be within the expectation of false positives with a threshold for significance of $\alpha=0.05$, as the p-values are uniform under the null hypothesis. Thus, if we have a small number of significant results, this indicates that the correlations we found might be spurious.

In addition to the above analysis, we also provide general data about the distributions of the variables both within the set of bugs and the control group to also enable a general understanding of the data composition.

\begin{mdframed}[linewidth=1pt]
    \rev{
        We analyze pairs of variables by grouping them according to their relation witin the bug and control group, and regarding the relation between the groups. The groups are \textbf{G1} (unrelated), \textbf{G2} (same relation), \textbf{G3} (distinct relation), \textbf{G31} (distinct relation or unrelated), \textbf{G32} (distinct or same relation), and \textbf{E} (error group). Nominal/numeric pairs were grouped using visual analysis of boxplots, while nominal pairs were grouped using cross tabulation and Fisher's exact test, and numeric pairs were grouped using Spearman's $\rho$.
    }
\end{mdframed}

% #########################################################
% 
% #########################################################

\subsubsection{Operationalization of the Groupings}
\label{sub-sub-sec:nom-num}

\begin{figure}
    \centering
    \includegraphics[width=\textwidth]{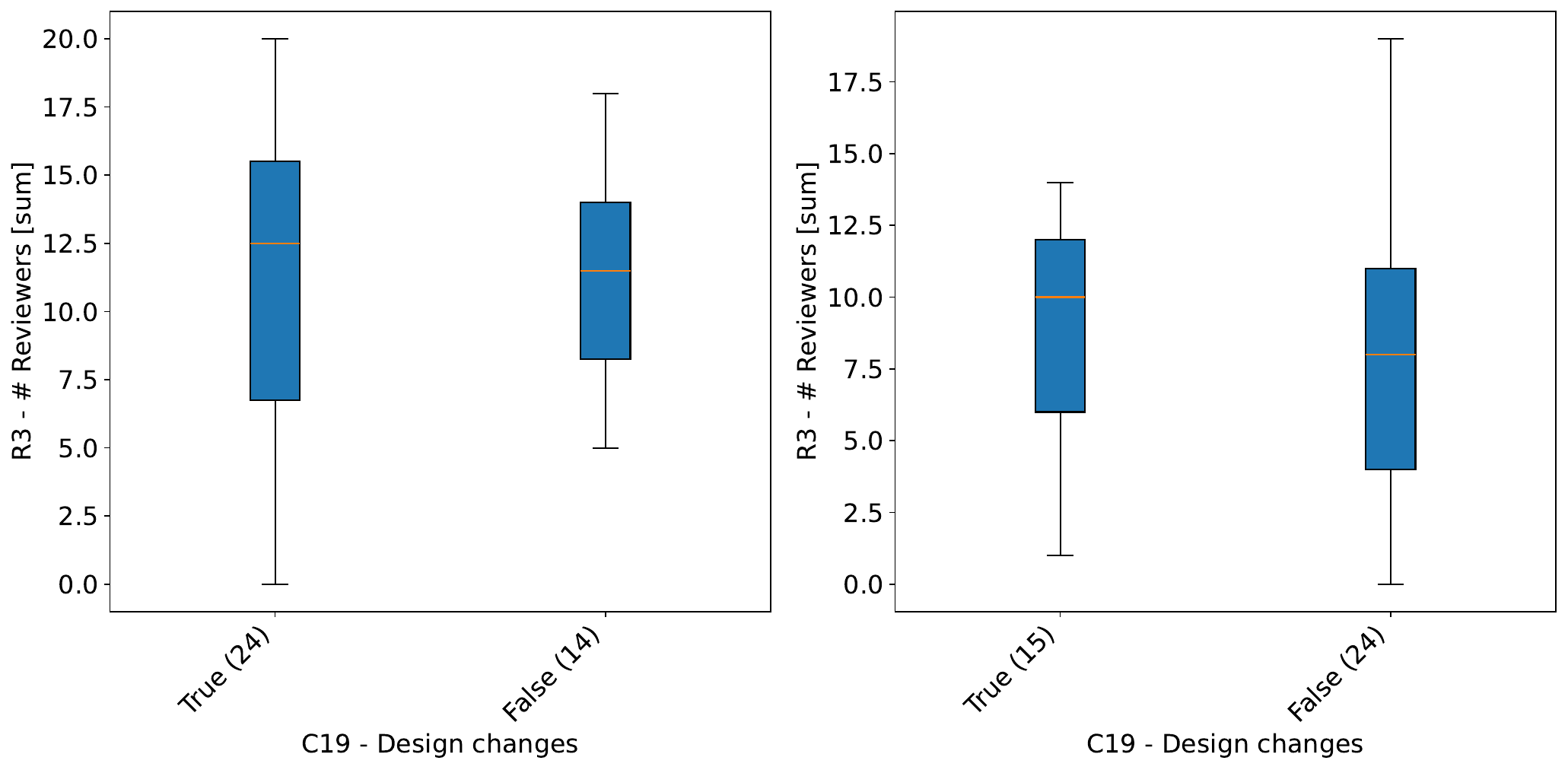}
    \caption{G1: C19 - Design changes / R3 - \# Reviewers. The left panel shows the bug group (BG, 38 valid pairs), the right panel shows the control group (CG, 39 valid pairs). Outliers are hidden.}
    \label{fig:C19_R3_histogram_boxplot-cropped}
\end{figure}

\begin{figure}
    \centering
    \includegraphics[width=\textwidth]{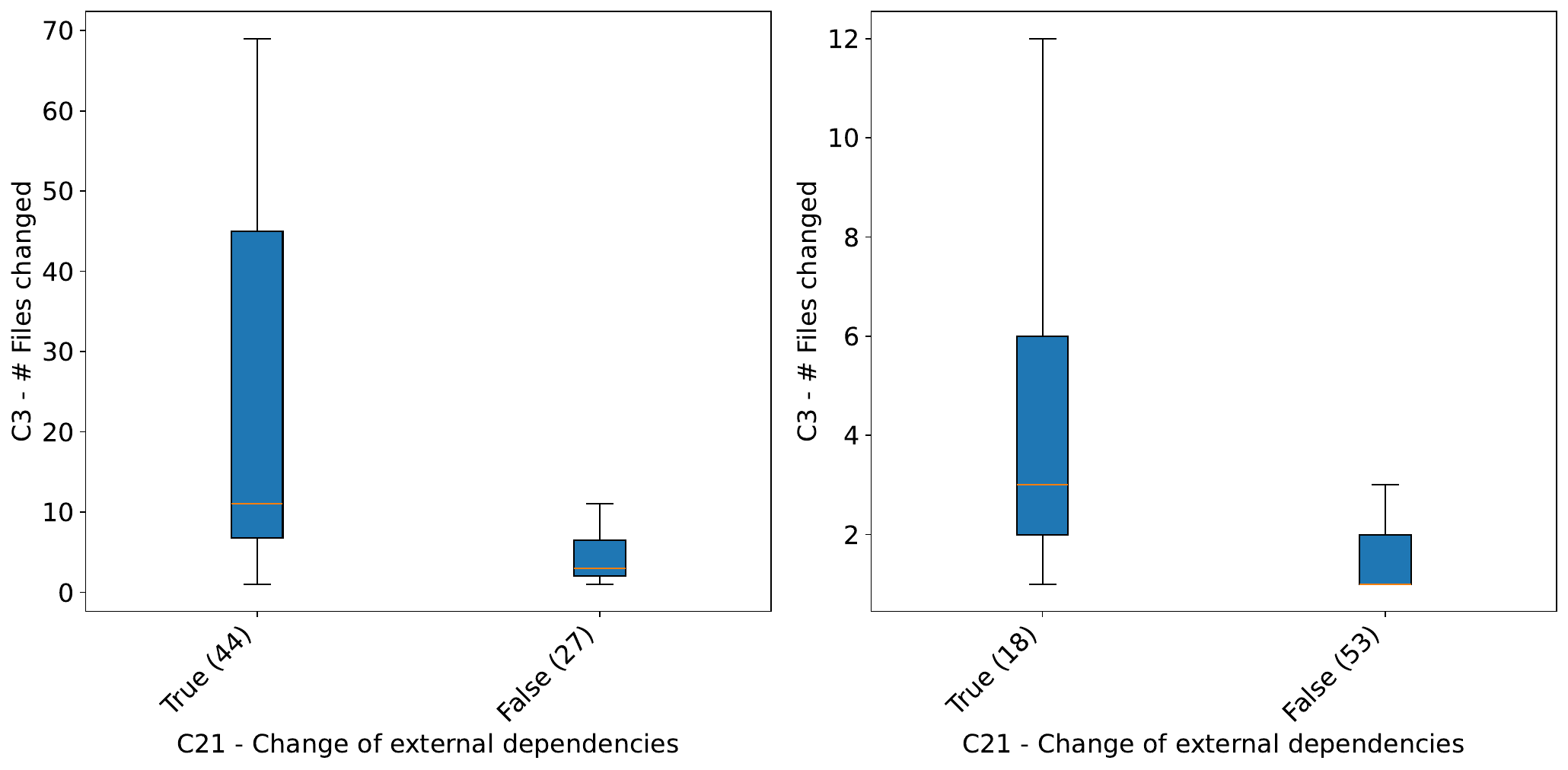}
    \caption{G2: C21 - Change of external dependencies / C3 - \# Files changed. The left panel shows the bug group (BG, 71 valid pairs), the right panel shows the control group (CG, 71 valid pairs). Outliers are hidden.}
    \label{fig:C21_C3_histogram_boxplot-cropped}
\end{figure}

\begin{figure}
    \centering
    \includegraphics[width=\textwidth]{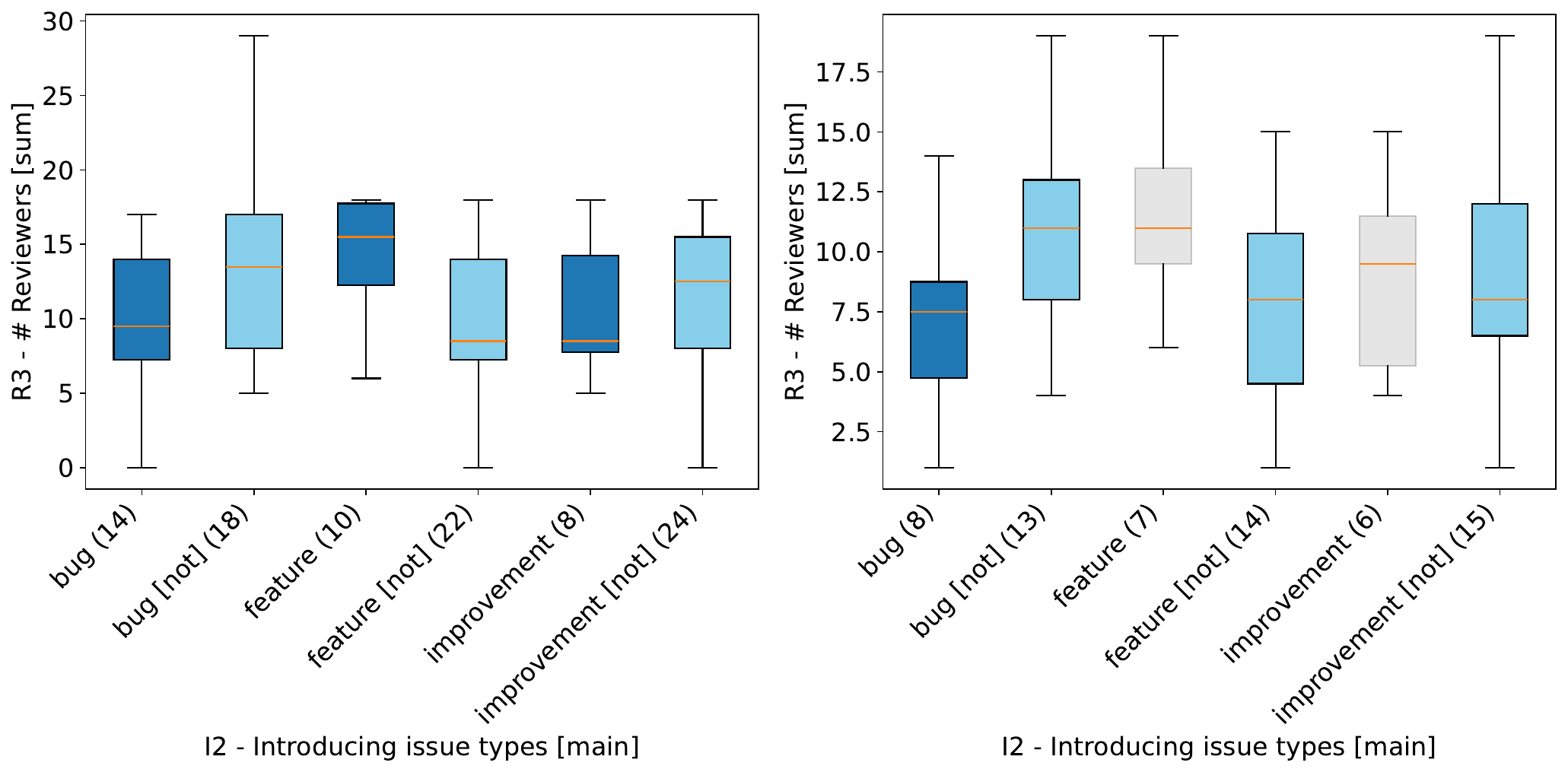}
    \caption{G2 (one versus rest): I2 - Introducing issue types / R3 - \# Reviewers. The left panel shows the bug group (BG, 22 valid pairs), the right panel shows the control group (CG, 21 valid pairs). Outliers are hidden.}
    \label{fig:I2_R3_histogram_boxplot_cropped}
\end{figure}

\begin{figure}
    \centering
    \includegraphics[width=\textwidth]{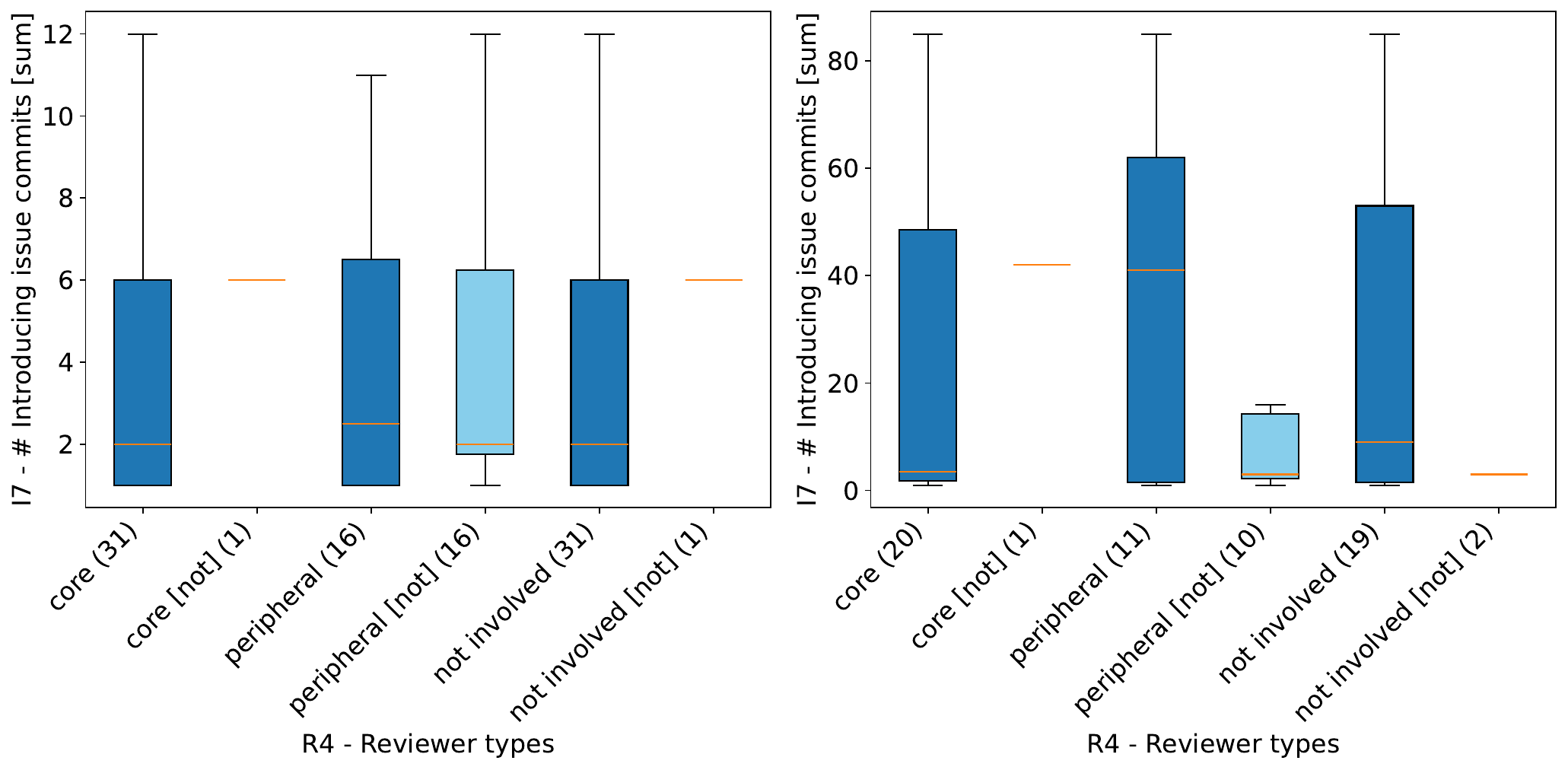}
    \caption{G3 (one versus rest): R4 - Reviewer types / I7 - \# Introducing issue comments. The left panel shows the bug group (BG, 32 valid pairs), the right panel shows the control group (CG, 21 valid pairs). Outliers are hidden.}
    \label{fig:R4_I7_histogram_boxplot_cropped}
\end{figure}

In order to better explain how exactly we operationalized the grouping of nominal/numeric pairs with visual comparison, the following section outlines this process. All visual comparisons were conducted with box plots and guided by the abstract overview in Fig. \ref{fig:visual-groups-abstract}.
\revv{
    We applied the following criteria to identify relevant relationships within sample groups:
    \begin{itemize}
        \item The mean of at least one category is outside or at the boundary of the interquartile ranges of another set of categories.
        \item The range of at least one category (excluding outliers) is visually distinct from other categories.
    \end{itemize}
}
\revv{
    If data was sparse, inverse labels were considered as well. Further, trends were considered the same between the sample groups if both of the following conjunctive criteria were met:
    \begin{itemize}
        \item Trend identifying categories have a substantial overlap between the bug and the control group.
        \item Trend identifying categories show the same trend direction.
    \end{itemize}
}
Fig. \ref{fig:C19_R3_histogram_boxplot-cropped} presents an example of a G2 assignment. The box plots have similar means and interquartile ranges within the sample groups. The pair is assigned to group G1, as there is no clear relationship between the binary variable \texttt{C19 - Design changes} (\textit{True}/\textit{False}) and the numeric scale \texttt{R3 - \# Reviewers}, in either the bug or control group.

Fig. \ref{fig:C21_C3_histogram_boxplot-cropped} displays another pair of nominal and numeric variables which is assigned to group G2. The left side shows the box plots of the distribution of the two variables for the bug group. Each box plot represents one label of the nominal variable \texttt{C21 - Change of external dependencies}, with \texttt{C3 - \# Files changed} represented by the y-axis. Here, through a higher mean and higher upper quartile, commits that change external dependencies (\textit{True}) relate to more files changed, while other commits (\textit{False}) relate to fewer files changed through a lower mean and lower interquartile ranges. The right side shows the control group, where we observe the same relation.

The distribution in Fig. \ref{fig:I2_R3_histogram_boxplot_cropped} is more complex, since we have multiple labels for the nominal variable \texttt{I2 - Introducing issue types}. Further, two labels in the control group do not meet the frequency threshold of at least 8 observations. The final assessment is therefore only possible through One-versus-Rest comparison using inverse labels (e.g., all instances that are labeled \textit{bug}, versus the rest of the instances - \textit{bug [not]}).
In the bug group panel on the left, we observe that the introducing issue types \textit{bug} and \textit{improvement} relate to fewer reviewers than the type \textit{feature}. In the control group panel, however, the infrequent observations of \textit{feature} (7 observations) and \textit{improvement} (6 observations), only allow us to compare the label \textit{bug} with the rest (i.e., the sum of the other labels). Here we observe the same relation as in the bug group, namely that the label \textit{bug} relates to fewer reviewers. We therefore assign this combination to group G2.

The variable pair \texttt{R4 - Reviewer type}s / \texttt{I7 - \#Introducing issue comments}, displayed in Fig. \ref{fig:R4_I7_histogram_boxplot_cropped}, was assigned to group G3 through a One-versus-Rest comparison. While data was not infrequent in this case, a direct comparison was hindered by the fact that the R4 variable is of type list of nominal. This means that multiple labels apply per observed instance, i.e., in this case, per review. One review can have multiple reviewers of different types. In the left panel, for example, there were 32 reviews, of which 31 had a reviewer of type \textit{core}, while at the same time 31 reviews also had reviewers of type \textit{not involved}. Since the inverse of those labels did not meet the frequency threshold, the comparison was therefore made on the reviewer type \textit{peripheral} alone, for both bug and control group. There, we observe that no relevant relation in the bug group, since mean and quartile ranges are similar, while in the control group, the presence of peripheral reviewers is related to more introducing issue commits.

% #########################################################
% 
% #########################################################

\subsection{Summary of Deviations}
\label{sec:deviations}

During the execution of the study, we encountered situations that required us to deviate from the pre-registered research protocol. Those deviations are listed below.

\begin{itemize}
    \item We adjusted the research question for more clarity.
    \item We initially planned to randomly select the control group commits within a 61-day time window of the respective bug group commits ($\pm$30 days and day of commit), with the option to increase this time window if necessary. Since the 61-day time window was sufficient in all cases, the mentioning of the time window increase was removed from the research protocol.
    \item We excluded one Nova bug (NOVA 1481164) that is present in the dataset by \cite{rodriguez-perezHowBugsAre2020a}, since the revision hash of the bug-introducing commit was identical to the bug-fixing commit. The total number of bugs is therefore 71, not 72, and the size of the control group was adjusted accordingly.
    \item We extended the set of variables during the coding process to cover important aspects that had not been previously considered. The following 19 variables were added:
          \begin{itemize}
              \item \texttt{F4.1  Bug discussant roles}
              \item \texttt{F4.2  Bug reporter role}
              \item \texttt{F4.3  Bug assignee role}
              \item \texttt{F4.4  Bug creator role}
              \item \texttt{F9    \#Bug comments}
              \item \texttt{I5.1	Introducing issue discussant roles}
              \item \texttt{I5.2	Introducing issue reporter role}
              \item \texttt{I5.3	Introducing issue assignee role}
              \item \texttt{I5.4	Introducing issue creator role}
              \item \texttt{I8    Introducing issues has wiki/specification}
              \item \texttt{I9    Introducing issue reopened}
              \item \texttt{I10	\#Introducing issue comments}
              \item \texttt{ML2.1 ML bug discussant roles}
              \item \texttt{ML5.1 ML bug introducing discussant roles}
              \item \texttt{O3.1	Discussant roles in other media}
              \item \texttt{R11	Review branch}
          \end{itemize}
    \item We split the variable \texttt{C2} into three:
          \begin{itemize}
              \item \texttt{C2.1	Commit type (current)}
              \item \texttt{C2.2	Commit types (children)}
              \item \texttt{C2.3	Commit types (parents)}
          \end{itemize}
    \item We removed the boolean variable \texttt{C14 - Bug-covering test failures}. It was supposed to tell whether there are any test failures reported that cover the bug-introducing change. However, neither of the studied projects had available records of relevant past test-runs that could identify whether the run covered the bug.
    \item We removed lists and nested lists from the scope of variables during the coding process so that they better cover the information at hand. For this, we joined lists where appropriate, as outlined in the respective variable descriptions. For the following variables, we removed list layers:
          \begin{itemize}
              \item \texttt{I4 - Introducing issue labels:} list of lists of nominals was changed to list of nominal.
              \item \texttt{I5 - \#Introducing issue discussants:} list of \rev{integers} was changed to integer.
              \item \texttt{I6 - Aspects of introducing issue discussion:} list of lists of nominals was changed to list of nominal.
              \item \texttt{I7 - \#Introducing issue comments:} list of \rev{integers} was changed to integer.
              \item \texttt{O3 - \#Discussants in other media:} list of \rev{integers} was changed to integer.
              \item \texttt{O4 - Aspects of discussion in other media:} list of lists of nominal was changed to list of nominal.
              \item \texttt{R2 - Review tool:} list of lists of nominal was changed to nominal.
              \item \texttt{R3 - \#Reviewers:} list of \rev{integers} was changed to integer.
              \item \texttt{R4 - Reviewer type:} list of lists of nominal was changed to list of nominal.
          \end{itemize}
    \item We added groups G31 and G32, as well as the Error group for pairwise combinations, to account for data availability issues. \rev{We introduce these in Section \ref{sec:study-design}}.
    \item After executing the cross tabulation, we tested the assumptions of the $\chi2$ test. Due to data sparsity, the frequency of the expected variables lower than 5 did more often than not turn out higher than 20\%, breaking one of the test's assumptions. We therefore use \rev{Fisher's} exact test instead.
    \item We extended our reporting for the amount of significant results to include those of the Spearman $\rho$ correlations.
\end{itemize}

% #########################################################
% 
% #########################################################

\section{Results}
\label{sec:results}

This section presents the results of our study, organized as follows:
First, we describe the outcomes of the data collection and coding process, including data availability (Section~\ref{subsub-sec:var-ds-limitations}), data distribution (Section~\ref{subsub-sec:data-dist}), and inter-rater agreement (Section~\ref{subsub-sec:agreement}).
Next, we report the overall distribution of pairs assigned to groups G1 through G32 and the Error group (Section~\ref{sub-sec:group-distribution}), and the significance of statistical test results (Section~\ref{sub-sec:amount-sigificant-results}). We then provide an overview of related variables through circular network diagrams for groups G2 and G3 (Section~\ref{sub-sec:network-graphs}). After that, we describe the trends observed between group G3 variables and what these mean in terms of causation (Section \ref{sub-sec:g3-how-relate}). Finally, we present a framework of logical variable groups, modelled after \textit{Inputs} and \textit{Techniques} of software engineering practices, which can serve as a basis for determining potential effects between certain practices from one part of the software engineering process and others (Section \ref{sub-sec:logical-groups}).

% #########################################################
% 
% #########################################################

\subsection{Data Collection}
\label{sub-sec:data-collection}

This section outlines the results of the data collection, including limitations of data availability that were encountered, the distribution of labels within variables, and a summary of the reliability of codes and inter-rater agreement.

% #########################################################
% 
% #########################################################

\subsubsection{Data Availability Limitations}
\label{subsub-sec:var-ds-limitations}

\begin{table}
    \begin{tabular}{r|rrrr}
        \toprule
                            & \shortstack{Data Sparsity                     \\~} & \shortstack{Incomparability\\~} & \shortstack{Project\\exclusivity} & \shortstack{Project\\dependence} \\
        \midrule
        Data Sparsity       & \rev{2488}                & ~    & ~    & ~   \\
        Incomparability     & 1676                      & 1676 & ~    & ~   \\
        Project exclusivity & 1126                      & 546  & 1608 & ~   \\
        Project dependence  & 236                       & 172  & 110  & 410 \\
        \bottomrule
    \end{tabular}
    \caption{Number of pairs for which limitations were observed. The diagonal shows the limitation exclusive counts. The lower triangle displays the number of pairs affected by both the row and column limitation types, indicating overlap between limitations.}
    \label{tab:data-limitations}
\end{table}

The following paragraphs introduce the terms for data availability limitations encountered during the mining and coding process, as well as their implications: \textit{data sparsity}, \textit{incomparability}, \textit{project dependence}, and \textit{project exclusivity}.

\textit{Data sparsity} is observed when single variables or whole data sources provide less than required data or data that has too little variance in at least one of the sample groups. The limitation becomes more pronounced when pairs of variables are considered, since both variable observations of a pair need to be valid for it to be considered for group assignment. Sparse pairs are therefore grouped into G31, G32 or the Error group.
For instance, \texttt{I5 - \# Introducing issue discussants} suffers from \textit{data sparsity} since not all commits were linked to introducing issues. In the bug group there are only 36 introducing issues and in the control group there are 28. Another example is \texttt{C15 - Documentation changes}, where in the bug group only 9 commits did not include a documentation change. And considering a pair of sparse variables, the number of interpretable observations is decreased even further. For \texttt{I5/C15}, the bug group code \textit{False} (no documentation change) only had 5 observations that were successfully paired with an issue.

\textit{Incomparability} is observed when a variable does not exist for either bug or control group. Examples are the \texttt{F} (fixing issue) variables, which by definition do not exist in the control group, since control group commits were not bug-introducing and therefore were never linked to a fix. Other examples include variables that capture bug-related characteristics, such as \texttt{C4 - Commit file bugginess}.

\textit{Project exclusivity} is observed when data sources or variables only exist for one of the projects. One example is the set of \texttt{ML}-variables (mailing list variables), which only exist for the Nova project, since the repositories of mailing lists of the Elasticsearch project were inaccessible. Or the continuous integration system logs, that were also no longer available for the Elasticsearch project.

\textit{Project dependence} is observed when certain codes only exist for one of the projects and could not be mapped to a common code between the projects. An example is \texttt{B1 - Build tools}, which is strongly project-dependent since Nova is written in Python and Elasticsearch in Java.

Table \ref{tab:data-limitations} shows the frequencies at which the limitations were observed in variable pairs. We observe that there is substantial overlap between the limitations. Logically, \textit{incomparability} is a sufficient cause for \textit{data sparsity}, i.e., all incomparable pairs are sparse. But also most \textit{project exclusive} and more than half of the \textit{project dependent} variable pairs are sparse.

% #########################################################
% 
% #########################################################

\subsubsection{Data Distribution}
\label{subsub-sec:data-dist}

This section reports the distribution of labels per variable. For all variables, we report the number of valid instances, i.e., the number of introducing commits for which we found instances of the variable. For numeric variables (Appendix~Table~\ref{tbl:data-dist-num}), we report the mean and standard deviation. For nominal variables (Appendix~Table~\ref{tbl:data-dist-nom}), we report the number of unique labels used, and for list variables (Appendix~Table~\ref{tbl:data-dist-list}) additionally the absolute number of labels.

Wherever the number of valid instances is zero, the tables highlight \textit{incomparability} of the sample groups. They also hint towards \textit{data sparsity} and \textit{project exclusivity}, since for many variables the number of valid instances does not reach the maximum of 71 (i.e., instances linked to the 71 bug-introducing commits). As a result, many variables are imbalanced between sample groups.

% #########################################################
% 
% #########################################################

\subsubsection{Reliability of Codes and Inter-Rater Agreement}
\label{subsub-sec:agreement}

\begin{table}[]
    \begin{tabular}{>{\ttfamily}lr|>{\ttfamily}lr|>{\ttfamily}lr|>{\ttfamily}lr}
        \toprule
        \multicolumn{1}{l}{Variable ID} & $\alpha$                                            & \multicolumn{1}{l}{Variable ID} & $\alpha$                                            & \multicolumn{1}{l}{Variable ID} & $\alpha$                                            & \multicolumn{1}{l}{Variable ID} & $\alpha$                                            \\ \midrule
        B1                              & \cellcolor[HTML]{FFCCCC}{\color[HTML]{CC0000} 0.57} & C12                             & 0.73                                                & F5                              & \cellcolor[HTML]{FFCCCC}{\color[HTML]{CC0000} 0.57} & O1                              & 0.84                                                \\
        B2                              & 0.99                                                & C13                             & 0.90                                                & F6                              & \cellcolor[HTML]{FFCCCC}{\color[HTML]{CC0000} 0.62} & O2                              & 0.72                                                \\
        B3                              & \cellcolor[HTML]{FFCCCC}{\color[HTML]{CC0000} 0.48} & C15                             & 0.92                                                & F7                              & 0.97                                                & O4                              & \cellcolor[HTML]{FFCCCC}{\color[HTML]{CC0000} 0.49} \\
        C2.1                            & 0.95                                                & C16                             & 0.83                                                & I2                              & 0.78                                                & O5                              & 0.93                                                \\
        C2.2                            & 0.97                                                & C17                             & 0.71                                                & I6                              & 0.67                                                & R2                              & 0.98                                                \\
        C2.3                            & 0.99                                                & C18                             & \cellcolor[HTML]{FFCCCC}{\color[HTML]{CC0000} 0.57} & I8                              & 0.96                                                & R7                              & 0.82                                                \\
        C8                              & 0.81                                                & C19                             & 0.76                                                & ML1                             & 1.00                                                & R8                              & \cellcolor[HTML]{FFCCCC}{\color[HTML]{CC0000} 0.53} \\
        C9                              & 0.73                                                & C20                             & 0.70                                                & ML3                             & \cellcolor[HTML]{FFCCCC}{\color[HTML]{CC0000} 0.54} & R9                              & 0.80                                                \\
        C10                             & 0.98                                                & C21                             & 0.95                                                & ML4                             & 0.99                                                & R10                             & \cellcolor[HTML]{FFCCCC}{\color[HTML]{CC0000} 0.63} \\
        C11                             & 0.70                                                & C22                             & 0.90                                                & ML6                             & \cellcolor[HTML]{FFCCCC}{\color[HTML]{CC0000} 0.46} & R11                             & 0.89                                                \\ \bottomrule
    \end{tabular}
    \caption{Krippendorff's alpha ($\alpha$) for all manually coded variables. Low agreement (less than $0.667$ \citep{krippendorffContentAnalysisIntroduction2019}) highlighted in red.}
    \label{tab:krippendorff}
\end{table}

We evaluated the agreement between the two raters on each manually coded variable using Krippendorff's $\alpha$ \citep{krippendorffContentAnalysisIntroduction2019}. The results can be found in Table \ref{tab:krippendorff}. We find that the overall agreement, after merging of codebooks but before the resolving of disagreements, is high in most cases. However, we observe a lower agreement for the more complex variables. Specifically, those variables describing the aspects of discussion (i.e., \texttt{}{F5}, \texttt{I6}, \texttt{ML3}, \texttt{ML6}, \texttt{O4}, and \texttt{R8}), those describing challenging to find information like build tools used (\texttt{B1}), build practices used (\texttt{B3})\rev{, or whether review changes covered the bug (\texttt{R10})}, and those with room for interpretation, like whether there are bug-introducing refactorings (\texttt{C18}) or the type of bug (\texttt{F6}). \rev{Except \texttt{I6}, all} of these fall below the threshold of $0.667$, which marks the point at which rater agreement is considered poor.

All disagreements were resolved involving both raters and their notes made during the coding, ensuring a high-quality of the final data. Here we observed that most disagreements were actually describing related practices from different viewpoints. For example, in a review, a comment provides additional information about how a feature should be implemented to match the pre-defined use case. One rater might interpret the topic as \textit{solution design}, focusing on the implementation aspect, while the other labels it as \textit{intended usage}, with a focus on the usage. Neither is necessarily incorrect and depending on the subsequent review of the context, one or both were accepted.

% #########################################################
% 
% #########################################################

\subsection{Group Distribution Overview}
\label{sub-sec:group-distribution}

\begin{figure}
    \centering
    \rev{\includegraphics[width=\textwidth]{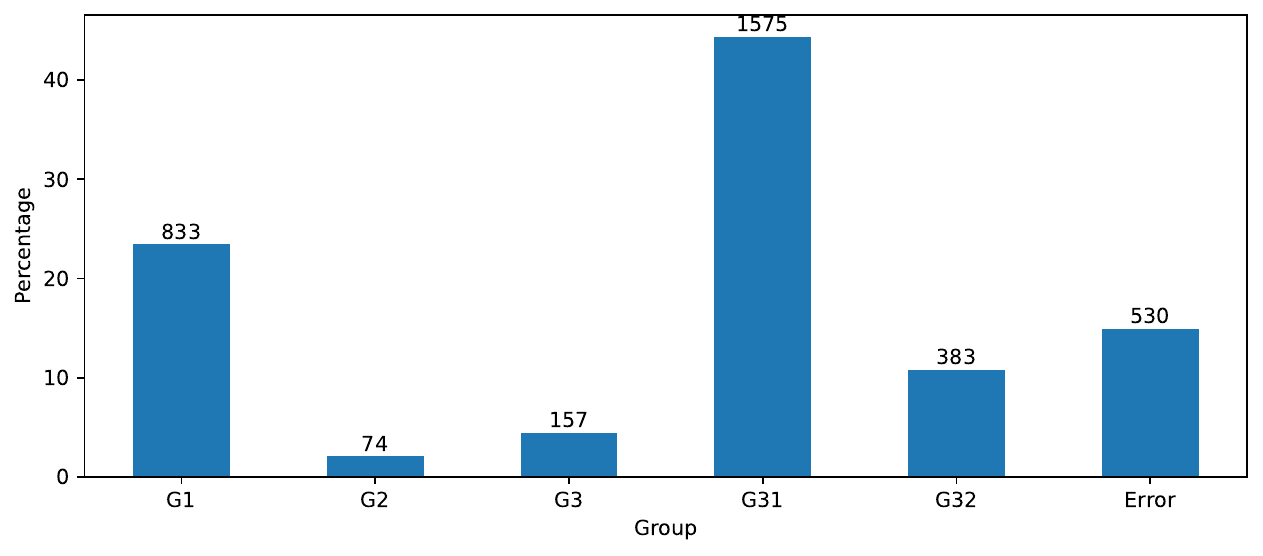}}
    \caption{Distribution of all 3552 variable pairs in the groups.}
    \label{fig:variable-pair-dist}
\end{figure}

An overview of the results of the grouping of variable pairs into groups G1 through G32 and the Error group is presented in Fig. \ref{fig:variable-pair-dist}.

In total, \rev{1958 pairs (55.1\%)} contained one error component, i.e., insufficient data for either bug or control group. They were therefore assigned to groups G31 or G32. Additionally, \rev{530} pairs (14.9\%) were assigned to the Error group since they had at best insufficient data for both sample groups. While we cannot make any claims regarding pairs in the Error group, for groups G31 and G32 we can at least say that they would not be assigned to G2 or G1, respectively, even if we had data in both sample groups.

Out of all valid pairs, most were assigned to G1, meaning that there is no relation between those variables, not within and also not between sample groups. This is expected with a broad set of variables, where variables were selected to describe the whole software engineering process and not handpicked for specific aspects based on hypothesized relationships.

In the following sections, we focus on the variable pairs that exhibit relations, namely pairs of G2 and G3, since these hint towards potentially causal effects within the software development process regarding bug introduction. We have \rev{74 pairs in group G2 (2.1\% of all possible pairs) and 157 pairs (4.4\%)} in group G3. We provide a full table of all variable pairs and their relation in the replication kit linked in Appendix~\ref{appendix:replication_kit}. Additionally, in Appendix~\ref{appendix:table_g2g3}, Table \ref{tbl-appendix:full_pairs_g2} and Table \ref{tbl-appendix:full_pairs_g3} contain the complete lists of the G2 and G3 relations.

\subsubsection{Frequency Threshold Sensitivity Analysis}
\label{subsub-sec:frequency-sensitivity}

\rev{To evaluate the robustness of our manual grouping procedure, we analyzed how nominal/numeric pair assignments change when the frequency threshold is varied. The contingency tables (Table \ref{tbl:freq6_8} and Table \ref{tbl:freq8_10}) summarize the transitions between groups as the threshold increases from 6 to 8 and from 8 to 10, respectively.}

\rev{As expected, raising the frequency threshold introduces additional \textit{data sparsity}. Pairs transition mostly from fully valid groups ($G1$, $G2$, $G3$) to partially valid groups ($G31$, $G32$) or to the Error group ($E$), which is expected with having less data available. With stricter \revv{(higher) thresholds,} less frequent pairs are excluded. However, the instability in group assignments is higher between thresholds $6$ and $8$ ($14$\% of pairs reassigned) than between $8$ and $10$ ($<9$\% reassigned). This suggests that the threshold of $8$ used in this paper is the preferred choice, reducing data sparsity while maintaining greater stability compared to lower thresholds.}

\begin{table}[]
    \centering
    \begin{minipage}{0.48\textwidth}
        \centering
        \begin{tabular}{@{}l|rrrrrrr@{}}
            \toprule
            \diagbox{from}{to} & G1 & G2 & G3 & G31 & G32 & E  & same \\ \midrule
            G1                 & -  & 0  & 0  & 4   & 0   & 0  & 0    \\
            G2                 & 0  & -  & 0  & 2   & 6   & 1  & 0    \\
            G3                 & 3  & 1  & -  & 11  & 4   & 1  & 0    \\
            G31                & 0  & 0  & 0  & -   & 2   & 44 & 0    \\
            G32                & 0  & 0  & 0  & 26  & -   & 39 & 0    \\
            E                  & 0  & 0  & 0  & 0   & 0   & -  & 0    \\
            same               & 0  & 0  & 0  & 0   & 0   & 0  & 906  \\ \bottomrule
        \end{tabular}
        \caption{Contingency table displaying the number of nominal/numeric pairs that are grouped differently between a frequency threshold of $6$ and $8$}
        \label{tbl:freq6_8}
    \end{minipage}
    % \hspace{0.75cm}
    \begin{minipage}{0.48\textwidth}
        \centering
        \begin{tabular}{@{}l|rrrrrrr@{}}
            \toprule
            \diagbox{from}{to} & G1 & G2 & G3 & G31 & G32 & E  & same \\ \midrule
            G1                 & -  & 0  & 0  & 3   & 0   & 1  & 0    \\
            G2                 & 0  & -  & 0  & 0   & 7   & 4  & 0    \\
            G3                 & 1  & 2  & -  & 5   & 12  & 1  & 0    \\
            G31                & 0  & 0  & 0  & -   & 0   & 26 & 0    \\
            G32                & 0  & 0  & 0  & 8   & -   & 20 & 0    \\
            E                  & 0  & 0  & 0  & 0   & 0   & -  & 0    \\
            same               & 0  & 0  & 0  & 0   & 0   & 0  & 960  \\ \bottomrule
        \end{tabular}
        \caption{Contingency table displaying the number of nominal/numeric pairs that are grouped differently between a frequency threshold of $8$ and $10$}
        \label{tbl:freq8_10}
    \end{minipage}
\end{table}

% #########################################################
% 
% #########################################################

\subsection{Significance of Results of the Statistical Tests}
\label{sub-sec:amount-sigificant-results}

\rev{After performing all possible comparisons of valid pairs with Fisher's exact test as well as Spearman's $\rho$, we evaluated the number of significant results against a significance threshold of $\alpha=5\%$. Of the $5680$ Fisher's exact tests conducted, $559$ ($9.8\%$) yielded significant results. Similarly, out of the $113$ Spearman's $\rho$ tests, 20 ($17.7\%$) were significant. Since the number of significant results clearly exceeds the threshold $\alpha$ for both tests, we conclude that our findings are unlikely to be spurious.}

% #########################################################
% 
% #########################################################

\subsection{Circular Network Diagrams of Related Variables}
\label{sub-sec:network-graphs}

\begin{figure}
    \centering
    \includegraphics[width=\textwidth]{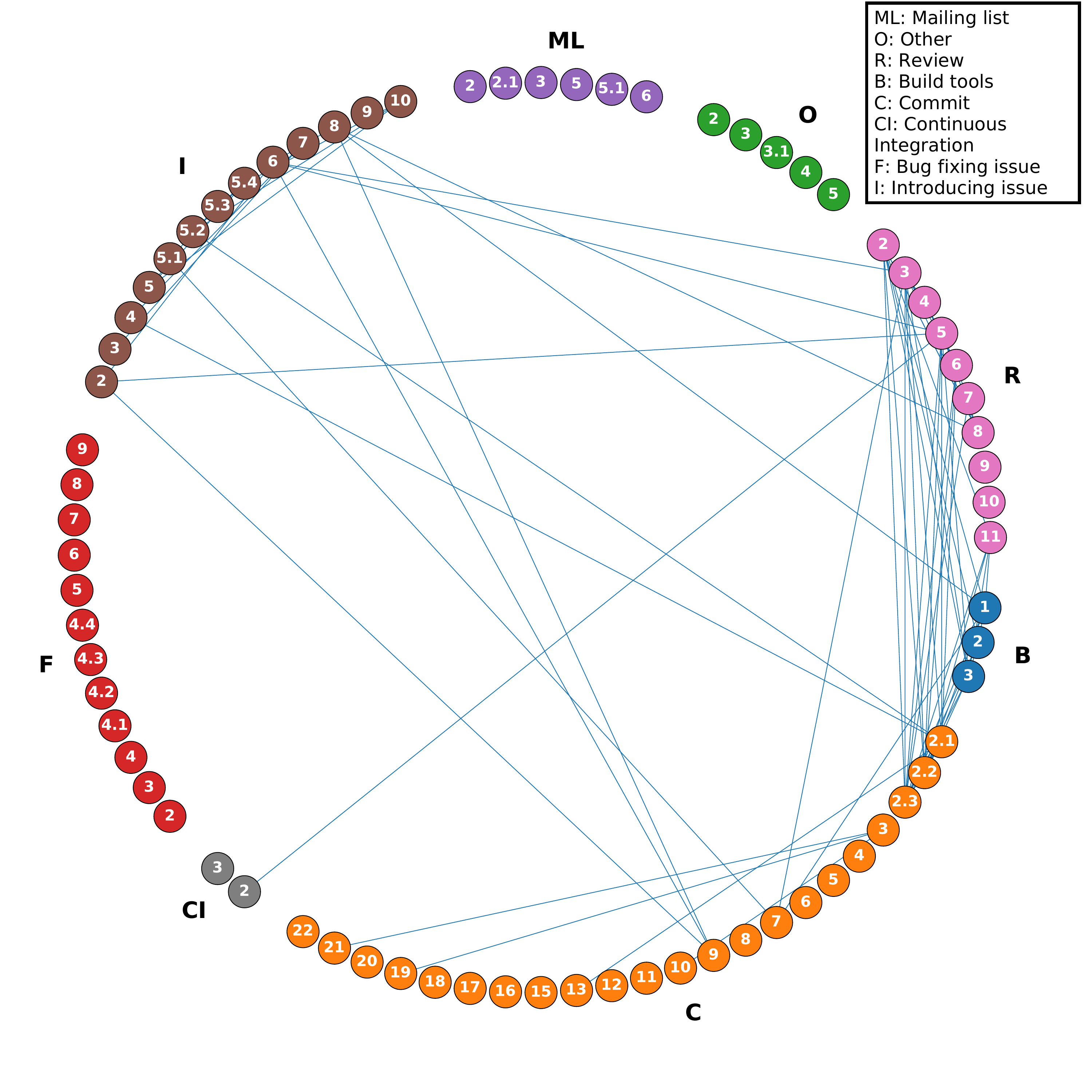}
    \caption{Network of group G2. Nodes are color-grouped by source, edges represent G2 relationships between pairs.}
    \label{fig:g2-network}
\end{figure}

\begin{figure}
    \centering
    \includegraphics[width=\textwidth]{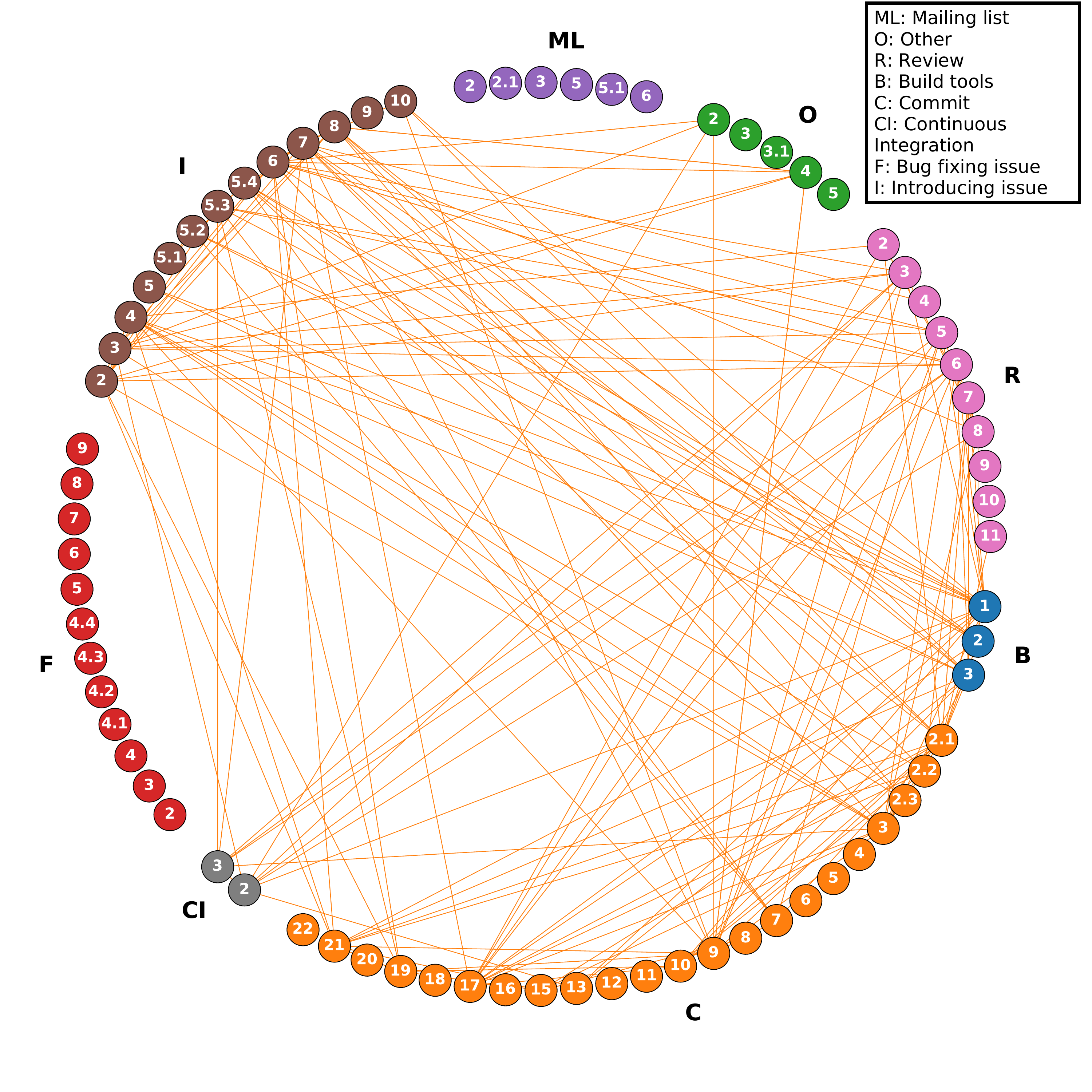}
    \caption{Network of group G3. Nodes are color-grouped by source, edges represent G3 relationships between pairs.}
    \label{fig:g3-network}
\end{figure}

In this section, we present two circular network diagrams: one for the variable pairs that are likely related and exhibit a relation that is the same between bug and control group (G2 - same relation, Fig. \ref{fig:g2-network}), and another for the variable pairs that exhibit a distinct relationship in either the bug or control group that is not present in the other (G3 - distinct relation, Fig. \ref{fig:g3-network}). The graphs present variables as nodes, colored by data source, and their respective relationships through edges. \revv{These edges can represent either behavioral relationships (reflecting developer actions or decisions), structural relationships (reflecting inherent properties or dependencies in the code or process), or contextual relationships (edges that describe interactions between behavioral and structural aspects).} From this high-level overview, we can make a number of initial observations that will be presented in this section. These relationships, G2 and G3, serve as a basis for all further results, \revv{and while our core contribution will be based on logical variable groups (Section \ref{sub-sec:logical-groups}), the logical groups can always be split up and investigated for their underlying individual relations}.

Fig. \ref{fig:g2-network} shows group G2 relations, in which we observe a cluster of tightly connected variables made up of \texttt{R}, \texttt{B}, and certain \texttt{C} variables, i.e., \texttt{C2.1}, \texttt{C2.2}, and \texttt{C2.3}. Further, the \texttt{I} variables are sparsely connected with each other. Another observation is that \texttt{ML}, \texttt{O}, and \texttt{F} variables are not connected on this graph at all.
The isolation of \texttt{ML} and \texttt{F} variables can be explained by data sparsity, i.e., they were mostly assigned to G31, G32, and the Error group. The \texttt{ML} variables are \textit{project-dependent} and \textit{exclusive}. In addition, the \texttt{F} variables are \textit{incomparable} since they only exist within the bug group and not in the control group. The isolation of \texttt{O} variables is not exclusively an effect of data sparsity. While \texttt{O} variables are infrequent and are not grouped into G2, they do show group G3 relations.

Fig. \ref{fig:g3-network} presents the G3 relations. We observe that \texttt{I}, \texttt{R}, \texttt{B}, and certain \texttt{C} variables frequently exhibit relations, i.e., distinct relations between bug and control group as per the definition of G3. With 18 edges, \texttt{B1 - Build tools} is the variable most frequently connected in G3. It is closely followed by \texttt{C2.1 - Commit type (current)}, \texttt{C3 - \# Files changed} and \rev{\texttt{I6 - Aspects of introducing issue discussion}} with 16 edges each. Overall, there is a rather dense network of variables where pairs exhibit signals through differences between bugs and the control group. We also observe the isolation of \texttt{ML} and \texttt{F} variables again, due to data sparsity, similar to the G2 graph.

% #########################################################
% 
% #########################################################

\subsection{Trends between Group G3 Variables}
\label{sub-sec:g3-how-relate}

\begin{table}[]
    \centering
    \begin{tabular}{@{}lr|rrr@{}}
        \toprule
        Type of Trend    & \#Pairs (Total) & Nominal  & Nominal/Numeric & Numeric \\ \midrule
        Different trends & 31              & 14       & 17              & 0       \\
        Opposing trends  & 4               & 0        & 4               & 0       \\
        Only BG          & \rev{79}        & \rev{56} & \rev{22}        & 1       \\
        Only CG          & \rev{43}        & \rev{22} & \rev{20}        & 1       \\ \bottomrule
    \end{tabular}
    \caption{Number of group G3 variable pairs that follow each trend. Opposing trends are also counted towards different trends.}
    \label{tab:g3-trend-count}
\end{table}

During the evaluation of group G3 variable pairs, we observed four trends, which we report by example. Fig. \ref{tab:g3-trend-count} reports the number of observations of each trend.

\subsubsection{Different Trends}
\label{subsubsec:different-trends}

Different trends are observed if both sample groups report a relation and the comparison between them reports a significant difference, yet a direction cannot be inferred. This can be the case for nominal pairs, since the Fisher's exact test does not provide insight into what constitutes the difference, but also for other pairs where the difference is complex (i.e., multiple labels constituting the difference in a nominal/numeric pair).
One example is the nominal pair \texttt{C15 - Documentation changes} and \texttt{C21 - changes of external dependencies}. It exhibits different trends between the sample groups and while the Fisher's exact test reported significant relations within both the bug and the control group, the comparison between them also yielded a p-value below the threshold of $\alpha = 0.05$. Therefore, we assume that the relationship is distinctly different.

\subsubsection{Opposing Trends}
\label{subsubsec:opposing-trends}

Opposing trends are a sub category of different trends. They were observed if a relation is different, and the direction can be inferred between the bug and control group. This can be the case for nominal and nominal/numeric pairs, but can potentially also exist for purely numeric pairs. One example is the nominal/numeric pair \texttt{I5.3 - Introducing issue assignee role} and \texttt{I7 - \# Introducing issue commits} exhibit opposing trends. Here, the bug group reports a relationship between the assignee role \rev{\textit{core}} and more \textit{commits} than the rest, while the control group reports a relation of \rev{\textit{core}} with fewer \textit{commits} being made to implement the introducing issue.
Regarding bug-introduction, these opposing trends are a hint towards a potential treatment, as identifying the cause of this trend can lead towards the identification of software engineering practices that reduce the introduction of bugs.

\subsubsection{Significant Trend Only in the Bug Group}
\label{subsubsec:trend-only-bg}

A significant trend in the bug group hints towards an issue unique to changes that are later identified as bug-introducing.
The nominal/numeric pair \rev{\texttt{C15 - Documentation changes} and \texttt{C3 - \# Files changed} is one example. Here, in the bug group, more files are changed if the commit contains documentation changes than if the documentation is not changed. }
This pattern can hint towards a bad practice or the absence of a good practice that affects the commits of the bug group and contributed to them becoming bug-introducing. If we can identify such a practice, we might be able to mitigate bug-introduction through countermeasures.

\subsubsection{Significant Trend Only in the Comparison Group}
\label{subsubsec:trend-only-cg}

A significant trend in the control group hints towards a practice that only affects those changes that are later found to be non-bug-introducing.
The numeric pair of \texttt{C3 - \# Files changed} and \texttt{R6 - \# Rounds of code review} exhibits such a significant trend only in the control group. We observe a clearly positive correlation coefficient reported by the Spearman's $\rho$ and a low p-value, while the relation in the bug group is insignificant.
This pattern can hint towards a good practice or the absence of a bad practice, that affects the commits of the control group and might contribute to them not introducing bugs. If we can identify such a practice, we might be able to encourage the application of this practice to reduce bugs.

% #########################################################
% 
% #########################################################

\subsection{Logical Variable Groups}
\label{sub-sec:logical-groups}

In previous sections, we outlined the trends observed between variable pairs and their relation to bug-introducing changes. However, related work (e.g, \cite{querelWarningIntroducingCommitsVs2021a, kononenkoInvestigatingCodeReview2015b, ghafariTestabilityFirst2019a}) usually considers groups of variables depending on the aspect of the bug-introducing software engineering process that they cover. To identify potential confounders, mediators, and backdoors for such studies, we map our variables to a suitable abstraction level. We achieve this by creating 19 logical groups, structured by conceptual or logical similarities of variables representing \textit{Inputs} and \textit{Techniques} in software engineering practices. This grouping process involved \revv{iterative} discussions among the three authors\revv{, where each variable was evaluated against the definitions of an initial set of logical groups. Disagreements were resolved through consensus, and logical groups were adjusted, added, or dropped accordingly. The final assignments of variables are documented in Appendix~\ref{appendix:table_associations} (see Table~\ref{tbl-appendix:logically-associated-variables}). The distinct group definitions, each representing areas of the software development process with potential relevance for past and future studies, are listed below.}

\begin{itemize}
    \item \textbf{Issue management (basic)}: Variables associated with basic functionality of issue management systems, i.e., issue labels, types, and severities.
    \item \textbf{Issue management (advanced)}: Variables associated with advanced functionality of issue management systems, e.g., duplicate tracking or commit linking.
    \item \textbf{Issue management roles}: Roles of developers involved in the issue creation.
    \item \textbf{Discussion aspects (action)}: \revv{Actions of discussion in any recorded discussion, (e.g., requests, answers, statements).}
    \item \textbf{Discussion aspects (target)}: Targets of any recorded discussion.
    \item \textbf{Discussion aspects (topic)}: Topics of any recorded discussion.
    \item \textbf{Discussion frequency}: Frequency metrics of any recorded discussion.
    \item \textbf{Discussant roles}: Roles of discussants in any recorded discussion.
    \item \textbf{Discussion tools}: Tools, i.e., types of media, in which discussants take place.
    \item \textbf{Review Scope}: \revv{Factors describing the extend of the review, e.g., the s}ize of changes reviewed, as well as on which branch the review was conducted.
    \item \textbf{Commit Type}: Variables describing the type and scope of the introducing commit.
    \item \textbf{Continuous Integration}: Variables associated with CI and building \revv{of introducing commits}.
    \item \textbf{Testing}: Variables associated to testing of introducing commits.
    \item \textbf{Refactorings}: Variables associated to refactorings in introducing commits.
    \item \textbf{Design Changes}: Variables associated to design changes in introducing commits.
    \item \textbf{Dependencies}: Variables associated to dependency changes in introducing commits.
    \item \textbf{Documentation}: Variables associated to documentation changes in introducing commits.
    \item \textbf{Bug type}: Variables associated with the type and scope of the bug.
    \item \textbf{Bug-covering quality assurance measures}: Variables associated to quality assurance (QA) of bug-introducing commits.
\end{itemize}

For the analysis of these logical groups, we counted the G2 and G3 relations between each logical group through their underlying variables. This count was then normalized against the maximum number of possible relations, excluding those with an error component (i.e., groups G31, G32 and Error). For better interpretability, the resulting fractions were mapped to a scale:

\begin{itemize}
    \item \textbf{Weak}: values in the interval (0, 0.33]
    \item \textbf{Medium}: values in the interval (0.33, 0.66]
    \item \textbf{Strong}: values in the interval (0.66, 1]
\end{itemize}

\begin{figure}
    \centering
    \includegraphics[width=\textwidth]{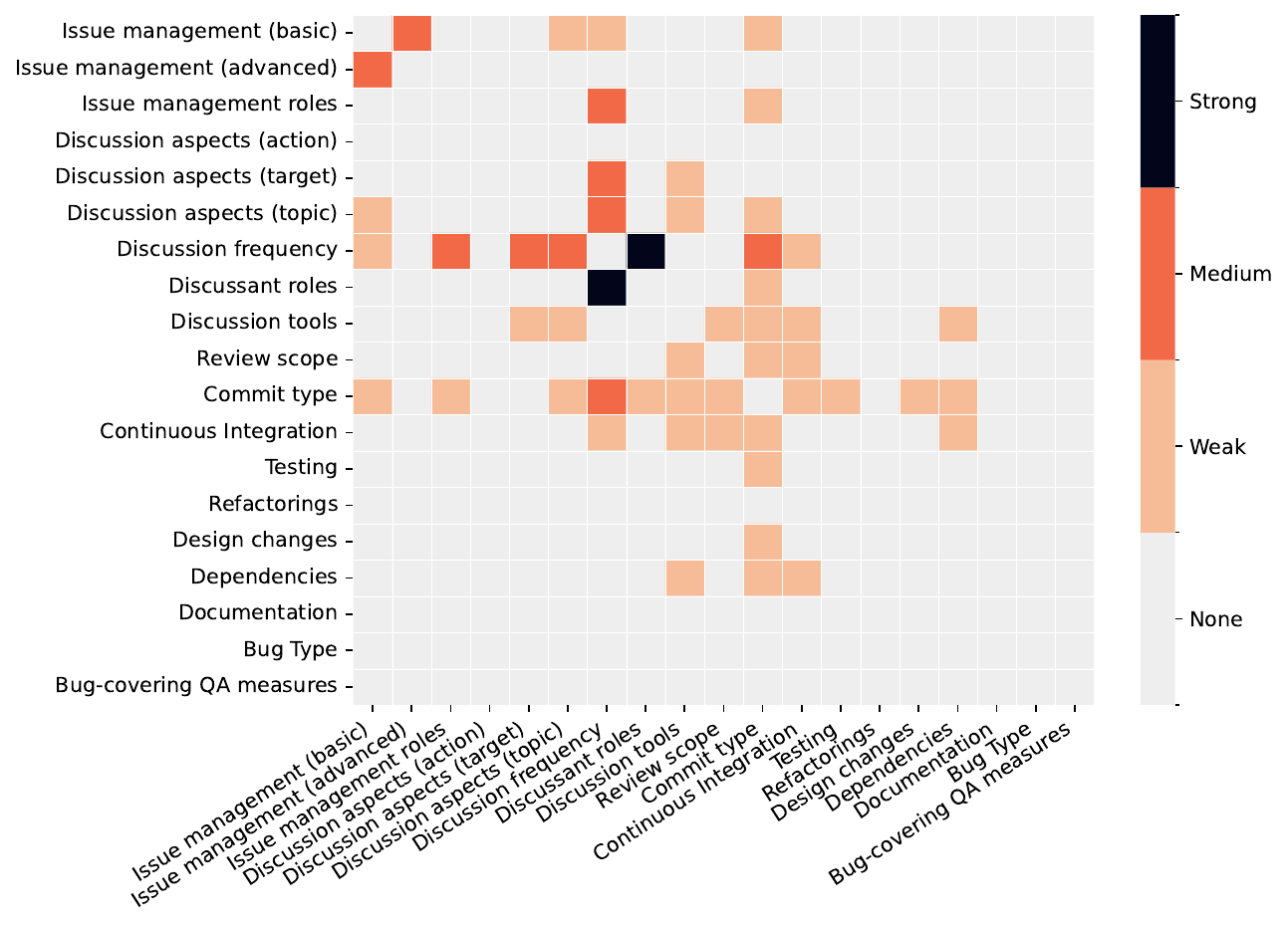}
    \caption{Strength of the G2 relations between the logical groups. }
    \label{fig:association_g2_relations_heatmap}
\end{figure}

\begin{figure}
    \centering
    \includegraphics[width=\textwidth]{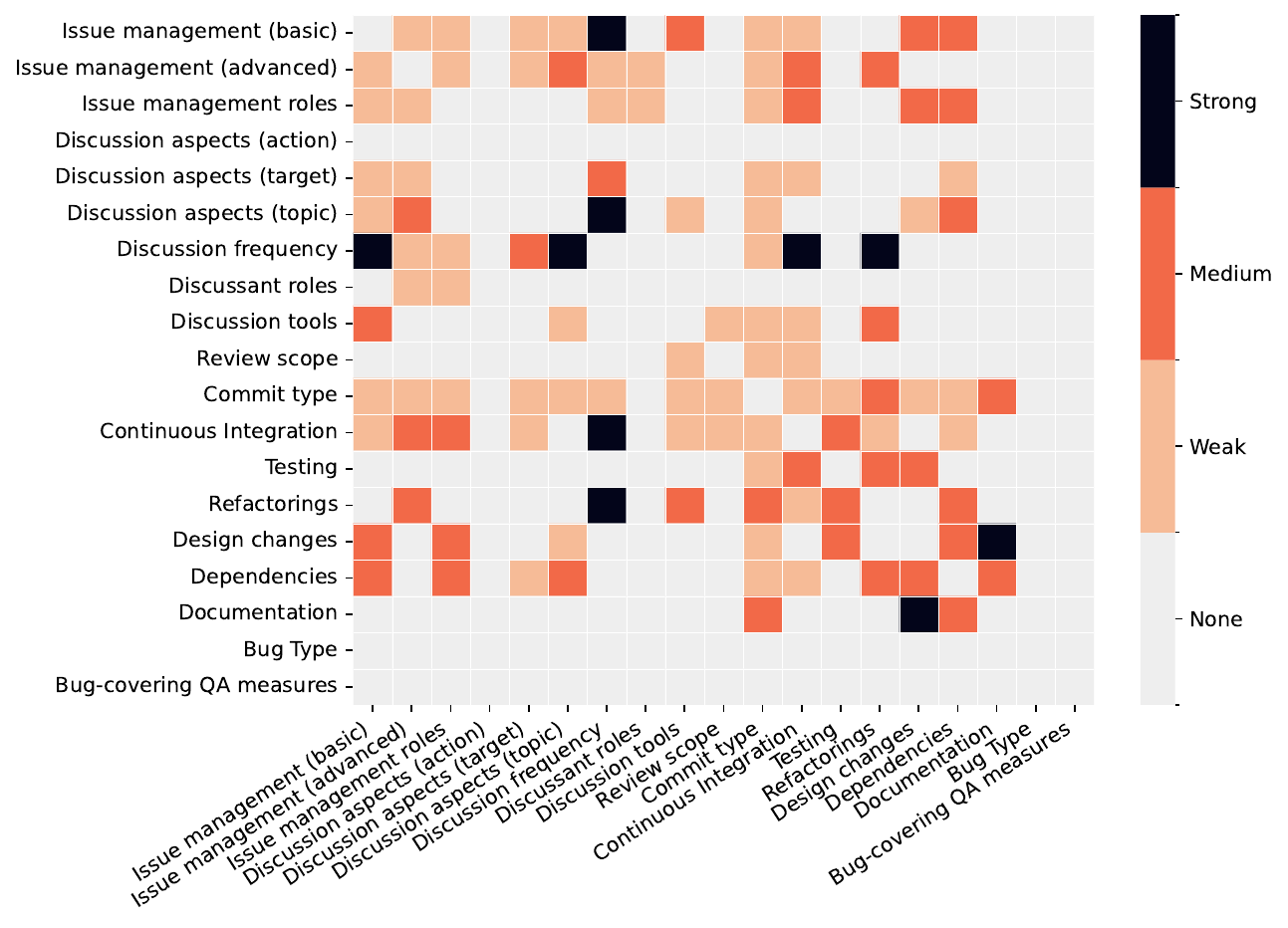}
    \caption{Strength of the  G3 relations between the logical groups.}
    \label{fig:association_g3_relations_heatmap}
\end{figure}

A value of 0 indicates no relation and is shown as \textbf{None}. The relations are reported through the heatmaps in Fig. \ref{fig:association_g2_relations_heatmap} and Fig. \ref{fig:association_g3_relations_heatmap}. Since the logical groups \textit{bug type} and \textit{bug-covering: QA measures} only contain variables that are \textit{incomparable}, we do not observe any relations towards them. These will not be considered further.

Many logical groups are not connected by G2 relations. Those that are connected may exhibit relations within the software development process that are not linked to bug-introduction. These relations can take the shape of mediator paths, forges, or confounding factors independent of bug-introduction and need to be considered when creating causal models. However, they are unlikely to have a significant effect on the outcome.

We observe more connections between logical groups for G3 relations. In total, five pairs exhibit strong connections, and all \rev{but one} involve \textit{discussion frequency}. \textit{Discussion frequency} is strongly linked to \textit{issue management (basic)}, \textit{discussion aspects (topic)}, \textit{continuous integration}, and \textit{refactorings}. \rev{Further, \textit{documentation} is strongly linked to \textit{design changes}.} For all the existing connections, it is important to be aware of potential effects between variables of the logical groups and to control for confounding factors to get reliable estimates of the causal effects towards bug introduction.

% #########################################################
% 
% #########################################################

\section{Discussion}
\label{sec:discussion}

In this section, we discuss the results of our study. We first discuss how our logical groups and their relations can be used as a framework for forming hypotheses on the origin of potential confounders in related work and future studies. We then address our research question on the software engineering practices used during the development of bug-introducing changes. Finally, we provide an outlook on future work.

% #########################################################
% 
% #########################################################

\subsection{Hypotheses on the impact of our Findings}

Our results, specifically the relations between logical groups in Section \ref{sub-sec:logical-groups}, provide researchers with a framework by which they can assess the scope of their studies regarding the origin of potential intermediate effects and confounders.
In Section \ref{subsubsec:impact-related-work} we revisit selected related work from Section \ref{sec:relatedWork} and apply that framework to these examples by mapping their variables to our logical groups. We emphasize that our findings do not imply that there are mistakes in the prior work. Instead, we propose two hypotheses on the broader implications of our findings: one for G2 relations and one for G3 relations.
In Section \ref{subsubsec:impact-future-studies} we then explore how future studies can use the framework to identify areas of interest to complement the area of the software engineering process they may be investigating.

% #########################################################
% 
% #########################################################

\subsubsection{Impact on Related Work}
\label{subsubsec:impact-related-work}

In this section, we discuss the impact of our results on related work. For this, we use the relations between our logical groups as a framework for assessing the origin of potential intermediate effects and confounders on the scope of related work. As examples, we choose the works of \cite{querelWarningIntroducingCommitsVs2021a}, \cite{ghafariTestabilityFirst2019a}, and \cite{kononenkoInvestigatingCodeReview2015b} introduced in Section \ref{sec:relatedWork}. To apply the framework, we map their scope (i.e., their variables) to our logical groups. We then analyze the relations between the mapped logical groups and the other logical groups to identify potential areas of interest regarding causal relations.

\begin{figure}
    \centering
    \begin{minipage}[b]{0.9\textwidth}
        \centering
        \textbf{G2 relations of logical groups}
        \includegraphics[width=\textwidth]{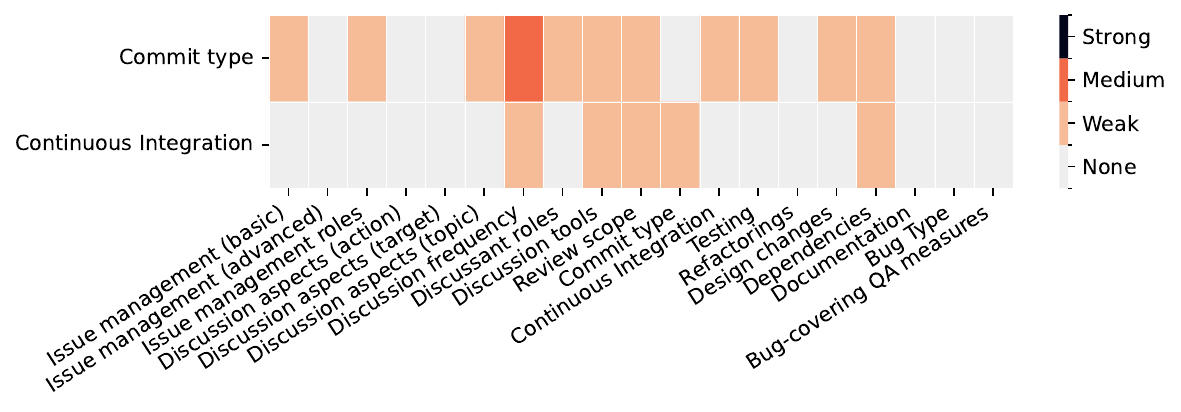}
    \end{minipage}
    \vfill
    \begin{minipage}[b]{0.9\textwidth}
        \centering
        \textbf{G3 relations of logical groups}
        \includegraphics[width=\textwidth]{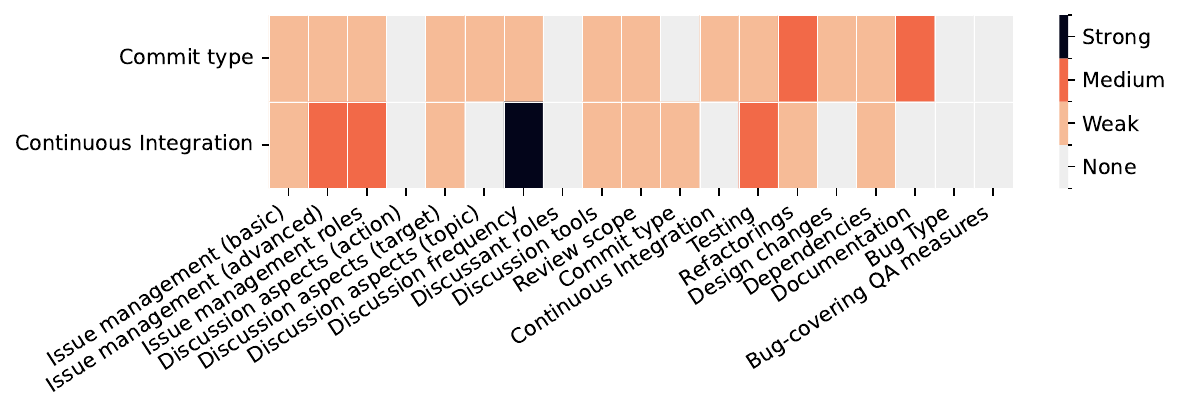}
    \end{minipage}
    \caption{Relations of logical groups mapped to the variables of \cite{querelWarningIntroducingCommitsVs2021a}.}
    \label{fig:logical-groups-querel}
\end{figure}

\cite{querelWarningIntroducingCommitsVs2021a} studied the effects of code changes, developer experience and build warnings on bug-introduction. In Fig. \ref{fig:logical-groups-querel}, we map the independent variables of their logistic regression models to two of our logical groups: \textit{commit type} and \textit{continuous integration}, shown on the y-axis. The x-axis shows all the logical groups introduced in Section \ref{sub-sec:logical-groups}.
The G2 relations indicate that there is a medium intermediate effect within the software engineering process towards their variables from \textit{discussion frequency}. Further, there are \rev{eleven} more logical groups that relate at least weakly.
The G3 relations hint towards most logical groups being linked with their variables at a medium or weak degree, and one strong link with \textit{discussion frequency}. All G3 relations are potential confounders and are of high relevance as well as potential threats to validity.

\begin{figure}
    \centering
    \begin{minipage}[b]{0.9\textwidth}
        \centering
        \textbf{G2 relations of logical groups}
        \includegraphics[width=\textwidth]{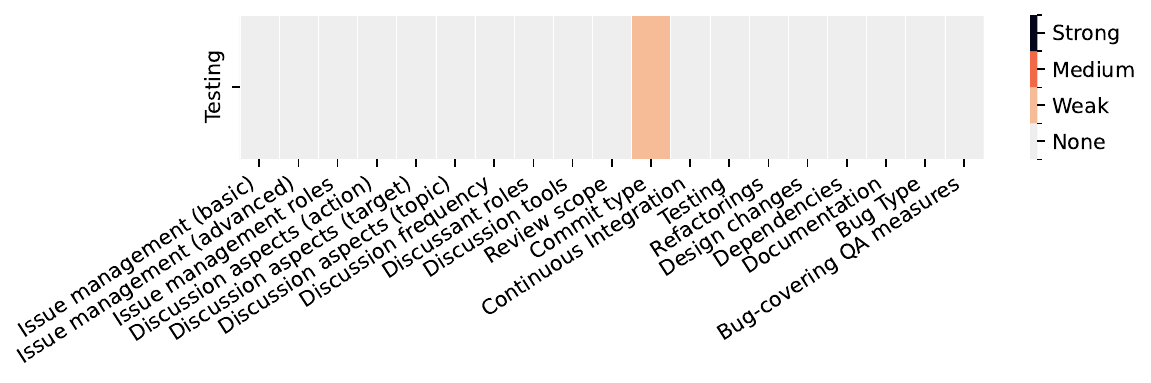}
    \end{minipage}
    \vfill
    \begin{minipage}[b]{0.9\textwidth}
        \centering
        \textbf{G3 relations of logical groups}
        \includegraphics[width=\textwidth]{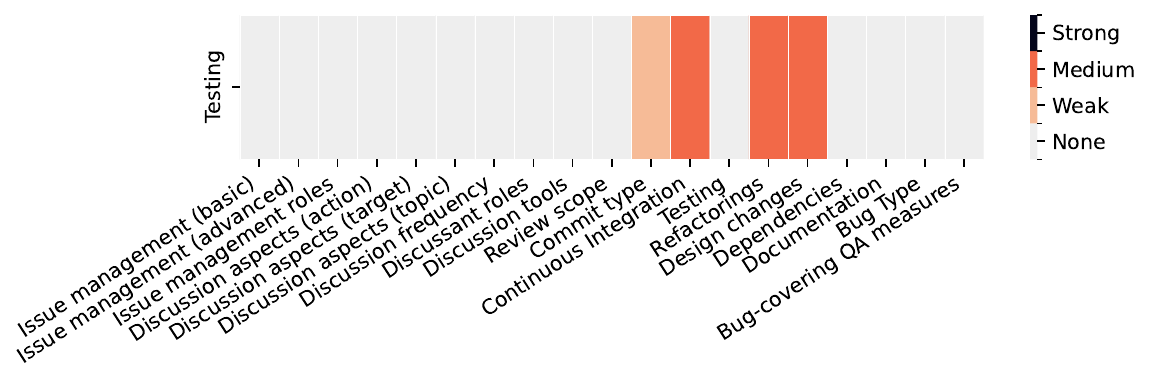}
    \end{minipage}
    \caption{Relations of logical groups mapped to the variables of \cite{ghafariTestabilityFirst2019a}.}
    \label{fig:logical-groups-ghafari}
\end{figure}

\cite{ghafariTestabilityFirst2019a} studied the effects of test coverage on bug-introducing changes. We map their subject, test coverage, to our logical group \textit{testing}, as depicted in Fig. \ref{fig:logical-groups-ghafari}. We observe only one weak G2 link with \textit{commit type}, \rev{one} weak G3 link with \textit{commit type}, and three medium G3 links with \textit{continuous integration}, \textit{refactorings} and \textit{design changes}. Based on this, it is therefore advisable to evaluate controlling for at least the presence of refactorings and design changes in commits, as well as the outcome of CI runs.

\begin{figure}
    \centering
    \begin{minipage}[b]{0.9\textwidth}
        \centering
        \textbf{G2 relations of logical groups}
        \includegraphics[width=\textwidth]{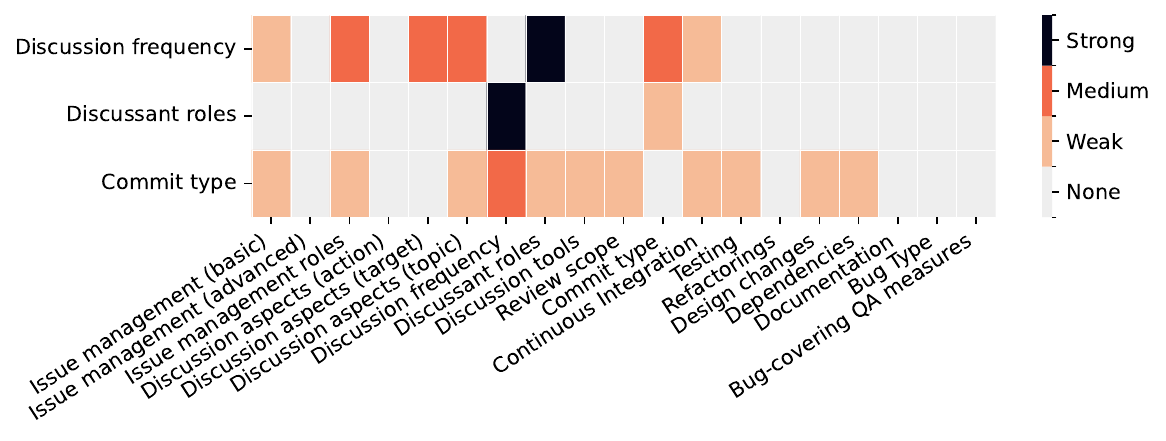}
    \end{minipage}
    \vfill
    \begin{minipage}[b]{0.9\textwidth}
        \centering
        \textbf{G3 relations of logical groups}
        \includegraphics[width=\textwidth]{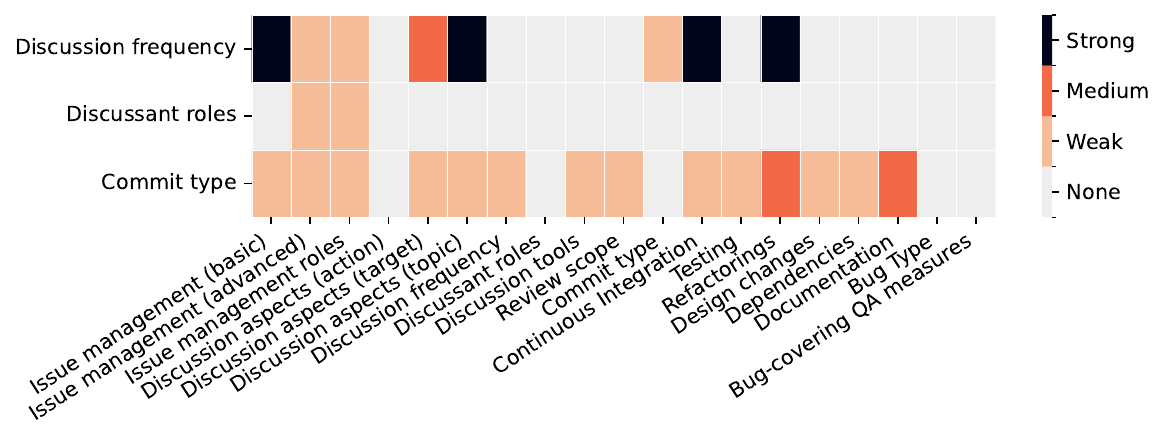}
    \end{minipage}
    \caption{Relations of logical groups mapped to the variables of \cite{kononenkoInvestigatingCodeReview2015b}.}
    \label{fig:logical-groups-kononenko}
\end{figure}

\cite{kononenkoInvestigatingCodeReview2015b} studied code review quality and its effect on bug-introduction. Since they consider a wide range of variables, we map these to three of our logical groups: \textit{discussion frequency}, \textit{discussant roles}, and \textit{commit type}. From Fig. \ref{fig:logical-groups-kononenko} we make the following observations: there are strong intermediate effects between \textit{discussion frequency} and \textit{discussant roles}. However, their variables also cover these logical groups at least partially. All G2 relations that they do not at least partially account for are either medium or weak, but should still be considered.
\rev{Regarding potential confounders, we suspect strong G3 connections of their scope with \textit{issue management (basic)}, \textit{discussion aspects (topic)}, as well as \textit{continuous integration} and \textit{refactorings}, and multiple additional medium and weak connections.}

\begin{mdframed}[linewidth=1pt]
    We summarize the impact of our findings on related work with two hypotheses:

    \vspace{5px}

    \noindent\textbf{H1-G2:} If related work bases their analysis on variables that are associated with one logical group, but does not consider those linked with G2 relations, we hypothesize that they may miss causal aspects within the software development process that can manifest as mediators, forges, or confounders, independent of bug-introduction.

    \noindent\textbf{H1-G3:} If related work bases their analysis on variables that are associated with one logical group, but does not consider those linked with G3 relations, we hypothesize that they may miss to consider potential confounders on bug-introduction, which likely threatens the validity of their results.
\end{mdframed}

% #########################################################
% 
% #########################################################

\subsubsection{Impact on Future Studies}
\label{subsubsec:impact-future-studies}

Our findings do not only allow us to interpret the results of previous related work, but also enable future studies to consider potential causal factors. We anticipate that our study allows studies with a more narrowly focused research to quickly identify areas of interest outside their scope, particularly potential confounders and intermediate effects for their variables, and consequently address them. Either by directly controlling for them, or at least by treating them as threats to validity.

Further, our study will also guide our own plans for causal modeling outlined in Section \ref{sub-sec:future-work}.

\begin{mdframed}[linewidth=1pt]
    Future work can use the relationships that we have identified, either on a variable or on a logical group level, to identify areas of interest regarding intermediate variable relations and confounders.
\end{mdframed}

% #########################################################
% 
% #########################################################

\subsection{Answer to the Research Question}
\label{sub-sec:discussion-research-question}

This section summarizes our findings and discussion to address the research question that drives this study.

\begin{itemize}
    \item[] \textbf{RQ:} Which software engineering practices are applied during the development process of bug-introducing code of open-source projects, and how does this relate to non-bug-introducing changes?
\end{itemize}

To investigate practices that were applied during the development process of bug-introducing code, we used our 81 variables to create 19 logical groups capturing different practices. These logical groups resemble practices through combinations of variables for \textit{Inputs} and \textit{Techniques}. We find that 17 of those logical groups have valid relationships within the variables and describe the practices used in the development process. Two logical groups were invalid, due to \textit{Incomparability} of their underlying variables.

We further find that all valid logical groups exhibit potentially causal relationships, which are identified in our study by group G2 and G3 relations. Therefore, none of the logical groups are fully independent, which is critical for any study that does not consider factors outside its primary scope. While G2 relations between practices are only suspect of intermediate effects, G3 relations pose direct threats to validity due to potential confounding that needs to be addressed. Therefore, we argue that not considering these relations misrepresents the effects within the development process (G2) and may even falsify results (G3).

\begin{mdframed}[linewidth=1pt]
    The software engineering practices applied during the development process of bug-introducing code are interdependent. All investigated practices exhibit distinct relations (G3) to other practices that may confound the underlying variables. This highlights the need for careful management of the relations of variables to address threats to validity.
\end{mdframed}

% #########################################################
% 
% #########################################################

\subsection{Future Work}
\label{sub-sec:future-work}

Our results lay the groundwork for future studies on the causal relationships between software engineering practices and bug-introducing changes. We can use our sets of variables as a foundation for modeling causal Directed Acyclic Graphs (DAGs), which in turn will enable us to quantify the relationships between the variables through causal inference techniques.
Beyond manual causal modeling by enriching the information about the relationships with expert knowledge, the availability of large amounts of data in software repositories opens up the possibility for automated approaches, i.e., causal discovery. Examples for causal discovery include the PC algorithm \citep{spirtesCausationPredictionSearch2000a} and its performance optimized variants.
With a detailed DAG, we can then apply regression-based causal inference techniques, data-intensive matching, e.g., based on a propensity score, or approaches like double machine learning, to quantify the causal effects encoded by our model \citep{huntington-kleinEffectIntroductionResearch2022a}.

\begin{figure}
    \centering
    \includegraphics[width=0.8\textwidth]{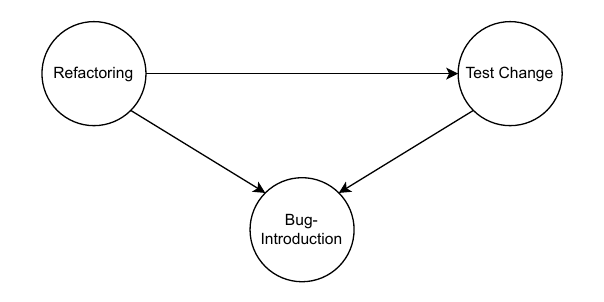}
    \caption{Example for a (simplified) causal model between refactorings, test changes and bug-introduction.}
    \label{fig:small-causal-model}
\end{figure}

One example is the relationship between \texttt{C10 - Test changes} and \texttt{C17 - Refactorings} in non-bug-introducing commits. Given their assignment to group G3, we hypothesize that both have an effect on the bug introduction. With causal modeling, we aim to pinpoint the direction of this effect. Do test changes cause refactorings, or do refactorings cause tests to change? Alternatively, did we observe a spurious correlation? We can argue that refactorings trigger tests being changed, since adjustments in function implementation (e.g., renaming variables or simplifying loops) require corresponding updates in, especially low-level, tests. Our causal model, for the aforementioned variables, is depicted in Fig. \ref{fig:small-causal-model}.

By gathering the correct observational data, future work can estimate the causal effect of refactorings and test changes on bug introduction through causal inference. In this case, test changes are confounded by refactorings, creating an open backdoor path that requires adjustment. To accurately estimate the effects, we can employ the aforementioned matching techniques, such as propensity score matching, to control for the confounder. In contrast, refactorings have a direct causal effect on bug introduction, which can also be quantified using regression models or other direct estimation methods.

Although simplified, this example outlines the potential of causal modeling, discovery, and inference, which we plan to explore further in future work.

\begin{mdframed}[linewidth=1pt]
    \rev{Our results enable future studies to make informed decisions about potential treatment practices. By modeling causal graphs (DAGs), researchers can systematically explore how confounders might influence the relationship between development practices and bug-introduction. The application of causal inference (e.g., methods such as propensity score matching and regression adjustment) can further support the understanding of the effect of observed and unobserved confounders. The robustness of these inferred relationships can then be tested through simulation or, where feasible, qualitative studies (e.g., developer interviews) to validate assumptions about unmeasured factors.}
\end{mdframed}

% #########################################################
% 
% #########################################################

\section{Threats to Validity}
\label{sec:threats}

We report the threats to validity from the perspective of construct, internal, and external validity as well as the reliability of our results.

\subsection{Construct Validity}
Construct validity refers to the extent to which the measures used in the study accurately represent the theoretical constructs they are intended to measure.
The variables used in the study were reviewed by all authors and were further peer-reviewed in the pre-registration. We therefore assume that the chosen variables adequately represent the development process regarding bug-introducing changes. However, due to the data availability limitations discussed in Section \ref{subsub-sec:var-ds-limitations}, not all were considered during the final analyses. For the control group, we further assume that since the sampled commits are not bug-introducing, data points obtained from directly related issues, reviews, etc., are also representative of non-bug-introducing changes. \rev{We acknowledge the limitation that, due to the limited sample size of 71 bug-introducing commits in the dataset provided by \cite{rodriguez-perezHowBugsAre2020a}, effects of underrepresented subgroups in either sample group may be missed}.
We \rev{also} acknowledge that this study measures correlations, which may not be causal, and our methods do not allow us to rule out confounding factors. Therefore, we do not make claims in our study that imply evidence for causality. Our observations are intended to guide future work in establishing causal theories through more qualified techniques.

\subsection{Internal Validity}
Internal validity refers to the extent to which the causal conclusions drawn from the study are warranted.
As an exploratory study, our work has yet to confirm the relations that we observed through qualified techniques, such as causal modeling and inference. However, our analysis of statistical significance suggests that the correlations we identified are unlikely to be due to chance and therefore merit further investigation.
Furthermore, we identified the following scenarios, in which individual group assignments can be based on spurious correlations: a pair was assigned to group G2, but our variables share a confounder that needs to be controlled for. This can mean that A) we measured spurious correlations within the sample groups, and the pair should have been assigned to group G1; or B) the influence of the confounder hides a distinct relation, and the pair should have been assigned to group G3.

\subsection{External Validity}
External validity refers to the extent to which the results of the study can be generalized to other settings.
Since our study is limited to two open source projects, Elasticsearch and Nova, our results may not generalize to other projects and especially not to other types of projects. This includes closed-source projects, projects that use other programming languages than Python and Java, and projects that rely on \revv{different toolchains and practices. We further acknowledge that the use of toolchains is not standardized between Elasticsearch and Nova, which means that opposing trends between the two may cancel each other out or that some trends are only present in one project but not in the other (e.g., mailing lists, which are not available for Elasticsearch).} Future research should validate these findings in a broader range of projects to assess their generalizability.

\subsection{Reliability}
Reliability refers to the consistency and stability of the results.
The subjective nature of multiple variables poses a non-negligible threat. Particularly for the aspects of discussion and other open-code variables, but also for predefined variables with ambiguous context. Different coders might arrive at different codes. We mitigated this by using inductive coding with two raters. In addition, we report the inter-rater agreement for transparency in Section \ref{subsub-sec:agreement} and resolved all disagreements involving both raters and a third rater if needed. Similarly, the assignment of variables to logical groups is subject to the individual experience of the authors, and different researchers might propose different groupings.

% #########################################################
% 
% #########################################################

\section{Conclusion}
\label{sec:conclusion}

With the research conducted in this study, we introduced a novel dataset of relationships between variables that describe the software engineering process around bug-introducing changes. We further analyzed the relation encoded in that dataset and discussed patterns of practices that are related to bug introduction. Moreover, we identified logical associations between the variables to be able to evaluate the effects between the resulting logical groups and to translate those to software engineering practices.

We conclude that software engineering practices applied during the development process of bug-introducing code are interdependent. Additionally, we established that our findings have an impact both on prior related work and future work by setting up a framework through which the scope of studies can be checked against potential outside influence of intermediate effects and confounding. With this framework, we analyzed selected related work, identified areas of interest for this influence, and formulated two hypotheses regarding the impact on them.

The work conducted in this paper lays the foundation for future work in which we will attempt to understand \textit{why} bugs are introduced, through the application of causal modelling, causal discovery, and causal inference.

\section*{Declarations}

\subsection*{Ethical Approval}

This study adheres to the ethical guidelines of academic publications. The data analyzed in our study is based on publicly available information from software repositories. To ensure the protection of individual privacy and the responsible use of the data aggregated by this study, our additional datasets are published in a pseudonymized form. Furthermore, our research does not evaluate the quality of work of individual developers. Instead, we focus on analyzing anonymous and aggregated activities within software repositories.

\subsection*{Informed Consent}

This study does not involve direct human participants and instead relies on publicly accessible data from software repositories. To respect individual privacy, our additional datasets are published in a pseudonymized form.

\subsection*{Author Contributions}

S.H. provided the initial idea for the study, including the methods. All authors contributed to the improvement of the methods and the final design of the study. A.MH. implemented solutions for manual labeling and collected the required metadata. L.S. extended the SmartSHARK platform for automatic data collection. Both A.MH. and L.S. conducted the manual labeling and analyses of the data. S.H. mediated in case of disagreements during the labeling process and advised on the execution of the analyses. S.H. wrote the initial manuscript of the registered report. L.S. wrote the initial manuscript of the main paper. All authors gave critical feedback on the registered report and main text.

\subsection*{Funding}

Open Access funding enabled and organized by Projekt DEAL. The authors did not receive support from any organization for the submitted work.

\subsection*{Conflicts of Interest}

The authors have no competing interests to declare that are relevant to the content of this article.

\subsection*{Clinical Trail Number}

Not applicable.

\subsection*{Data Availability Statement}

The datasets generated and/or analyzed during the current study are available in the Zenodo repositories referenced in Appendix~\ref{appendix:replication_kit}.

%%===========================================================================================%%
%% If you are submitting to one of the Nature Portfolio journals, using the eJP submission   %%
%% system, please include the references within the manuscript file itself. You may do this  %%
%% by copying the reference list from your .bbl file, paste it into the main manuscript .tex %%
%% file, and delete the associated \verb+\bibliography+ commands.                            %%
%%===========================================================================================%%

\bibliography{sn-bibliography}% common bib file

@inproceedings{bellerDichotomyDebuggingBehavior2018,
  title = {On the Dichotomy of Debugging Behavior among Programmers},
  booktitle = {Proceedings of the 40th {{International Conference}} on {{Software Engineering}}},
  author = {Beller, Moritz and Spruit, Niels and Spinellis, Diomidis and Zaidman, Andy},
  year = {2018},
  month = may,
  pages = {572--583},
  publisher = {ACM},
  address = {Gothenburg Sweden},
  doi = {10.1145/3180155.3180175},
  urldate = {2025-03-21},
  isbn = {978-1-4503-5638-1},
  langid = {english}
}

@article{bockAutomaticCoreDeveloperIdentification2023a,
  title = {Automatic {{Core-Developer Identification}} on {{GitHub}}: {{A Validation Study}}},
  shorttitle = {Automatic {{Core-Developer Identification}} on {{GitHub}}},
  author = {Bock, Thomas and Alznauer, Nils and Joblin, Mitchell and Apel, Sven},
  year = {2023},
  month = nov,
  journal = {ACM Transactions on Software Engineering and Methodology},
  volume = {32},
  number = {6},
  pages = {1--29},
  issn = {1049-331X, 1557-7392},
  doi = {10.1145/3593803},
  urldate = {2025-03-21},
  langid = {english}
}

@article{borleAnalyzingEffectsTest2018a,
  title = {Analyzing the Effects of Test Driven Development in {{GitHub}}},
  author = {Borle, Neil C. and Feghhi, Meysam and Stroulia, Eleni and Greiner, Russell and Hindle, Abram},
  year = {2018},
  month = aug,
  journal = {Empirical Software Engineering},
  volume = {23},
  number = {4},
  pages = {1931--1958},
  issn = {1382-3256, 1573-7616},
  doi = {10.1007/s10664-017-9576-3},
  urldate = {2025-03-21},
  langid = {english}
}

@article{catolinoNotAllBugs2019,
  title = {Not All Bugs Are the Same: {{Understanding}}, Characterizing, and Classifying Bug Types},
  shorttitle = {Not All Bugs Are the Same},
  author = {Catolino, Gemma and Palomba, Fabio and Zaidman, Andy and Ferrucci, Filomena},
  year = {2019},
  month = jun,
  journal = {Journal of Systems and Software},
  volume = {152},
  pages = {165--181},
  issn = {01641212},
  doi = {10.1016/j.jss.2019.03.002},
  urldate = {2025-03-21},
  langid = {english}
}

@inproceedings{femmerImpactPassiveVoice2014,
  title = {On the Impact of Passive Voice Requirements on Domain Modelling},
  booktitle = {Proceedings of the 8th {{ACM}}/{{IEEE International Symposium}} on {{Empirical Software Engineering}} and {{Measurement}}},
  author = {Femmer, Henning and Ku{\v c}era, Jan and Vetr{\`o}, Antonio},
  year = {2014},
  month = sep,
  series = {{{ESEM}} '14},
  pages = {1--4},
  publisher = {Association for Computing Machinery},
  address = {New York, NY, USA},
  doi = {10.1145/2652524.2652554},
  urldate = {2025-03-21},
  isbn = {978-1-4503-2774-9}
}

@article{frattiniApplyingBayesianData2025,
  title = {Applying Bayesian Data Analysis for Causal Inference about Requirements Quality: A Controlled Experiment},
  shorttitle = {Applying Bayesian Data Analysis for Causal Inference about Requirements Quality},
  author = {Frattini, Julian and Fucci, Davide and Torkar, Richard and Montgomery, Lloyd and Unterkalmsteiner, Michael and Fischbach, Jannik and Mendez, Daniel},
  year = {2025},
  month = jan,
  journal = {Empirical Software Engineering},
  volume = {30},
  number = {1},
  pages = {29},
  issn = {1382-3256, 1573-7616},
  doi = {10.1007/s10664-024-10582-1},
  urldate = {2025-03-21},
  langid = {english}
}

@article{furiaCausalAnalysisEmpirical2024a,
  title = {Towards {{Causal Analysis}} of {{Empirical Software Engineering Data}}: {{The Impact}} of {{Programming Languages}} on {{Coding Competitions}}},
  shorttitle = {Towards {{Causal Analysis}} of {{Empirical Software Engineering Data}}},
  author = {Furia, Carlo A. and Torkar, Richard and Feldt, Robert},
  year = {2024},
  month = jan,
  journal = {ACM Transactions on Software Engineering and Methodology},
  volume = {33},
  number = {1},
  pages = {1--35},
  issn = {1049-331X, 1557-7392},
  doi = {10.1145/3611667},
  urldate = {2025-03-21},
  langid = {english}
}

@misc{furiaMitigatingOmittedVariable2025a,
  title = {Mitigating {{Omitted Variable Bias}} in {{Empirical Software Engineering}}},
  author = {Furia, Carlo A. and Torkar, Richard},
  year = {2025},
  publisher = {arXiv},
  doi = {10.48550/ARXIV.2501.17026},
  urldate = {2025-03-21},
  copyright = {arXiv.org perpetual, non-exclusive license}
}

@inproceedings{ghafariTestabilityFirst2019a,
  title = {Testability {{First}}!},
  booktitle = {2019 {{ACM}}/{{IEEE International Symposium}} on {{Empirical Software Engineering}} and {{Measurement}} ({{ESEM}})},
  author = {Ghafari, Mohammad and Eggiman, Markus and Nierstrasz, Oscar},
  year = {2019},
  month = sep,
  pages = {1--6},
  publisher = {IEEE},
  address = {Porto de Galinhas, Recife, Brazil},
  doi = {10.1109/ESEM.2019.8870170},
  urldate = {2025-03-21},
  copyright = {https://ieeexplore.ieee.org/Xplorehelp/downloads/license-information/IEEE.html},
  isbn = {978-1-7281-2968-6}
}

@article{herboldFinegrainedDataSet2022a,
  title = {A Fine-Grained Data Set and Analysis of Tangling in Bug Fixing Commits},
  author = {Herbold, Steffen and Trautsch, Alexander and Ledel, Benjamin and Aghamohammadi, Alireza and Ghaleb, Taher A. and Chahal, Kuljit Kaur and Bossenmaier, Tim and Nagaria, Bhaveet and Makedonski, Philip and Ahmadabadi, Matin Nili and Szabados, Kristof and Spieker, Helge and Madeja, Matej and Hoy, Nathaniel and Lenarduzzi, Valentina and Wang, Shangwen and {Rodr{\'i}guez-P{\'e}rez}, Gema and {Colomo-Palacios}, Ricardo and Verdecchia, Roberto and Singh, Paramvir and Qin, Yihao and Chakroborti, Debasish and Davis, Willard and Walunj, Vijay and Wu, Hongjun and Marcilio, Diego and Alam, Omar and Aldaeej, Abdullah and Amit, Idan and Turhan, Burak and Eismann, Simon and Wickert, Anna-Katharina and Malavolta, Ivano and Sul{\'i}r, Mat{\'u}{\v s} and Fard, Fatemeh and Henley, Austin Z. and Kourtzanidis, Stratos and Tuzun, Eray and Treude, Christoph and Shamasbi, Simin Maleki and Pashchenko, Ivan and Wyrich, Marvin and Davis, James and Serebrenik, Alexander and Albrecht, Ella and Aktas, Ethem Utku and Str{\"u}ber, Daniel and Erbel, Johannes},
  year = {2022},
  month = nov,
  journal = {Empirical Software Engineering},
  volume = {27},
  number = {6},
  pages = {125},
  issn = {1382-3256, 1573-7616},
  doi = {10.1007/s10664-021-10083-5},
  urldate = {2025-03-21},
  langid = {english}
}

@article{herboldProblemsSZZFeatures2022b,
  title = {Problems with {{SZZ}} and Features: {{An}} Empirical Study of the State of Practice of Defect Prediction Data Collection},
  shorttitle = {Problems with {{SZZ}} and Features},
  author = {Herbold, Steffen and Trautsch, Alexander and Trautsch, Fabian and Ledel, Benjamin},
  year = {2022},
  month = jan,
  journal = {Empirical Software Engineering},
  volume = {27},
  number = {2},
  pages = {42},
  issn = {1573-7616},
  doi = {10.1007/s10664-021-10092-4},
  urldate = {2025-03-21},
  langid = {english}
}

@book{huntington-kleinEffectIntroductionResearch2022a,
  title = {The Effect: An Introduction to Research Design and Causality},
  shorttitle = {The Effect},
  author = {{Huntington-Klein}, Nick},
  year = {2022},
  series = {A {{Chapman}} \& {{Hall}} Book},
  edition = {First edition},
  publisher = {CRC Press, Taylor \& Francis Group},
  address = {Boca Raton London New York},
  isbn = {978-1-032-12578-7 978-1-032-12745-3},
  langid = {english}
}

@article{huPracticalApproachExplaining2023a,
  title = {A Practical Approach to Explaining Defect Proneness of Code Commits by Causal Discovery},
  author = {Hu, Yamin and Luo, Wenjian and Hu, Zongyao},
  year = {2023},
  month = aug,
  journal = {Engineering Applications of Artificial Intelligence},
  volume = {123},
  pages = {106187},
  issn = {09521976},
  doi = {10.1016/j.engappai.2023.106187},
  urldate = {2025-03-21},
  langid = {english}
}

@inproceedings{joblinClassifyingDevelopersCore2017a,
  title = {Classifying {{Developers}} into {{Core}} and {{Peripheral}}: {{An Empirical Study}} on {{Count}} and {{Network Metrics}}},
  shorttitle = {Classifying {{Developers}} into {{Core}} and {{Peripheral}}},
  booktitle = {2017 {{IEEE}}/{{ACM}} 39th {{International Conference}} on {{Software Engineering}} ({{ICSE}})},
  author = {Joblin, Mitchell and Apel, Sven and Hunsen, Claus and Mauerer, Wolfgang},
  year = {2017},
  month = may,
  pages = {164--174},
  issn = {1558-1225},
  doi = {10.1109/ICSE.2017.23},
  urldate = {2025-03-21}
}

@inproceedings{justDefects4JDatabaseExisting2014,
  title = {{{Defects4J}}: A Database of Existing Faults to Enable Controlled Testing Studies for {{Java}} Programs},
  shorttitle = {{{Defects4J}}},
  booktitle = {Proceedings of the 2014 {{International Symposium}} on {{Software Testing}} and {{Analysis}}},
  author = {Just, Ren{\'e} and Jalali, Darioush and Ernst, Michael D.},
  year = {2014},
  month = jul,
  pages = {437--440},
  publisher = {ACM},
  address = {San Jose CA USA},
  doi = {10.1145/2610384.2628055},
  urldate = {2025-03-21},
  isbn = {978-1-4503-2645-2},
  langid = {english}
}

@article{kameiLargescaleEmpiricalStudy2013,
  title = {A Large-Scale Empirical Study of Just-in-Time Quality Assurance},
  author = {Kamei, Yasutaka and Shihab, Emad and Adams, Bram and Hassan, Ahmed E. and Mockus, Audris and Sinha, Anand and Ubayashi, Naoyasu},
  year = {2013},
  month = jun,
  journal = {IEEE Transactions on Software Engineering},
  volume = {39},
  number = {6},
  pages = {757--773},
  issn = {1939-3520},
  doi = {10.1109/TSE.2012.70},
  urldate = {2025-03-21}
}

@inproceedings{kazmanCausalModelingDiscovery2017a,
  title = {Causal {{Modeling}}, {{Discovery}}, \& {{Inference}} for {{Software Engineering}}},
  booktitle = {2017 {{IEEE}}/{{ACM}} 39th {{International Conference}} on {{Software Engineering Companion}} ({{ICSE-C}})},
  author = {Kazman, Rick and Stoddard, Robert and Danks, David and Cai, Yuanfang},
  year = {2017},
  month = may,
  pages = {172--174},
  doi = {10.1109/ICSE-C.2017.138},
  urldate = {2025-03-21}
}

@inproceedings{kononenkoInvestigatingCodeReview2015b,
  title = {Investigating Code Review Quality: {{Do}} People and Participation Matter?},
  shorttitle = {Investigating Code Review Quality},
  booktitle = {2015 {{IEEE International Conference}} on {{Software Maintenance}} and {{Evolution}} ({{ICSME}})},
  author = {Kononenko, Oleksii and Baysal, Olga and Guerrouj, Latifa and Cao, Yaxin and Godfrey, Michael W.},
  year = {2015},
  month = sep,
  pages = {111--120},
  publisher = {IEEE},
  address = {Bremen, Germany},
  doi = {10.1109/ICSM.2015.7332457},
  urldate = {2025-03-21},
  isbn = {978-1-4673-7532-0}
}

@book{krippendorffContentAnalysisIntroduction2019,
  title = {Content Analysis: An Introduction to Its Methodology},
  shorttitle = {Content Analysis},
  author = {Krippendorff, Klaus},
  year = {2019},
  edition = {Fourth edition},
  publisher = {SAGE},
  address = {Los Angeles London New Delhi Singapore Washington DC Melbourne},
  isbn = {978-1-5063-9566-1},
  langid = {english}
}

@misc{mojica-hankePerspectiveSoftwareEngineering2024a,
  title = {Perspective of {{Software Engineering Researchers}} on {{Machine Learning Practices Regarding Research}}, {{Review}}, and {{Education}}},
  author = {{Mojica-Hanke}, Anamaria and Palacio, David Nader and Poshyvanyk, Denys and {Linares-V{\'a}squez}, Mario and Herbold, Steffen},
  year = {2024},
  publisher = {arXiv},
  doi = {10.48550/ARXIV.2411.19304},
  urldate = {2025-03-21},
  copyright = {arXiv.org perpetual, non-exclusive license}
}

@inproceedings{ogarrioHybridCausalSearch2016a,
  title = {A {{Hybrid Causal Search Algorithm}} for {{Latent Variable Models}}},
  booktitle = {Proceedings of the {{Eighth International Conference}} on {{Probabilistic Graphical Models}}},
  author = {Ogarrio, Juan Miguel and Spirtes, Peter and Ramsey, Joe},
  year = {2016},
  month = aug,
  pages = {368--379},
  publisher = {PMLR},
  issn = {1938-7228},
  urldate = {2025-03-21},
  langid = {english}
}

@article{perscheidStudyingAdvancementDebugging2017,
  title = {Studying the Advancement in Debugging Practice of Professional Software Developers},
  author = {Perscheid, Michael and Siegmund, Benjamin and Taeumel, Marcel and Hirschfeld, Robert},
  year = {2017},
  month = mar,
  journal = {Software Quality Journal},
  volume = {25},
  number = {1},
  pages = {83--110},
  issn = {0963-9314, 1573-1367},
  doi = {10.1007/s11219-015-9294-2},
  urldate = {2025-03-21},
  langid = {english}
}

@inproceedings{querelWarningIntroducingCommitsVs2021a,
  title = {Warning-{{Introducing Commits}} vs {{Bug-Introducing Commits}}: {{A}} Tool, Statistical Models, and a Preliminary User Study},
  shorttitle = {Warning-{{Introducing Commits}} vs {{Bug-Introducing Commits}}},
  booktitle = {2021 {{IEEE}}/{{ACM}} 29th {{International Conference}} on {{Program Comprehension}} ({{ICPC}})},
  author = {Querel, Louis-Philippe and Rigby, Peter C.},
  year = {2021},
  month = may,
  pages = {433--443},
  publisher = {IEEE},
  address = {Madrid, Spain},
  doi = {10.1109/ICPC52881.2021.00051},
  urldate = {2025-03-21},
  copyright = {https://ieeexplore.ieee.org/Xplorehelp/downloads/license-information/IEEE.html},
  isbn = {978-1-6654-1403-6}
}

@article{rodriguez-perezHowBugsAre2020a,
  title = {How Bugs Are Born: A Model to Identify How Bugs Are Introduced in Software Components},
  shorttitle = {How Bugs Are Born},
  author = {{Rodr{\'i}guez-P{\'e}rez}, Gema and Robles, Gregorio and Serebrenik, Alexander and Zaidman, Andy and Germ{\'a}n, Daniel M. and {Gonzalez-Barahona}, Jesus M.},
  year = {2020},
  month = mar,
  journal = {Empirical Software Engineering},
  volume = {25},
  number = {2},
  pages = {1294--1340},
  issn = {1382-3256, 1573-7616},
  doi = {10.1007/s10664-019-09781-y},
  urldate = {2025-03-21},
  langid = {english}
}

@misc{schulteExploratoryStudyBugintroducing2023,
  title = {An Exploratory Study of Bug-Introducing Changes: What Happens When Bugs Are Introduced in Open Source Software?},
  shorttitle = {An Exploratory Study of Bug-Introducing Changes},
  author = {Schulte, Lukas and {Mojica-Hanke}, Anamaria and {Linares-V{\'a}squez}, Mario and Herbold, Steffen},
  year = {2023},
  month = apr,
  number = {arXiv:2304.05358},
  eprint = {2304.05358},
  primaryclass = {cs},
  publisher = {arXiv},
  doi = {10.48550/arXiv.2304.05358},
  urldate = {2025-03-21},
  archiveprefix = {arXiv}
}

@article{sliwerskiWhenChangesInduce2005a,
  title = {When Do Changes Induce Fixes?},
  author = {{\'S}liwerski, Jacek and Zimmermann, Thomas and Zeller, Andreas},
  year = {2005},
  month = jul,
  journal = {ACM SIGSOFT Software Engineering Notes},
  volume = {30},
  number = {4},
  pages = {1--5},
  issn = {0163-5948},
  doi = {10.1145/1082983.1083147},
  urldate = {2025-03-21},
  langid = {english}
}

@article{souzaDevelopersViewpointsAvoid2022,
  title = {Developers' Viewpoints to Avoid Bug-Introducing Changes},
  author = {Souza, Jairo and Lima, Rodrigo and Fonseca, Baldoino and Cartaxo, Bruno and Ribeiro, M{\'a}rcio and Pinto, Gustavo and Gheyi, Rohit and Garcia, Alessandro},
  year = {2022},
  month = mar,
  journal = {Information and Software Technology},
  volume = {143},
  pages = {106766},
  issn = {09505849},
  doi = {10.1016/j.infsof.2021.106766},
  urldate = {2025-03-21},
  langid = {english}
}

@book{spirtesCausationPredictionSearch2000a,
  title = {Causation, Prediction, and Search},
  author = {Spirtes, Peter and Glymour, Clark N. and Scheines, Richard},
  year = {2000},
  series = {Adaptive Computation and Machine Learning},
  edition = {Second edition},
  publisher = {The MIT Press},
  address = {Cambridge, Massachussetts London},
  collaborator = {Heckerman, David},
  isbn = {978-0-262-52792-7 978-0-262-19440-2},
  langid = {english}
}

@inproceedings{swansonDimensionsMaintenance1976a,
  title = {The Dimensions of Maintenance},
  booktitle = {Proceedings of the 2nd International Conference on {{Software}} Engineering},
  author = {Swanson, E. Burton},
  year = {1976},
  month = oct,
  series = {{{ICSE}} '76},
  pages = {492--497},
  publisher = {IEEE Computer Society Press},
  address = {Washington, DC, USA},
  urldate = {2025-03-21}
}

@inproceedings{taoHowSoftwareEngineers2012,
  title = {How Do Software Engineers Understand Code Changes?: An Exploratory Study in Industry},
  shorttitle = {How Do Software Engineers Understand Code Changes?},
  booktitle = {Proceedings of the {{ACM SIGSOFT}} 20th {{International Symposium}} on the {{Foundations}} of {{Software Engineering}}},
  author = {Tao, Yida and Dang, Yingnong and Xie, Tao and Zhang, Dongmei and Kim, Sunghun},
  year = {2012},
  month = nov,
  pages = {1--11},
  publisher = {ACM},
  address = {Cary North Carolina},
  doi = {10.1145/2393596.2393656},
  urldate = {2025-03-21},
  isbn = {978-1-4503-1614-9},
  langid = {english}
}

@article{thomasGeneralInductiveApproach2006,
  title = {A {{General Inductive Approach}} for {{Analyzing Qualitative Evaluation Data}}},
  author = {Thomas, David R.},
  year = {2006},
  month = jun,
  journal = {American Journal of Evaluation},
  volume = {27},
  number = {2},
  pages = {237--246},
  issn = {1098-2140, 1557-0878},
  doi = {10.1177/1098214005283748},
  urldate = {2025-03-21},
  copyright = {https://journals.sagepub.com/page/policies/text-and-data-mining-license},
  langid = {english}
}

@article{trautschAddressingProblemsReplicability2018,
  title = {Addressing Problems with Replicability and Validity of Repository Mining Studies through a Smart Data Platform},
  author = {Trautsch, Fabian and Herbold, Steffen and Makedonski, Philip and Grabowski, Jens},
  year = {2018},
  month = apr,
  journal = {Empirical Software Engineering},
  volume = {23},
  number = {2},
  pages = {1036--1083},
  issn = {1573-7616},
  doi = {10.1007/s10664-017-9537-x},
  urldate = {2025-03-21},
  langid = {english}
}

@inproceedings{trautschSmartSHARKEcosystemSoftware2020,
  title = {The {{SmartSHARK}} Ecosystem for Software Repository Mining},
  booktitle = {Proceedings of the {{ACM}}/{{IEEE}} 42nd {{International Conference}} on {{Software Engineering}}: {{Companion Proceedings}}},
  author = {Trautsch, Alexander and Trautsch, Fabian and Herbold, Steffen and Ledel, Benjamin and Grabowski, Jens},
  year = {2020},
  month = oct,
  series = {{{ICSE}} '20},
  pages = {25--28},
  publisher = {Association for Computing Machinery},
  address = {New York, NY, USA},
  doi = {10.1145/3377812.3382139},
  urldate = {2025-03-21},
  isbn = {978-1-4503-7122-3}
}

@inproceedings{trautschStaticSourceCode2020a,
  title = {Static Source Code Metrics and Static Analysis Warnings for Fine-Grained Just-in-Time Defect Prediction},
  booktitle = {2020 {{IEEE International Conference}} on {{Software Maintenance}} and {{Evolution}} ({{ICSME}})},
  author = {Trautsch, Alexander and Herbold, Steffen and Grabowski, Jens},
  year = {2020},
  month = sep,
  pages = {127--138},
  publisher = {IEEE},
  address = {Adelaide, Australia},
  doi = {10.1109/ICSME46990.2020.00022},
  urldate = {2025-03-21},
  copyright = {https://ieeexplore.ieee.org/Xplorehelp/downloads/license-information/IEEE.html},
  isbn = {978-1-7281-5619-4}
}
%% if required, the content of .bbl file can be included here once bbl is generated
%%\input sn-article.bbl

\begin{appendices}

    % #########################################################
    % 
    % #########################################################

    \FloatBarrier
    \section{Replication Kit}
    \label{appendix:replication_kit}

    The datasets generated and/or analyzed during the study are available in the following Zenodo repositories:

    \begin{itemize}
        \item \rev{Code, graphs, and tables: \url{https://doi.org/10.5281/zenodo.17207031}}
        \item SmartSHARK and repository archives: \url{https://doi.org/10.5281/zenodo.15363704}
    \end{itemize}

    % #########################################################
    % 
    % #########################################################

    \FloatBarrier
    \section{Data Distribution}
    \label{appendix:table_datadist}

\begin{table}[h]
\centering
\begin{tabular}{c|cc|cc|cc}
\toprule
\multirow2{*}{ID} & \multicolumn{2}{c|}{Valid Instances} & \multicolumn{2}{c|}{Mean} & \multicolumn{2}{c}{Standard Deviation} \\
\cmidrule{2-7}
& BG & CG & BG & CG & BG & CG \\\midrule
\texttt{F4} & 71 & 0 & 2.72 & - & 1.93 & - \\
\texttt{F9} & 71 & 0 & 4.06 & - & 6.18 & - \\
\midrule
\texttt{I5} & 36 & 17 & 3.64 & 2.12 & 3.73 & 1.05 \\
\texttt{I7} & 51 & 28 & 6.18 & 20.18 & 12.71 & 29.11 \\
\texttt{I10} & 36 & 17 & 7.11 & 2.65 & 7.55 & 3.0 \\
\midrule
\texttt{ML2} & 4 & 0 & 1.75 & - & 0.96 & - \\
\texttt{ML5} & 13 & 8 & 6.31 & 5.0 & 6.79 & 5.81 \\
\midrule
\texttt{O3} & 1 & 0 & 2.0 & - & - & - \\
\midrule
\texttt{C3} & 71 & 71 & 166.92 & 2.56 & 1037.62 & 2.44 \\
\texttt{C4} & 71 & 0 & 0.25 & - & 0.27 & - \\
\texttt{C5} & 71 & 0 & 0.34 & - & 0.31 & - \\
\texttt{C6} & 71 & 0 & 0.4 & - & 0.41 & - \\
\midrule
\texttt{R3} & 38 & 39 & 11.5 & 8.31 & 5.78 & 4.53 \\
\texttt{R5} & 38 & 39 & 119.97 & 32.56 & 159.49 & 29.72 \\
\texttt{R6} & 38 & 39 & 14.39 & 3.31 & 17.92 & 3.1 \\
\bottomrule
\end{tabular}
\caption{Distribution of numeric variables between bug group (BG) and comparison group (CG). A dash (-) indicates that the variable is not present in the sample group.}
\label{tbl:data-dist-num}
\end{table}

\begin{table}[h]
\centering
\begin{tabular}[t]{c|cc|cc}
\toprule
\multirow2{*}{ID} & \multicolumn{2}{c|}{Valid Instances} & \multicolumn{2}{c}{\#Unique Labels} \\
\cmidrule{2-5}
& BG & CG & BG & CG \\\midrule
\texttt{F2} & 33 & 0 & 5 & - \\
\texttt{F4.2} & 71 & 0 & 3 & - \\
\texttt{F4.3} & 54 & 0 & 3 & - \\
\texttt{F4.4} & 71 & 0 & 3 & - \\
\texttt{F7} & 71 & 0 & 2 & - \\
\texttt{F8} & 71 & 0 & 2 & - \\
\midrule
\texttt{I2} & 51 & 28 & 3 & 3 \\
\texttt{I3} & 26 & 18 & 5 & 5 \\
\texttt{I5.2} & 51 & 28 & 3 & 3 \\
\texttt{I5.3} & 35 & 22 & 3 & 3 \\
\texttt{I5.4} & 48 & 26 & 3 & 3 \\
\texttt{I8} & 51 & 28 & 2 & 2 \\
\texttt{I9} & 51 & 28 & 2 & 1 \\
\midrule
\texttt{O5} & 71 & 0 & 5 & - \\
\midrule
\texttt{CI2} & 31 & 31 & 2 & 2 \\
\texttt{CI3} & 31 & 31 & 2 & 2 \\
\midrule
\texttt{B2} & 71 & 70 & 2 & 2 \\
\midrule
\texttt{C2.1} & 71 & 71 & 9 & 8 \\
\texttt{C7} & 71 & 71 & 2 & 2 \\
\texttt{C8} & 71 & 71 & 2 & 2 \\
\texttt{C10} & 71 & 71 & 2 & 2 \\
\texttt{C11} & 71 & 0 & 4 & - \\
\texttt{C13} & 33 & 33 & 2 & 2 \\
\texttt{C15} & 71 & 71 & 2 & 2 \\
\texttt{C16} & 70 & 0 & 2 & - \\
\texttt{C17} & 71 & 71 & 2 & 2 \\
\texttt{C18} & 70 & 0 & 2 & - \\
\texttt{C19} & 71 & 71 & 2 & 2 \\
\texttt{C20} & 70 & 0 & 2 & - \\
\texttt{C21} & 71 & 71 & 2 & 2 \\
\texttt{C22} & 70 & 0 & 2 & - \\
\midrule
\texttt{R2} & 38 & 39 & 2 & 2 \\
\texttt{R7} & 40 & 39 & 2 & 2 \\
\texttt{R9} & 40 & 0 & 2 & - \\
\texttt{R10} & 39 & 0 & 2 & - \\
\bottomrule
\end{tabular}
\caption{Distribution of nominal variables between bug group (BG) and comparison group (CG). A dash (-) indicates that the variable is not present in the sample group.}
\label{tbl:data-dist-nom}
\end{table}

\begin{table}[h]
\centering
\begin{tabular}{c|cc|cc|cc}
\toprule
\multirow{2}{*}{ID} & \multicolumn{2}{c|}{Valid Instances} & \multicolumn{2}{c|}{\#Absolute Labels} & \multicolumn{2}{c}{\#Unique Labels} \\
\cmidrule{2-7}
& BG & CG & BG & CG & BG & CG \\\midrule
\texttt{F3} & 65 & 0 & 128 & - & 8 & - \\
\texttt{F4.1} & 71 & 0 & 193 & - & 3 & - \\
\texttt{F5} & 71 & 0 & 451 & - & 52 & - \\
\texttt{F6} & 71 & 0 & 75 & - & 5 & - \\
\midrule
\texttt{I4} & 32 & 13 & 65 & 27 & 9 & 9 \\
\texttt{I5.1} & 36 & 17 & 131 & 36 & 3 & 3 \\
\texttt{I6} & 51 & 28 & 331 & 148 & 60 & 39 \\
\midrule
\texttt{ML2.1} & 4 & 0 & 7 & - & 3 & - \\
\texttt{ML3} & 4 & 0 & 27 & - & 21 & - \\
\texttt{ML5.1} & 13 & 8 & 28 & 16 & 3 & 3 \\
\texttt{ML6} & 12 & 8 & 121 & 87 & 50 & 43 \\
\midrule
\texttt{O2} & 11 & 6 & 14 & 9 & 5 & 4 \\
\texttt{O3.1} & 1 & 0 & 2 & - & 2 & - \\
\texttt{O4} & 11 & 5 & 77 & 33 & 19 & 9 \\
\midrule
\texttt{B1} & 71 & 70 & 1365 & 1408 & 64 & 72 \\
\texttt{B3} & 71 & 70 & 887 & 861 & 27 & 27 \\
\midrule
\texttt{C2.2} & 71 & 71 & 79 & 79 & 9 & 8 \\
\texttt{C2.3} & 71 & 71 & 71 & 71 & 9 & 10 \\
\texttt{C9} & 71 & 71 & 80 & 78 & 3 & 3 \\
\texttt{C12} & 71 & 0 & 77 & - & 5 & - \\
\midrule
\texttt{R4} & 37 & 38 & 95 & 89 & 3 & 3 \\
\texttt{R8} & 39 & 39 & 1037 & 461 & 181 & 110 \\
\texttt{R11} & 40 & 39 & 42 & 41 & 2 & 2 \\
\bottomrule
\end{tabular}
\caption{Distribution of list of nominal variables between bug group (BG) and comparison group (CG). A dash (-) indicates that the variable is not present in the sample group.}
\label{tbl:data-dist-list}
\end{table}

    % #########################################################
    % 
    % #########################################################

    \FloatBarrier
    \begin{landscape}

        \section{Group G2 and G3 Relation Pairs}
        \label{appendix:table_g2g3}

        \begin{longtable}{l|l}
\caption{Pairs with group G2 relations.} \label{tbl-appendix:full_pairs_g2} \\
\toprule
Variable A & Variable B \\
\midrule
\endfirsthead
\caption[]{Pairs with group G2 relations.} \\
\toprule
Variable A & Variable B \\
\midrule
\endhead
\midrule
\multicolumn{2}{r}{Continued on next page} \\
\midrule
\endfoot
\bottomrule
\endlastfoot
B1 - Build tools & B2 - Dependency resolution \\
B1 - Build tools & C2.2 - Commit types (children) \\
B1 - Build tools & C2.3 - Commit types (parents) \\
B1 - Build tools & C7 - Committer type \\
B1 - Build tools & R2 - Review tool \\
B1 - Build tools & R11 - Review branch \\
B2 - Dependency resolution & B3 - Build practices \\
B2 - Dependency resolution & C2.2 - Commit types (children) \\
B2 - Dependency resolution & C2.3 - Commit types (parents) \\
B2 - Dependency resolution & R2 - Review tool \\
B3 - Build practices & C2.2 - Commit types (children) \\
B3 - Build practices & C2.3 - Commit types (parents) \\
B3 - Build practices & R2 - Review tool \\
B3 - Build practices & R3 - \# Reviewers \\
\rev{B3 - Build practices} & \rev{R5 - \# Review comments} \\
B3 - Build practices & R11 - Review branch \\
C2.1 - Commit type (current) & C2.2 - Commit types (children) \\
C2.1 - Commit type (current) & C2.3 - Commit types (parents) \\
C2.1 - Commit type (current) & C13 - Test failures \\
\rev{C2.1 - Commit type (current)} & \rev{R3 - \# Reviewers} \\
C2.1 - Commit type (current) & R5 - \# Review comments \\
C2.1 - Commit type (current) & R6 - \# Rounds of code review \\
C2.2 - Commit types (children) & C2.3 - Commit types (parents) \\
C2.2 - Commit types (children) & R2 - Review tool \\
C2.2 - Commit types (children) & R3 - \# Reviewers \\
C2.2 - Commit types (children) & R5 - \# Review comments \\
C2.2 - Commit types (children) & R6 - \# Rounds of code review \\
C2.2 - Commit types (children) & R11 - Review branch \\
C2.3 - Commit types (parents) & R2 - Review tool \\
C2.3 - Commit types (parents) & R3 - \# Reviewers \\
C2.3 - Commit types (parents) & R5 - \# Review comments \\
C2.3 - Commit types (parents) & R6 - \# Rounds of code review \\
C2.3 - Commit types (parents) & R7 - Caused changes \\
C2.3 - Commit types (parents) & R11 - Review branch \\
C7 - Committer type & R3 - \# Reviewers \\
C10 - Test changes & C3 - \# Files changed \\
C19 - Design changes & C3 - \# Files changed \\
C21 - Change of external dependencies & C3 - \# Files changed \\
CI2 - CI build status & R5 - \# Review comments \\
I2 - Introducing issue types & C9 - Commit type Swanson \\
I2 - Introducing issue types & I6 - Aspects of introducing issue discussion: topic \\
I2 - Introducing issue types & R5 - \# Review comments \\
I3 - Introducing issue severity & I7 - \# Introducing issue commits \\
I4 - Introducing issue labels & C2.1 - Commit type (current) \\
I4 - Introducing issue labels & I7 - \# Introducing issue commits \\
I5 - \# Introducing issue discussants & I10 - \# Introducing issue comments \\
I5.1 - Introducing issue discussant roles & C7 - Committer type \\
I5.2 - Introducing issue reporter role & C2.1 - Commit type (current) \\
I5.2 - Introducing issue reporter role & I5 - \# Introducing issue discussants \\
I5.2 - Introducing issue reporter role & I5.3 - Introducing issue assignee role \\
I5.2 - Introducing issue reporter role & I5.4 - Introducing issue creator role \\
I5.2 - Introducing issue reporter role & I10 - \# Introducing issue comments \\
I5.3 - Introducing issue assignee role & I5.4 - Introducing issue creator role \\
I5.4 - Introducing issue creator role & I5 - \# Introducing issue discussants \\
I5.4 - Introducing issue creator role & I10 - \# Introducing issue comments \\
I6 - Aspects of introducing issue discussion: topic & C9 - Commit type Swanson \\
I6 - Aspects of introducing issue discussion: topic & I8 - Introducing issue has wiki/specification \\
I6 - Aspects of introducing issue discussion: topic & R3 - \# Reviewers \\
I6 - Aspects of introducing issue discussion: topic & R5 - \# Review comments \\
I8 - Introducing issue has wiki/specification & B1 - Build tools \\
\rev{I8 - Introducing issue has wiki/specification} & \rev{C9 - Commit type Swanson} \\
\rev{I8 - Introducing issue has wiki/specification} & \rev{R8 - Aspects of review comments: target} \\
\rev{R2 - Review tool} & \rev{R8 - Aspects of review comments: target} \\
R2 - Review tool & R8 - Aspects of review comments: topic \\
R2 - Review tool & R11 - Review branch \\
R3 - \# Reviewers & R5 - \# Review comments \\
R3 - \# Reviewers & R6 - \# Rounds of code review \\
R4 - Reviewer types & R3 - \# Reviewers \\
R4 - Reviewer types & R5 - \# Review comments \\
R4 - Reviewer types & R6 - \# Rounds of code review \\
R5 - \# Review comments & R6 - \# Rounds of code review \\
R8 - Aspects of review comments: target & R3 - \# Reviewers \\
R8 - Aspects of review comments: target & R5 - \# Review comments \\
R8 - Aspects of review comments: target & R6 - \# Rounds of code review   \\ \bottomrule
\end{longtable}
        \begin{longtable}{l|l}
\caption{Pairs with group G3 relations.} \label{tbl-appendix:full_pairs_g3} \\
\toprule
Variable A & Variable B \\
\midrule
\endfirsthead
\caption[]{Pairs with group G3 relations.} \\
\toprule
Variable A & Variable B \\
\midrule
\endhead
\midrule
\multicolumn{2}{r}{Continued on next page} \\
\midrule
\endfoot
\bottomrule
\endlastfoot
BB1 - Build tools & C2.1 - Commit type (current) \\
B1 - Build tools & C3 - \# Files changed \\
B1 - Build tools & C10 - Test changes \\
B1 - Build tools & C13 - Test failures \\
B1 - Build tools & C17 - Refactorings \\
B1 - Build tools & C21 - Change of external dependencies \\
B1 - Build tools & I5 - \# Introducing issue discussants \\
B1 - Build tools & I7 - \# Introducing issue commits \\
B1 - Build tools & I10 - \# Introducing issue comments \\
B1 - Build tools & R3 - \# Reviewers \\
B1 - Build tools & R5 - \# Review comments \\
B1 - Build tools & R6 - \# Rounds of code review \\
B1 - Build tools & R7 - Caused changes \\
B2 - Dependency resolution & C2.1 - Commit type (current) \\
B2 - Dependency resolution & C17 - Refactorings \\
B2 - Dependency resolution & R8 - Aspects of review comments: target \\
B2 - Dependency resolution & R8 - Aspects of review comments: topic \\
B3 - Build practices & C2.1 - Commit type (current) \\
B3 - Build practices & C3 - \# Files changed \\
B3 - Build practices & C10 - Test changes \\
B3 - Build practices & C13 - Test failures \\
B3 - Build practices & C21 - Change of external dependencies \\
B3 - Build practices & I5 - \# Introducing issue discussants \\
B3 - Build practices & I7 - \# Introducing issue commits \\
B3 - Build practices & I10 - \# Introducing issue comments \\
B3 - Build practices & R6 - \# Rounds of code review \\
B3 - Build practices & R7 - Caused changes \\
C2.1 - Commit type (current) & C3 - \# Files changed \\
C2.1 - Commit type (current) & C7 - Committer type \\
C2.1 - Commit type (current) & C9 - Commit type Swanson \\
C2.1 - Commit type (current) & C10 - Test changes \\
C2.1 - Commit type (current) & C15 - Documentation changes \\
C2.1 - Commit type (current) & C17 - Refactorings \\
C2.1 - Commit type (current) & C21 - Change of external dependencies \\
C2.1 - Commit type (current) & I7 - \# Introducing issue commits \\
C2.1 - Commit type (current) & R2 - Review tool \\
C2.1 - Commit type (current) & R7 - Caused changes \\
C2.1 - Commit type (current) & R11 - Review branch \\
C2.2 - Commit types (children) & C3 - \# Files changed \\
C2.2 - Commit types (children) & C10 - Test changes \\
C2.2 - Commit types (children) & C17 - Refactorings \\
C2.2 - Commit types (children) & C21 - Change of external dependencies \\
C2.2 - Commit types (children) & I7 - \# Introducing issue commits \\
C2.2 - Commit types (children) & R7 - Caused changes \\
C2.3 - Commit types (parents) & C3 - \# Files changed \\
C3 - \# Files changed & R6 - \# Rounds of code review \\
C7 - Committer type & C9 - Commit type Swanson \\
C7 - Committer type & R5 - \# Review comments \\
C8 - Multiple concerns & C15 - Documentation changes \\
C9 - Commit type Swanson & C3 - \# Files changed \\
C9 - Commit type Swanson & C19 - Design changes \\
C9 - Commit type Swanson & C21 - Change of external dependencies \\
C9 - Commit type Swanson & I7 - \# Introducing issue commits \\
C9 - Commit type Swanson & I10 - \# Introducing issue comments \\
C9 - Commit type Swanson & R3 - \# Reviewers \\
C9 - Commit type Swanson & R5 - \# Review comments \\
C9 - Commit type Swanson & R6 - \# Rounds of code review \\
C10 - Test changes & C17 - Refactorings \\
C10 - Test changes & C19 - Design changes \\
C15 - Documentation changes & C3 - \# Files changed \\
C15 - Documentation changes & C19 - Design changes \\
C15 - Documentation changes & C21 - Change of external dependencies \\
C17 - Refactorings & C3 - \# Files changed \\
C17 - Refactorings & I7 - \# Introducing issue commits \\
C17 - Refactorings & R2 - Review tool \\
\rev{C17 - Refactorings} & \rev{R3 - \# Reviewers} \\
C17 - Refactorings & R5 - \# Review comments \\
C17 - Refactorings & R6 - \# Rounds of code review \\
C19 - Design changes & C21 - Change of external dependencies \\
CI2 - CI build status & B1 - Build tools \\
CI2 - CI build status & C13 - Test failures \\
CI2 - CI build status & CI3 - CI build status change \\
CI2 - CI build status & R3 - \# Reviewers \\
CI2 - CI build status & R6 - \# Rounds of code review \\
\rev{CI2 - CI build status} & \rev{R8 - Aspects of review comments: target} \\
CI3 - CI build status change & C3 - \# Files changed \\
CI3 - CI build status change & I7 - \# Introducing issue commits \\
CI3 - CI build status change & R3 - \# Reviewers \\
CI3 - CI build status change & R5 - \# Review comments \\
\rev{CI3 - CI build status change} & \rev{R6 - \# Rounds of code review} \\
I2 - Introducing issue types & C3 - \# Files changed \\
I2 - Introducing issue types & C19 - Design changes \\
I2 - Introducing issue types & C21 - Change of external dependencies \\
I2 - Introducing issue types & I4 - Introducing issue labels \\
I2 - Introducing issue types & I5.3 - Introducing issue assignee role \\
I2 - Introducing issue types & I6 - Aspects of introducing issue discussion: target \\
I2 - Introducing issue types & I7 - \# Introducing issue commits \\
I2 - Introducing issue types & I8 - Introducing issue has wiki/specification \\
I2 - Introducing issue types & O4 - Aspects of discussions in other media: topic \\
I2 - Introducing issue types & R3 - \# Reviewers \\
I2 - Introducing issue types & R6 - \# Rounds of code review \\
I3 - Introducing issue severity & C3 - \# Files changed \\
I3 - Introducing issue severity & CI2 - CI build status \\
\rev{I3 - Introducing issue severity} & \rev{I6 - Aspects of introducing issue discussion: topic} \\
I3 - Introducing issue severity & I8 - Introducing issue has wiki/specification \\
I3 - Introducing issue severity & O2 - Type of other media in which bug was discussed \\
I3 - Introducing issue severity & O4 - Aspects of discussions in other media: topic \\
I3 - Introducing issue severity & R3 - \# Reviewers \\
I3 - Introducing issue severity & R5 - \# Review comments \\
I3 - Introducing issue severity & R6 - \# Rounds of code review \\
I4 - Introducing issue labels & B1 - Build tools \\
I4 - Introducing issue labels & B3 - Build practices \\
I4 - Introducing issue labels & C2.2 - Commit types (children) \\
I4 - Introducing issue labels & C2.3 - Commit types (parents) \\
I4 - Introducing issue labels & C3 - \# Files changed \\
I4 - Introducing issue labels & C9 - Commit type Swanson \\
I4 - Introducing issue labels & C21 - Change of external dependencies \\
I4 - Introducing issue labels & R2 - Review tool \\
I5.1 - Introducing issue discussant roles & I5.2 - Introducing issue reporter role \\
I5.1 - Introducing issue discussant roles & I5.4 - Introducing issue creator role \\
I5.2 - Introducing issue reporter role & B1 - Build tools \\
I5.2 - Introducing issue reporter role & B2 - Dependency resolution \\
I5.2 - Introducing issue reporter role & C7 - Committer type \\
I5.3 - Introducing issue assignee role & B1 - Build tools \\
I5.3 - Introducing issue assignee role & C7 - Committer type \\
I5.3 - Introducing issue assignee role & C19 - Design changes \\
I5.3 - Introducing issue assignee role & CI3 - CI build status change \\
I5.3 - Introducing issue assignee role & I7 - \# Introducing issue commits \\
I5.3 - Introducing issue assignee role & R5 - \# Review comments \\
I5.3 - Introducing issue assignee role & R6 - \# Rounds of code review \\
I5.4 - Introducing issue creator role & B1 - Build tools \\
I5.4 - Introducing issue creator role & B2 - Dependency resolution \\
I5.4 - Introducing issue creator role & C2.1 - Commit type (current) \\
I5.4 - Introducing issue creator role & C2.3 - Commit types (parents) \\
I5.4 - Introducing issue creator role & C7 - Committer type \\
I6 - Aspects of introducing issue discussion: topic & B2 - Dependency resolution \\
I6 - Aspects of introducing issue discussion: target & C3 - \# Files changed \\
I6 - Aspects of introducing issue discussion: topic & C3 - \# Files changed \\
I6 - Aspects of introducing issue discussion: topic & C19 - Design changes \\
I6 - Aspects of introducing issue discussion: topic & C21 - Change of external dependencies \\
\rev{I6 - Aspects of introducing issue discussion: target} & \rev{I7 - \# Introducing issue commits} \\
I6 - Aspects of introducing issue discussion: topic & I7 - \# Introducing issue commits \\
I6 - Aspects of introducing issue discussion: target & I8 - Introducing issue has wiki/specification \\
I6 - Aspects of introducing issue discussion: topic & O2 - Type of other media in which bug was discussed \\
I6 - Aspects of introducing issue discussion: topic & O4 - Aspects of discussions in other media: topic \\
I6 - Aspects of introducing issue discussion: target & R3 - \# Reviewers \\
I6 - Aspects of introducing issue discussion: target & R5 - \# Review comments \\
I6 - Aspects of introducing issue discussion: target & R6 - \# Rounds of code review \\
I6 - Aspects of introducing issue discussion: topic & R6 - \# Rounds of code review \\
I7 - \# Introducing issue commits & I10 - \# Introducing issue comments \\
I8 - Introducing issue has wiki/specification & B2 - Dependency resolution \\
I8 - Introducing issue has wiki/specification & B3 - Build practices \\
I8 - Introducing issue has wiki/specification & C2.1 - Commit type (current) \\
I8 - Introducing issue has wiki/specification & C2.3 - Commit types (parents) \\
I8 - Introducing issue has wiki/specification & O4 - Aspects of discussions in other media: target \\
I8 - Introducing issue has wiki/specification & O4 - Aspects of discussions in other media: topic \\
O2 - Type of other media in which bug was discussed & C9 - Commit type Swanson \\
O2 - Type of other media in which bug was discussed & CI2 - CI build status \\
\rev{O4 - Aspects of discussions in other media: target} & \rev{C9 - Commit type Swanson} \\
O4 - Aspects of discussions in other media: topic & C9 - Commit type Swanson \\
R2 - Review tool & R7 - Caused changes \\
R4 - Reviewer types & I7 - \# Introducing issue commits \\
R8 - Aspects of review comments: topic & C3 - \# Files changed \\
\rev{R8 - Aspects of review comments: topic} & \rev{I7 - \# Introducing issue commits} \\
R8 - Aspects of review comments: topic & R3 - \# Reviewers \\
R8 - Aspects of review comments: topic & R5 - \# Review comments \\
R8 - Aspects of review comments: topic & R6 - \# Rounds of code review  \\ \bottomrule
\end{longtable}

% Removed 16.09.25:

% C8 - Multiple concerns & C9 - Commit type Swanson \\
% I5.2 - Introducing issue reporter role & C9 - Commit type Swanson \\
% I5.4 - Introducing issue creator role & C9 - Commit type Swanson \\
% I8 - Introducing issue has wiki/specification & R8 - Aspects of review comments: target \\
% R2 - Review tool & R8 - Aspects of review comments: target \\
% R8 - Aspects of review comments: action & R3 - \# Reviewers \\
% R8 - Aspects of review comments: action & R5 - \# Review comments \\
% R8 - Aspects of review comments: action & R6 - \# Rounds of code review

    \end{landscape}

    % #########################################################
    % 
    % #########################################################

    \FloatBarrier
    \section{Logically Associated Variables}
    \label{appendix:table_associations}
    \begin{longtable}{>{\ttfamily}l | p{0.15cm}p{0.15cm}p{0.15cm}p{0.15cm}p{0.15cm}p{0.15cm}p{0.15cm}p{0.15cm}p{0.15cm}p{0.15cm}p{0.15cm}p{0.15cm}p{0.15cm}p{0.15cm}p{0.15cm}p{0.15cm}p{0.15cm}p{0.15cm}p{0.15cm}}
\caption{Groups of logically associated variables.} \label{tbl-appendix:logically-associated-variables} \\
\toprule
\multicolumn{1}{l}{ID} & \rot{Issue management (basic)} & \rot{Issue management (advanced)} & \rot{Issue management roles} & \rot{Discussion aspects (action)} & \rot{Discussion aspects (target)} & \rot{Discussion aspects (topic)} & \rot{Discussion frequency} & \rot{Discussant roles} & \rot{Discussion tools} & \rot{Review scope} & \rot{Commit type} & \rot{Continuous Integration} & \rot{Testing} & \rot{Refactorings} & \rot{Design changes} & \rot{Dependencies} & \rot{Documentation} & \rot{Bug Type} & \rot{Bug-covering QA measures} \\
\midrule
\endfirsthead
\caption[]{Groups of logically associated variables.} \\
\toprule
\multicolumn{1}{l}{ID} & \rot{Issue management (basic)} & \rot{Issue management (advanced)} & \rot{Issue management roles} & \rot{Discussion aspects (action)} & \rot{Discussion aspects (target)} & \rot{Discussion aspects (topic)} & \rot{Discussion frequency} & \rot{Discussant roles} & \rot{Discussion tools} & \rot{Review scope} & \rot{Commit type} & \rot{Continuous Integration} & \rot{Testing} & \rot{Refactorings} & \rot{Design changes} & \rot{Dependencies} & \rot{Documentation} & \rot{Bug Type} & \rot{Bug-covering QA measures} \\
\midrule
\endhead
\midrule
\multicolumn{20}{r}{Continued on next page} \\
\midrule
\endfoot
\bottomrule
\endlastfoot
B1 & ~ & ~ & ~ & ~ & ~ & ~ & ~ & ~ & ~ & ~ & ~ & X & ~ & ~ & ~ & ~ & ~ & ~ & ~ \\
B2 & ~ & ~ & ~ & ~ & ~ & ~ & ~ & ~ & ~ & ~ & ~ & ~ & ~ & ~ & ~ & X & ~ & ~ & ~ \\
B3 & ~ & ~ & ~ & ~ & ~ & ~ & ~ & ~ & ~ & ~ & ~ & X & ~ & ~ & ~ & ~ & ~ & ~ & ~ \\
\midrule
C2.1 & ~ & ~ & ~ & ~ & ~ & ~ & ~ & ~ & ~ & ~ & X & ~ & ~ & ~ & ~ & ~ & ~ & ~ & ~ \\
C2.2 & ~ & ~ & ~ & ~ & ~ & ~ & ~ & ~ & ~ & ~ & X & ~ & ~ & ~ & ~ & ~ & ~ & ~ & ~ \\
C2.3 & ~ & ~ & ~ & ~ & ~ & ~ & ~ & ~ & ~ & ~ & X & ~ & ~ & ~ & ~ & ~ & ~ & ~ & ~ \\
C3 & ~ & ~ & ~ & ~ & ~ & ~ & ~ & ~ & ~ & ~ & X & ~ & ~ & ~ & ~ & ~ & ~ & ~ & ~ \\
C4 & ~ & ~ & ~ & ~ & ~ & ~ & ~ & ~ & ~ & ~ & X & ~ & ~ & ~ & ~ & ~ & ~ & ~ & ~ \\
C5 & ~ & ~ & ~ & ~ & ~ & ~ & ~ & ~ & ~ & ~ & X & ~ & ~ & ~ & ~ & ~ & ~ & ~ & ~ \\
C6 & ~ & ~ & ~ & ~ & ~ & ~ & ~ & ~ & ~ & ~ & X & ~ & ~ & ~ & ~ & ~ & ~ & ~ & ~ \\
C7 & ~ & ~ & ~ & ~ & ~ & ~ & ~ & ~ & ~ & ~ & X & ~ & ~ & ~ & ~ & ~ & ~ & ~ & ~ \\
C8 & ~ & ~ & ~ & ~ & ~ & ~ & ~ & ~ & ~ & ~ & X & ~ & ~ & ~ & ~ & ~ & ~ & ~ & ~ \\
C9 & ~ & ~ & ~ & ~ & ~ & ~ & ~ & ~ & ~ & ~ & X & ~ & ~ & ~ & ~ & ~ & ~ & ~ & ~ \\
C10 & ~ & ~ & ~ & ~ & ~ & ~ & ~ & ~ & ~ & ~ & ~ & ~ & X & ~ & ~ & ~ & ~ & ~ & ~ \\
C11 & ~ & ~ & ~ & ~ & ~ & ~ & ~ & ~ & ~ & ~ & ~ & ~ & ~ & ~ & ~ & ~ & ~ & ~ & X \\
C12 & ~ & ~ & ~ & ~ & ~ & ~ & ~ & ~ & ~ & ~ & ~ & ~ & ~ & ~ & ~ & ~ & ~ & ~ & X \\
C13 & ~ & ~ & ~ & ~ & ~ & ~ & ~ & ~ & ~ & ~ & ~ & ~ & X & ~ & ~ & ~ & ~ & ~ & ~ \\
C15 & ~ & ~ & ~ & ~ & ~ & ~ & ~ & ~ & ~ & ~ & ~ & ~ & ~ & ~ & ~ & ~ & X & ~ & ~ \\
C16 & ~ & ~ & ~ & ~ & ~ & ~ & ~ & ~ & ~ & ~ & ~ & ~ & ~ & ~ & ~ & ~ & X & ~ & ~ \\
C17 & ~ & ~ & ~ & ~ & ~ & ~ & ~ & ~ & ~ & ~ & ~ & ~ & ~ & X & ~ & ~ & ~ & ~ & ~ \\
C18 & ~ & ~ & ~ & ~ & ~ & ~ & ~ & ~ & ~ & ~ & ~ & ~ & ~ & X & ~ & ~ & ~ & ~ & ~ \\
C19 & ~ & ~ & ~ & ~ & ~ & ~ & ~ & ~ & ~ & ~ & ~ & ~ & ~ & ~ & X & ~ & ~ & ~ & ~ \\
C20 & ~ & ~ & ~ & ~ & ~ & ~ & ~ & ~ & ~ & ~ & ~ & ~ & ~ & ~ & X & ~ & ~ & ~ & ~ \\
C21 & ~ & ~ & ~ & ~ & ~ & ~ & ~ & ~ & ~ & ~ & ~ & ~ & ~ & ~ & ~ & X & ~ & ~ & ~ \\
C22 & ~ & ~ & ~ & ~ & ~ & ~ & ~ & ~ & ~ & ~ & ~ & ~ & ~ & ~ & ~ & X & ~ & ~ & ~ \\
\midrule
CI2 & ~ & ~ & ~ & ~ & ~ & ~ & ~ & ~ & ~ & ~ & ~ & X & ~ & ~ & ~ & ~ & ~ & ~ & ~ \\
CI3 & ~ & ~ & ~ & ~ & ~ & ~ & ~ & ~ & ~ & ~ & ~ & X & ~ & ~ & ~ & ~ & ~ & ~ & ~ \\
\midrule
F2 & X & ~ & ~ & ~ & ~ & ~ & ~ & ~ & ~ & ~ & ~ & ~ & ~ & ~ & ~ & ~ & ~ & ~ & ~ \\
F3 & X & ~ & ~ & ~ & ~ & ~ & ~ & ~ & ~ & ~ & ~ & ~ & ~ & ~ & ~ & ~ & ~ & ~ & ~ \\
F4 & ~ & ~ & ~ & ~ & ~ & ~ & X & ~ & ~ & ~ & ~ & ~ & ~ & ~ & ~ & ~ & ~ & ~ & ~ \\
F4.1 & ~ & ~ & ~ & ~ & ~ & ~ & ~ & X & ~ & ~ & ~ & ~ & ~ & ~ & ~ & ~ & ~ & ~ & ~ \\
F4.2 & ~ & ~ & X & ~ & ~ & ~ & ~ & ~ & ~ & ~ & ~ & ~ & ~ & ~ & ~ & ~ & ~ & ~ & ~ \\
F4.3 & ~ & ~ & X & ~ & ~ & ~ & ~ & ~ & ~ & ~ & ~ & ~ & ~ & ~ & ~ & ~ & ~ & ~ & ~ \\
F4.4 & ~ & ~ & X & ~ & ~ & ~ & ~ & ~ & ~ & ~ & ~ & ~ & ~ & ~ & ~ & ~ & ~ & ~ & ~ \\
F5 & ~ & ~ & ~ & X & ~ & ~ & ~ & ~ & ~ & ~ & ~ & ~ & ~ & ~ & ~ & ~ & ~ & ~ & ~ \\
F5 & ~ & ~ & ~ & ~ & X & ~ & ~ & ~ & ~ & ~ & ~ & ~ & ~ & ~ & ~ & ~ & ~ & ~ & ~ \\
F5 & ~ & ~ & ~ & ~ & ~ & X & ~ & ~ & ~ & ~ & ~ & ~ & ~ & ~ & ~ & ~ & ~ & ~ & ~ \\
F6 & ~ & ~ & ~ & ~ & ~ & ~ & ~ & ~ & ~ & ~ & ~ & ~ & ~ & ~ & ~ & ~ & ~ & X & ~ \\
F7 & ~ & X & ~ & ~ & ~ & ~ & ~ & ~ & ~ & ~ & ~ & ~ & ~ & ~ & ~ & ~ & ~ & ~ & ~ \\
F8 & ~ & X & ~ & ~ & ~ & ~ & ~ & ~ & ~ & ~ & ~ & ~ & ~ & ~ & ~ & ~ & ~ & ~ & ~ \\
F9 & ~ & ~ & ~ & ~ & ~ & ~ & X & ~ & ~ & ~ & ~ & ~ & ~ & ~ & ~ & ~ & ~ & ~ & ~ \\
\midrule
I2 & X & ~ & ~ & ~ & ~ & ~ & ~ & ~ & ~ & ~ & ~ & ~ & ~ & ~ & ~ & ~ & ~ & ~ & ~ \\
I3 & X & ~ & ~ & ~ & ~ & ~ & ~ & ~ & ~ & ~ & ~ & ~ & ~ & ~ & ~ & ~ & ~ & ~ & ~ \\
I4 & X & ~ & ~ & ~ & ~ & ~ & ~ & ~ & ~ & ~ & ~ & ~ & ~ & ~ & ~ & ~ & ~ & ~ & ~ \\
I5 & ~ & ~ & ~ & ~ & ~ & ~ & X & ~ & ~ & ~ & ~ & ~ & ~ & ~ & ~ & ~ & ~ & ~ & ~ \\
I5.1 & ~ & ~ & ~ & ~ & ~ & ~ & ~ & X & ~ & ~ & ~ & ~ & ~ & ~ & ~ & ~ & ~ & ~ & ~ \\
I5.2 & ~ & ~ & X & ~ & ~ & ~ & ~ & ~ & ~ & ~ & ~ & ~ & ~ & ~ & ~ & ~ & ~ & ~ & ~ \\
I5.3 & ~ & ~ & X & ~ & ~ & ~ & ~ & ~ & ~ & ~ & ~ & ~ & ~ & ~ & ~ & ~ & ~ & ~ & ~ \\
I5.4 & ~ & ~ & X & ~ & ~ & ~ & ~ & ~ & ~ & ~ & ~ & ~ & ~ & ~ & ~ & ~ & ~ & ~ & ~ \\
I6 & ~ & ~ & ~ & X & ~ & ~ & ~ & ~ & ~ & ~ & ~ & ~ & ~ & ~ & ~ & ~ & ~ & ~ & ~ \\
I6 & ~ & ~ & ~ & ~ & X & ~ & ~ & ~ & ~ & ~ & ~ & ~ & ~ & ~ & ~ & ~ & ~ & ~ & ~ \\
I6 & ~ & ~ & ~ & ~ & ~ & X & ~ & ~ & ~ & ~ & ~ & ~ & ~ & ~ & ~ & ~ & ~ & ~ & ~ \\
I7 & ~ & X & ~ & ~ & ~ & ~ & ~ & ~ & ~ & ~ & ~ & ~ & ~ & ~ & ~ & ~ & ~ & ~ & ~ \\
I8 & ~ & ~ & ~ & ~ & ~ & ~ & ~ & ~ & ~ & ~ & ~ & ~ & ~ & ~ & ~ & ~ & X & ~ & ~ \\
I9 & ~ & X & ~ & ~ & ~ & ~ & ~ & ~ & ~ & ~ & ~ & ~ & ~ & ~ & ~ & ~ & ~ & ~ & ~ \\
I10 & ~ & ~ & ~ & ~ & ~ & ~ & X & ~ & ~ & ~ & ~ & ~ & ~ & ~ & ~ & ~ & ~ & ~ & ~ \\
\midrule
ML2 & ~ & ~ & ~ & ~ & ~ & ~ & X & ~ & ~ & ~ & ~ & ~ & ~ & ~ & ~ & ~ & ~ & ~ & ~ \\
ML2.1 & ~ & ~ & ~ & ~ & ~ & ~ & ~ & X & ~ & ~ & ~ & ~ & ~ & ~ & ~ & ~ & ~ & ~ & ~ \\
ML3 & ~ & ~ & ~ & X & ~ & ~ & ~ & ~ & ~ & ~ & ~ & ~ & ~ & ~ & ~ & ~ & ~ & ~ & ~ \\
ML3 & ~ & ~ & ~ & ~ & X & ~ & ~ & ~ & ~ & ~ & ~ & ~ & ~ & ~ & ~ & ~ & ~ & ~ & ~ \\
ML3 & ~ & ~ & ~ & ~ & ~ & X & ~ & ~ & ~ & ~ & ~ & ~ & ~ & ~ & ~ & ~ & ~ & ~ & ~ \\
ML5 & ~ & ~ & ~ & ~ & ~ & ~ & X & ~ & ~ & ~ & ~ & ~ & ~ & ~ & ~ & ~ & ~ & ~ & ~ \\
ML5.1 & ~ & ~ & ~ & ~ & ~ & ~ & ~ & X & ~ & ~ & ~ & ~ & ~ & ~ & ~ & ~ & ~ & ~ & ~ \\
ML6 & ~ & ~ & ~ & X & ~ & ~ & ~ & ~ & ~ & ~ & ~ & ~ & ~ & ~ & ~ & ~ & ~ & ~ & ~ \\
ML6 & ~ & ~ & ~ & ~ & X & ~ & ~ & ~ & ~ & ~ & ~ & ~ & ~ & ~ & ~ & ~ & ~ & ~ & ~ \\
ML6 & ~ & ~ & ~ & ~ & ~ & X & ~ & ~ & ~ & ~ & ~ & ~ & ~ & ~ & ~ & ~ & ~ & ~ & ~ \\
\midrule
O2 & ~ & ~ & ~ & ~ & ~ & ~ & ~ & ~ & X & ~ & ~ & ~ & ~ & ~ & ~ & ~ & ~ & ~ & ~ \\
O3 & ~ & ~ & ~ & ~ & ~ & ~ & X & ~ & ~ & ~ & ~ & ~ & ~ & ~ & ~ & ~ & ~ & ~ & ~ \\
O3.1 & ~ & ~ & ~ & ~ & ~ & ~ & ~ & X & ~ & ~ & ~ & ~ & ~ & ~ & ~ & ~ & ~ & ~ & ~ \\
O4 & ~ & ~ & ~ & X & ~ & ~ & ~ & ~ & ~ & ~ & ~ & ~ & ~ & ~ & ~ & ~ & ~ & ~ & ~ \\
O4 & ~ & ~ & ~ & ~ & X & ~ & ~ & ~ & ~ & ~ & ~ & ~ & ~ & ~ & ~ & ~ & ~ & ~ & ~ \\
O4 & ~ & ~ & ~ & ~ & ~ & X & ~ & ~ & ~ & ~ & ~ & ~ & ~ & ~ & ~ & ~ & ~ & ~ & ~ \\
O5 & ~ & ~ & ~ & ~ & ~ & ~ & ~ & ~ & ~ & ~ & ~ & ~ & ~ & ~ & ~ & ~ & ~ & X & ~ \\
\midrule
R2 & ~ & ~ & ~ & ~ & ~ & ~ & ~ & ~ & X & ~ & ~ & ~ & ~ & ~ & ~ & ~ & ~ & ~ & ~ \\
R3 & ~ & ~ & ~ & ~ & ~ & ~ & X & ~ & ~ & ~ & ~ & ~ & ~ & ~ & ~ & ~ & ~ & ~ & ~ \\
R4 & ~ & ~ & ~ & ~ & ~ & ~ & ~ & X & ~ & ~ & ~ & ~ & ~ & ~ & ~ & ~ & ~ & ~ & ~ \\
R5 & ~ & ~ & ~ & ~ & ~ & ~ & X & ~ & ~ & ~ & ~ & ~ & ~ & ~ & ~ & ~ & ~ & ~ & ~ \\
R6 & ~ & ~ & ~ & ~ & ~ & ~ & X & ~ & ~ & ~ & ~ & ~ & ~ & ~ & ~ & ~ & ~ & ~ & ~ \\
R7 & ~ & ~ & ~ & ~ & ~ & ~ & ~ & ~ & ~ & X & ~ & ~ & ~ & ~ & ~ & ~ & ~ & ~ & ~ \\
R8 & ~ & ~ & ~ & X & ~ & ~ & ~ & ~ & ~ & ~ & ~ & ~ & ~ & ~ & ~ & ~ & ~ & ~ & ~ \\
R8 & ~ & ~ & ~ & ~ & X & ~ & ~ & ~ & ~ & ~ & ~ & ~ & ~ & ~ & ~ & ~ & ~ & ~ & ~ \\
R8 & ~ & ~ & ~ & ~ & ~ & X & ~ & ~ & ~ & ~ & ~ & ~ & ~ & ~ & ~ & ~ & ~ & ~ & ~ \\
R9 & ~ & ~ & ~ & ~ & ~ & ~ & ~ & ~ & ~ & ~ & ~ & ~ & ~ & ~ & ~ & ~ & ~ & ~ & X \\
R10 & ~ & ~ & ~ & ~ & ~ & ~ & ~ & ~ & ~ & ~ & ~ & ~ & ~ & ~ & ~ & ~ & ~ & ~ & X \\
R11 & ~ & ~ & ~ & ~ & ~ & ~ & ~ & ~ & ~ & X & ~ & ~ & ~ & ~ & ~ & ~ & ~ & ~ & ~ \\
\end{longtable}

    %%=============================================%%
    %% For submissions to Nature Portfolio Journals %%
    %% please use the heading ``Extended Data''.   %%
    %%=============================================%%

    %%=============================================================%%
    %% Sample for another appendix section			       %%
    %%=============================================================%%

    %% \section{Example of another appendix section}\label{secA2}%
    %% Appendices may be used for helpful, supporting or essential material that would otherwise 
    %% clutter, break up or be distracting to the text. Appendices can consist of sections, figures, 
    %% tables and equations etc.

\end{appendices}

\end{document}